\documentclass[runningheads]{llncs}

\usepackage{amsmath}
\usepackage{microtype}

\usepackage{quiver, stmaryrd}
\usepackage{graphicx}
\usepackage{xcolor}
\usepackage[capitalise]{cleveref}
\usepackage{listings}
\usepackage{booktabs}
\usepackage{subcaption}
\usepackage{soul}
\usepackage{xhfill}
\makeatletter
\@ifundefined{lhs2tex.lhs2tex.sty.read}%
  {\@namedef{lhs2tex.lhs2tex.sty.read}{}%
   \newcommand\SkipToFmtEnd{}%
   \newcommand\EndFmtInput{}%
   \long\def\SkipToFmtEnd#1\EndFmtInput{}%
  }\SkipToFmtEnd

\newcommand\ReadOnlyOnce[1]{\@ifundefined{#1}{\@namedef{#1}{}}\SkipToFmtEnd}
\usepackage{amstext}
\usepackage{amssymb}
\usepackage{stmaryrd}
\DeclareFontFamily{OT1}{cmtex}{}
\DeclareFontShape{OT1}{cmtex}{m}{n}
  {<5><6><7><8>cmtex8
   <9>cmtex9
   <10><10.95><12><14.4><17.28><20.74><24.88>cmtex10}{}
\DeclareFontShape{OT1}{cmtex}{m}{it}
  {<-> ssub * cmtt/m/it}{}

\DeclareFontShape{OT1}{cmtt}{bx}{n}
  {<5><6><7><8>cmtt8
   <9>cmbtt9
   <10><10.95><12><14.4><17.28><20.74><24.88>cmbtt10}{}
\DeclareFontShape{OT1}{cmtex}{bx}{n}
  {<-> ssub * cmtt/bx/n}{}

\newcommand{\Conid}[1]{\mathit{#1}}
\newcommand{\Varid}[1]{\mathit{#1}}
\newcommand{\anonymous}{\kern0.06em \vbox{\hrule\@width.5em}}
\newcommand{\plus}{\mathbin{+\!\!\!+}}
\newcommand{\bind}{\mathbin{>\!\!\!>\mkern-6.7mu=}}

\renewcommand{\geq}{\geqslant}
\usepackage{polytable}

\@ifundefined{mathindent}%
  {\newdimen\mathindent\mathindent\leftmargini}%
  {}%

\def\resethooks{%
  \global\let\SaveRestoreHook\empty
  \global\let\ColumnHook\empty}
\newcommand*{\savecolumns}[1][default]%
  {\g@addto@macro\SaveRestoreHook{\savecolumns[#1]}}
\newcommand*{\restorecolumns}[1][default]%
  {\g@addto@macro\SaveRestoreHook{\restorecolumns[#1]}}
\newcommand*{\aligncolumn}[2]%
  {\g@addto@macro\ColumnHook{\column{#1}{#2}}}

\resethooks

\newcommand{\onelinecommentchars}{\quad-{}- }
\newcommand{\commentbeginchars}{\enskip\{-}
\newcommand{\commentendchars}{-\}\enskip}

\newcommand{\visiblecomments}{%
  \let\onelinecomment=\onelinecommentchars
  \let\commentbegin=\commentbeginchars
  \let\commentend=\commentendchars}

\newcommand{\invisiblecomments}{%
  \let\onelinecomment=\empty
  \let\commentbegin=\empty
  \let\commentend=\empty}

\visiblecomments

\newlength{\blanklineskip}
\newcommand{\hsindent}[1]{\quad}
\let\hspre\empty
\let\hspost\empty

\EndFmtInput
\makeatother
\ReadOnlyOnce{polycode.fmt}%
\makeatletter

\newcommand{\hsnewpar}[1]%
  {{\parskip=0pt\parindent=0pt\par\vskip #1\noindent}}

\newcommand{\hscodestyle}{}

\newcommand{\sethscode}[1]%
  {\expandafter\let\expandafter\hscode\csname #1\endcsname
   \expandafter\let\expandafter\endhscode\csname end#1\endcsname}

  {\par\noindent
   \advance\leftskip\mathindent
   \hscodestyle
   \let\\=\@normalcr
   \let\hspre\(\let\hspost\)%
   \pboxed}%
  {\endpboxed\)%
   \par\noindent
   \ignorespacesafterend}

  {\hsnewpar\abovedisplayskip
   \advance\leftskip\mathindent
   \hscodestyle
   \let\hspre\(\let\hspost\)%
   \pboxed}%
  {\endpboxed%
   \hsnewpar\belowdisplayskip
   \ignorespacesafterend}

  {\hsnewpar\abovedisplayskip
   \advance\leftskip\mathindent
   \hscodestyle
   \let\\=\@normalcr
   \(\pboxed}%
  {\endpboxed\)%
   \hsnewpar\belowdisplayskip
   \ignorespacesafterend}

\newcommand{\plainhs}{\sethscode{plainhscode}}

\plainhs

  {\hsnewpar\abovedisplayskip
   \advance\leftskip\mathindent
   \hscodestyle
   \let\\=\@normalcr
   \(\parray}%
  {\endparray\)%
   \hsnewpar\belowdisplayskip
   \ignorespacesafterend}

  {\parray}{\endparray}

  {\(\parray}{\endparray\)}

\def\codeframewidth{\arrayrulewidth}
\RequirePackage{calc}

  {\parskip=\abovedisplayskip\par\noindent
   \hscodestyle
   \arrayrulewidth=\codeframewidth
   \tabular{@{}|p{\linewidth-2\arraycolsep-2\arrayrulewidth-2pt}|@{}}%
   \hline\framedhslinecorrect\\{-1.5ex}%
   \let\endoflinesave=\\
   \let\\=\@normalcr
   \(\pboxed}%
  {\endpboxed\)%
   \framedhslinecorrect\endoflinesave{.5ex}\hline
   \endtabular
   \parskip=\belowdisplayskip\par\noindent
   \ignorespacesafterend}

\newcommand{\framedhslinecorrect}[2]%
  {#1[#2]}

  {\(\def\column##1##2{}%
   \let\>\undefined\let\<\undefined\let\\\undefined
   \newcommand\>[1][]{}\newcommand\<[1][]{}\newcommand\\[1][]{}%
   \def\fromto##1##2##3{##3}%
   }{\) }%

  {\let\orighscode=\hscode
   \let\origendhscode=\endhscode
   \def\endhscode{\def\hscode{\endgroup\def\@currenvir{hscode}\\}\begingroup}
   \orighscode\def\hscode{\endgroup\def\@currenvir{hscode}}}%
  {\origendhscode
   \global\let\hscode=\orighscode
   \global\let\endhscode=\origendhscode}%

\makeatother
\EndFmtInput
\ReadOnlyOnce{forall.fmt}%
\makeatletter

\let\HaskellResetHook\empty
\newcommand*{\AtHaskellReset}[1]{%
  \g@addto@macro\HaskellResetHook{#1}}
\newcommand*{\HaskellReset}{\HaskellResetHook}

\newcommand\hsforall{\global\let\hsdot=\hsperiodonce}
\newcommand*\hsperiodonce[2]{#2\global\let\hsdot=\hscompose}
\newcommand*\hscompose[2]{#1}

\AtHaskellReset{\global\let\hsdot=\hscompose}

\HaskellReset

\makeatother
\EndFmtInput

\crefformat{section}{#2§#1#3}

\begin{document}
\title{Latent Effects for\\ Reusable Language Components}
\subtitle{Extended Version}
%
%

 \author{Birthe van den Berg\inst{1}\orcidID{0000-0002-0088-9546} \and
 Tom Schrijvers \inst{1}\orcidID{0000-0001-8771-5559} \and
 Casper Bach Poulsen\inst{2}\orcidID{0000-0003-0622-7639} \and
 Nicolas~Wu\inst{3}\orcidID{0000-0002-4161-985X}}
\authorrunning{van den Berg et al.}
%
\institute{KU Leuven, Belgium \email{\{birthe.vandenberg,tom.schrijvers\}@kuleuven.be} \and
Delft University of Technology, The Netherlands \email{c.b.poulsen@tudelft.nl} \and
Imperial College London, UK \email{n.wu@imperial.ac.uk}
}

\maketitle              
\begin{abstract}
  The development of programming languages can be quite complicated and
  costly. Hence, much effort has been devoted to the modular definition of
  language features that can be reused in various combinations to define new
  languages and experiment with their semantics. A notable outcome of these
  efforts is the algebra-based ``datatypes \`a la carte'' (DTC) approach.  When
  combined with algebraic effects, DTC can model a wide range of common
  language features.  Unfortunately, the current state of the art does not cover
  modular definitions of advanced control-flow mechanisms that defer execution
  to an appropriate point, such as call-by-name and call-by-need evaluation, as
  well as \mbox{(multi-)staging}.

  This paper defines \emph{latent effects}, a generic class of such
  control-flow mechanisms. We demonstrate how function abstractions, lazy
  computations and a MetaML-like staging can all be expressed in a \emph{modular}
  fashion using latent effects, and how they can be combined in various ways
  to obtain complex semantics.
  We provide a full Haskell implementation of our 
  effects and handlers with a range of examples.

%
%

\end{abstract}

\section{Introduction}
\label{introduction}


\newcommand{\lam}{\mathtt{fun}}
\newcommand{\call}{\mathtt{call}}
\newcommand{\var}{\mathtt{var}}
\newcommand{\rref}{\mathtt{ref}}
\newcommand{\deref}{\mathtt{deref}}
\newcommand{\rraise}{\mathtt{raise}}
\newcommand{\catch}{\mathtt{catch}}

\newcommand{\abs}{\mathsf{abs}}
\newcommand{\app}{\mathsf{app}}
\newcommand{\lookup}{\mathsf{lookup}}
\newcommand{\alloc}{\mathsf{alloc}}
\newcommand{\fetch}{\mathsf{fetch}}
\newcommand{\throw}{\mathsf{throw}}
\newcommand{\exchdl}{\mathsf{exchdl}}

Modern programming languages, be they general purpose or domain-specific, can be built in a flexible manner by composing simple, off-the-shelf language components.
It is attractive to build languages in this way as it is useful to study language components in isolation.
Furthermore, it reduces the cost of developing new and improved programming languages.
Indeed, reducing the effort of building languages to the effort of composing off-the-shelf language components for features such as function abstraction, exceptions or mutable state, is likely to enable language designers with limited
resources or expertise---e.g., domain experts---to build their own languages.
Providing a modular definition for these advanced language features enables a 
more widespread use of them, especially in the development of domain-specific 
languages (DSLs).

To build programming languages from reusable components, we need a framework for defining those components.
Two promising techniques for such a framework are \emph{Data Types \`{a} la Carte}~\cite{Swierstra08} (DTC) and \emph{Algebraic Effects \& Handlers}~\cite{PlotkinP09} (AE\&H).
Using these, language interpreters can be implemented in two steps:
\begin{equation*}
\ensuremath{\Conid{Syntax}} \xrightarrow{\ensuremath{\Varid{denote}}}
\ensuremath{\Conid{Effects}} \xrightarrow{\ensuremath{\Varid{handle}}}
\ensuremath{\Conid{Result}}
\end{equation*}%
The first step defines a \ensuremath{\Varid{denote}} function that maps the syntax of an object language onto effectful (\emph{monadic}~\cite{Moggi89a}) operations.
By implementing \ensuremath{\Varid{denote}} using DTC, we can seamlessly compose \ensuremath{\Varid{denote}} functions from isolated cases for different \ensuremath{\Conid{Syntax}} constructors.

In the second step, \ensuremath{\Varid{handle}} defines the semantics of effectful operations and their effect interactions.
Using AE\&H allows cases of \ensuremath{\Varid{denote}} to be defined in an effect polymorphic way; i.e., a case maps to a monad that has \emph{at least} the relevant operations, and possibly more.
Furthermore, we can define \ensuremath{\Varid{handle}} functions for different effects in isolation, and seamlessly compose them by nesting handlers.
Thus, DTC+AE\&H provides a powerful framework for composing programming languages from isolated components.

However, not all language fragments have an obvious modular definition in terms of AE\&H.
Traditional algebraic effects and scoped effects~\cite{WuSH14} are baking in assumptions about control-flow and data-flow.
These assumptions get in the way of expressing a modular semantics for some language features.
In particular, control-flow mechanisms that defer execution, such as lambda abstractions with effectful bodies, lazy evaluation strategies, or (multi-)staging, are neither \emph{algebraic}~\cite{PlotkinP03} nor scoped.
This implies that the only way to implement these mechanisms is by means of sophisticated encodings that are often relatively low-level and non-modular.
As such, these control-flow mechanisms present a severe challenge for allowing non-experts to build programming languages by composing off-the-shelf components.

This paper presents a novel form of effects and handlers that addresses this challenge: 
\emph{latent effects and handlers}.
Latent effect handlers defer the running of side effects, such that they are ``latent'' until handlers for all effects have been applied.
This notion of effects can be used to directly define handlers for language components that defer execution, such as lambda abstractions with effectful bodies, lazy evaluation, and (multi-)staging.

After introducing the background (Section \ref{motivation}), our main contributions are:
\begin{itemize}
\item We introduce \emph{latent effect trees}, implemented using Haskell~(Section \ref{latent_effects}).

\item We show how to encode lambda abstraction and application in terms of
latent effects. We illustrate how this allows us to define lambda abstraction
with effectful bodies as a composable language component (Section \ref{latent_effects}).

%

\item We illustrate how to compose languages from reusable components using
latent effects by showing how to encode lazy evaluation (call-by-need and
call-by-name), and MetaML-like multi-staging (Section \ref{case_studies}).

\item We provide an effect library\footnote{\url{https://github.com/birthevdb/Latent-Effect-and-Handlers.git}.} with syntax and semantics in Haskell for various simple and advanced language features. These features can be used as isolated components to construct languages (Section \ref{app:effect-lib}).

\end{itemize}
Finally, we discuss related work (Section \ref{related_work}) and conclude (Section \ref{conclusion}).

\section{Background and Motivation}
\label{motivation}


\newcommand{\MetaLanguage}{Haskell}
\newcommand{\OtherMetaLanguage}{Agda}

This section summarizes the state of the art in modular language
definition by means of DTC and AE\&H,
discusses the challenges and problems associated with treating lambdas in this
setting, and
sketches how our new latent effects enable the integration
of modular forms of advanced control-flow effects.


  
\subsection{Modular Syntax and Semantics with Data Types \`{a} la Carte}
\label{DTC}
Data types \`{a} la Carte~\cite{Swierstra08} (DTC) solves the core problem of assembling
language definitions from isolated components.
It solves this for both syntax and semantics to be composed in a modular way.
For an accessible introduction to DTC, we refer to \cite{Swierstra08}; we summarize the
key points here.

Firstly, the abstract syntax of a language is modularized into 
the syntax of each isolated feature. 
This is achieved by defining a recursive type \ensuremath{\Conid{Syntax}\;\Conid{S}} of
ASTs in terms of the shape \ensuremath{\Conid{S}} of its nodes. 
This shape \ensuremath{\Conid{S}} can be composed out of shapes \ensuremath{\Varid{S}_{1}},\ldots,\ensuremath{\Varid{S}_{n}} of the individual 
features: \ensuremath{\Conid{S}\mathrel{=}\Varid{S}_{1}\mathbin{+}\ldots\mathbin{+}\Varid{S}_{n}}.

Secondly, the semantic function \ensuremath{\Varid{denote}\mathbin{::}\Conid{Syntax}\;\Conid{S}\to \Conid{M}} that maps ASTs 
\ensuremath{\Conid{Syntax}\;\Conid{S}} to their meaning \ensuremath{\Conid{M}} is modularized into the separate
syntactic maps of the individual features. This is done by 
parameterizing the recursive semantic map of the AST with the semantic
mapping of the individual nodes, \ensuremath{\Varid{denote}\mathrel{=}\Varid{fold}\;\Varid{denote}_{S}} where \ensuremath{\Varid{denote}_{S}\mathbin{::}\Conid{S}\;\Conid{M}\to \Conid{M}} is composed out of the semantic maps: 

\noindent
\begin{minipage}[t][][t]{0.4\textwidth}
\vspace{-\abovedisplayskip}\begin{hscode}\SaveRestoreHook
\column{B}{@{}>{\hspre}l<{\hspost}@{}}%
\column{3}{@{}>{\hspre}l<{\hspost}@{}}%
\column{E}{@{}>{\hspre}l<{\hspost}@{}}%
\>[3]{}\Varid{denote}_{S_1}\mathbin{::}\Varid{S}_{1}\;\Conid{M}\to \Conid{M}{}\<[E]%
\\
\>[3]{}\Varid{denote}_{S_n}\mathbin{::}\Varid{S}_{n}\;\Conid{M}\to \Conid{M}{}\<[E]%
\ColumnHook
\end{hscode}\resethooks
\end{minipage}%
\begin{minipage}[t][][t]{0.6\textwidth}
\vspace{-\abovedisplayskip}\begin{hscode}\SaveRestoreHook
\column{B}{@{}>{\hspre}l<{\hspost}@{}}%
\column{3}{@{}>{\hspre}l<{\hspost}@{}}%
\column{E}{@{}>{\hspre}l<{\hspost}@{}}%
\>[3]{}\Varid{denote}_{S}\mathrel{=}\Varid{denote}_{S_1}\mathbin{+}\ldots\mathbin{+}\Varid{denote}_{S_n}{}\<[E]%
\ColumnHook
\end{hscode}\resethooks
\end{minipage}


\noindent
This modular approach affords great flexibility to quickly assemble a range of
different languages and to explore language design: features can be added or removed, and 
their semantic maps
can be changed. 
We use the DTC approach extensively when defining a library of effects in 
Section \ref{library-of-effects}.

Unfortunately, this approach comes with a serious limitation: to be able to
combine the semantic maps of different features, they must agree on the
semantic domain \ensuremath{\Conid{M}}.  This often prohibits the unanticipated combinations of
features, even due to small differences in their semantic domain, such as one
feature requiring access to an environment or store that the other feature does
not expect. 

Moggi~\cite{Moggi89a} observed that many of the differences in semantic domains of common
language features, usually called (side-)effects, can be captured in the
category-theoretical pattern of \emph{monads}.
Although not all monads compose, the state
of the art in modular monads, \emph{algebraic
effects and handlers} (AE\&H)~\cite{PlotkinP09}, is well-aligned
with the DTC approach of modular language features.

In fact, AE\&H tackles the problem of modularly defining a monad in much the
same way DTC tackles the problem of modularly defining a language. Indeed, the
API of the monad (its ``syntax'') is modularized into separate effects; the
category-theoretical concept of a free monad plays the role of an abstract
syntax of API calls. A modular meaning is assigned to the free monad by means of
semantic maps for the separate effects. The key difference with DTC is that the
different effects do not all have to be mapped to the same semantics.
Instead, by means of ``modular carriers'' their semantics can be layered~\cite{SchrijversPWJ19}.

Whereas AE\&H is used to define (simpler) language features, we use latent effects and handlers,
which we introduce in Section \ref{latent_effects},
to modularly define more complex language features.

\subsection{Advanced, Non-Modular Control-Flow Effects}
\label{advanced-non-modular-effects}
\label{non-modular-control-flow}

The AE\&H literature shows how to express a range of different control-flow
effects such as exceptions, non-determinism and even
call-with-current-continuation. However, as traditional effects and handlers rely
on assumptions about data-flow and control-flow, more advanced and complicated
control-flow effects are missing, such as call-by-name and call-by-need
evaluation or multi-staging. These features typically defer execution and are \emph{non-algebraic}.

An operation is \emph{algebraic} when it meets the following requirements: 
\begin{enumerate}
  \item It can be expressed as \ensuremath{\Varid{op}\mathbin{::}\Conid{D}\to \forall \Varid{a}\hsforall \hsdot{\circ }{.}\Conid{M}\;\Varid{a}\to \mathbin{...}\to \Conid{M}\;\Varid{a}} where \ensuremath{\Conid{D}} represents the non-computational parameters that the operation has, \ensuremath{\Conid{M}} is a monad with algebraic operations, and each \ensuremath{\Conid{M}\;\Varid{a}} typed parameter represents a possible continuation of the operation.

  \item It satisfies the \emph{algebraicity property}, which states that,
  no matter which possible continuation \ensuremath{\Varid{m}_\Varid{i}} we take, 
  the continuation \emph{immediately} transfers control to the current continuation \ensuremath{\Varid{k}}.
\begin{hscode}\SaveRestoreHook
\column{B}{@{}>{\hspre}l<{\hspost}@{}}%
\column{3}{@{}>{\hspre}l<{\hspost}@{}}%
\column{E}{@{}>{\hspre}l<{\hspost}@{}}%
\>[3]{}(\Varid{op}\;\Varid{d}\;\Varid{m}_{1}\;\ldots\;\Varid{m}_\Varid{n})\bind \Varid{k}\equiv \Varid{op}\;\Varid{d}\;(\Varid{m}_{1}\bind \Varid{k})\;\ldots\;(\Varid{m}_\Varid{n}\bind \Varid{k}){}\<[E]%
\ColumnHook
\end{hscode}\resethooks
\end{enumerate}

Although many operations are algebraic,
there are many common control-flow operations that are not.
For instance, \ensuremath{\Varid{catch}\mathbin{::}\forall \Varid{a}\hsforall \hsdot{\circ }{.}\Conid{M}\;\Varid{a}\to \Conid{M}\;\Varid{a}\to \Conid{M}\;\Varid{a}} is an operation that executes the computation in the first argument, and if that computation raises an exception, proceeds by executing the second argument.
The \ensuremath{\Varid{catch}} operation is not algebraic; i.e.,\begin{hscode}\SaveRestoreHook
\column{B}{@{}>{\hspre}l<{\hspost}@{}}%
\column{3}{@{}>{\hspre}l<{\hspost}@{}}%
\column{E}{@{}>{\hspre}l<{\hspost}@{}}%
\>[3]{}(\Varid{catch}\;\Varid{m}_{1}\;\Varid{m}_{2})\bind \Varid{k}\not\equiv \Varid{catch}\;(\Varid{m}_{1}\bind \Varid{k})\;(\Varid{m}_{2}\bind \Varid{k}){}\<[E]%
\ColumnHook
\end{hscode}\resethooks
since exceptions raised during evaluation of \ensuremath{\Varid{k}} should \emph{not} be handled by \ensuremath{\Varid{m}_{2}}.

The lack of support for control-flow effects such as exception handling, motivated the development of \emph{scoped
effects}~\cite{WuSH14}.
An operation is \emph{scoped} when it can be expressed as having the following type:
{\small\begin{hscode}\SaveRestoreHook
\column{B}{@{}>{\hspre}l<{\hspost}@{}}%
\column{3}{@{}>{\hspre}l<{\hspost}@{}}%
\column{E}{@{}>{\hspre}l<{\hspost}@{}}%
\>[3]{}\Varid{op}\mathbin{::}\Conid{D}\to \forall \Varid{a}\hsforall \hsdot{\circ }{.}\Conid{M}\;\Varid{a}\to \mathbin{...}\to \Conid{M}\;\Varid{a}\to \forall \Varid{b}\hsforall \hsdot{\circ }{.}(\Varid{a}\to \Conid{M}\;\Varid{b})\to \mathbin{...}\to (\Varid{a}\to \Conid{M}\;\Varid{b})\to \Conid{M}\;\Varid{b}{}\<[E]%
\ColumnHook
\end{hscode}\resethooks
}%

\noindent
The universal quantification over \ensuremath{\Varid{a}} restricts data flow: for a given operation
\ensuremath{\Varid{op}\;\Varid{d}\;\Varid{m}_{1}\mathbin{...}\Varid{m}_\Varid{n}\;\Varid{k}_{1}\mathbin{...}k_m}, it is \emph{only} the possible continuations \ensuremath{\Varid{k}_{1}\mathbin{...}k_m} that can inspect values yielded by computations
\ensuremath{m_i}. 
The allowable destinations of data produced by the computation are restricted
to those determined by the operation.
The \ensuremath{\Varid{catch}} operation is compatible with this restriction: it can be
implemented as a scoped operation \ensuremath{\Varid{catch'}\mathbin{::}\forall \Varid{a}\hsforall \hsdot{\circ }{.}\Conid{M}\;\Varid{a}\to \Conid{M}\;\Varid{a}\to \forall \Varid{b}\hsforall \hsdot{\circ }{.}(\Varid{a}\to \Conid{M}\;\Varid{b})\to \Conid{M}\;\Varid{b}}.

However, this pattern does not apply to more advanced control-flow effects for
which the data produced by a computation can be used outside of the operation.
For example, lambda abstractions delay the execution of
computations in the body of a function until the function is applied (or not,
depending on dynamic control- and data flow).
To support such deferred execution, the return type \ensuremath{\Conid{V}}
in the body of a lambda abstraction \ensuremath{\Varid{abstr}\mathbin{::}\Conid{String}\to \Conid{M}\;\Conid{V}\to \Conid{M}\;\Conid{V}} is
\emph{not} universally quantified.  Thus, \ensuremath{\Varid{abstr}} is not a scoped operation.  It
is also not algebraic, as the equation \ensuremath{\Varid{abstr}\;\Varid{x}\;\Varid{m}\bind \Varid{k}\equiv \Varid{abstr}\;\Varid{x}\;(\Varid{m}\bind \Varid{k})}
would cause \ensuremath{\Varid{k}} to be (wrongly) deferred, and could cause free variables in \ensuremath{\Varid{k}}
to be captured.  Other control-flow effects, such as call-by-need and
call-by-name evaluation and multi-staging annotations for (dynamically) staging
and unstaging code, defer execution in a similar way, and are similarly neither
scoped nor algebraic.

It is theoretically possible to define the control-flow effects discussed above, by making the control flow of all operations \emph{explicit}; e.g., by writing interpreters in continuation-passing style (CPS).
However, the relatively low-level nature of CPS and its non-modular style, make this approach fall short of our goal of composing languages from simple, off-the-shelf components.



\subsection{Our Approach: Latent Effects}
\label{our-approach}

We provide modular support for advanced control-flow effects such as function
abstraction, with its different evaluation semantics, and staging. Our solution
generalizes the approach of DTC and AE\&H outlined above. In fact, it
does not require any changes to the DTC approach for modular abstract syntax
trees and modular semantic mapping. It only impacts the second part of the
pipeline, replacing AE\&H with a more general notion of \emph{latent
effects and handlers} (LE\&H).

LE\&H is based on a different, more sophisticated structure than AE\&H's free
monad. This structure supports non-atomic operations (e.g., function
abstraction, thunking, quoting) that contain or delimit computations whose
execution may be deferred. Also, the layered handling is different. The idea is
still the same, to replace bit by bit the structure of the tree by its meaning.
Yet, while AE\&H grows the meaning around the shrinking tree, LE\&H grows
little ``pockets of meaning'' around the individual nodes remaining in the tree, and not just
around the root. The latter supports deferred effects because later handlers
can still re-arrange the semantic pockets created by earlier handlers.

LE\&H are the first to modularly express advanced
control-flow effects, such as staging and lambda abstractions, and provide
different handlers, e.g., for call-by-name and call-by-need evaluation.
Moreover, they combine with existing algebraic effects to express varying semantics
for  a
large range of languages.

\section{Latent Effects}
\label{latent_effects}

This section presents \emph{latent effects}.
Latent effects generalize algebraic effects to include control-flow mechanisms that defer computation.
As our running example we use lambda  abstraction, which---as discussed in
Section \ref{advanced-non-modular-effects}---is neither an algebraic nor a scoped
operation. We show that it can be defined as a latent effect.
We start from a non-modular version that we refine in two steps. First
we add  support for modular signatures, and then support for modular
handlers.

\subsection{Non-Modular Definition of Lambda Abstraction}

First we provide a non-modular definition of the lambda abstraction effect.

\paragraph{Monadic Syntax Tree}
The type \ensuremath{\Conid{LC}\;\Varid{v}\;\Varid{a}} is a non-modular monadic syntax tree that
supports three primitive operations for a de Bruijn indexed $\lambda$-calculus.
\begin{hscode}\SaveRestoreHook
\column{B}{@{}>{\hspre}l<{\hspost}@{}}%
\column{3}{@{}>{\hspre}l<{\hspost}@{}}%
\column{11}{@{}>{\hspre}l<{\hspost}@{}}%
\column{E}{@{}>{\hspre}l<{\hspost}@{}}%
\>[B]{}\mathbf{data}\;\Conid{LC}\;\Varid{v}\;\Varid{a}\;\mathbf{where}{}\<[E]%
\\
\>[B]{}\hsindent{3}{}\<[3]%
\>[3]{}\Conid{Return}{}\<[11]%
\>[11]{}\mathbin{::}\Varid{a}\to \Conid{LC}\;\Varid{v}\;\Varid{a}{}\<[E]%
\\
\>[B]{}\hsindent{3}{}\<[3]%
\>[3]{}\Conid{Var}{}\<[11]%
\>[11]{}\mathbin{::}\Conid{Int}\to (\Varid{v}\to \Conid{LC}\;\Varid{v}\;\Varid{a})\to \Conid{LC}\;\Varid{v}\;\Varid{a}{}\<[E]%
\\
\>[B]{}\hsindent{3}{}\<[3]%
\>[3]{}\Conid{App}{}\<[11]%
\>[11]{}\mathbin{::}\Varid{v}\to \Varid{v}\to (\Varid{v}\to \Conid{LC}\;\Varid{v}\;\Varid{a})\to \Conid{LC}\;\Varid{v}\;\Varid{a}{}\<[E]%
\\
\>[B]{}\hsindent{3}{}\<[3]%
\>[3]{}\Conid{Abs}{}\<[11]%
\>[11]{}\mathbin{::}\Conid{LC}\;\Varid{v}\;\Varid{v}\to (\Varid{v}\to \Conid{LC}\;\Varid{v}\;\Varid{a})\to \Conid{LC}\;\Varid{v}\;\Varid{a}{}\<[E]%
\ColumnHook
\end{hscode}\resethooks
Here, the \ensuremath{\Varid{v}} of \ensuremath{\Conid{LC}\;\Varid{v}\;\Varid{a}} is a value type parameter, and \ensuremath{\Varid{a}} represents the return type of the computation.
Thus \ensuremath{\Conid{Return}\;\Varid{x}} is a trivial computation that returns \ensuremath{\Varid{x}}.  \ensuremath{\Conid{Var}\;\Varid{i}\;\Varid{k}}
retrieves the value of type \ensuremath{\Varid{v}} associated with the \ensuremath{\Varid{i}}th variable and passes
it to the continuation \ensuremath{\Varid{k}}. The application \ensuremath{\Conid{App}\;\Varid{v}_{1}\;\Varid{v}_{2}\;\Varid{k}} applies the value
\ensuremath{\Varid{v}_{1}} to the value \ensuremath{\Varid{v}_{2}} and passes the result to the continuation. Finally, \ensuremath{\Conid{Abs}\;\Varid{e}\;\Varid{k}} builds a closure value out of the function body \ensuremath{\Varid{e}} and passes it to the
continuation.

For example, we can represent the lambda expression \ensuremath{(\lambda \Varid{x}\to \Varid{x})\;\mathrm{1}} as the \ensuremath{\Conid{LC}}
expression \ensuremath{\Conid{Abs}\;(\Conid{Var}\;\mathrm{0}\;\Conid{Return})\;(\lambda \Varid{v}\to \Conid{App}\;\Varid{v}\;\mathrm{1}\;\Conid{Return})}. This computation
constructs an abstraction and passes it to the continuation as a (closure) value
\ensuremath{\Varid{v}}. The continuation applies \ensuremath{\Varid{v}} to \ensuremath{\mathrm{1}} and passes the result to \ensuremath{\Conid{Return}}. The
closure retrieves the the value of variable with index \ensuremath{\mathrm{0}} (i.e., \ensuremath{\Varid{x}}) and
passes it to \ensuremath{\Conid{Return}}.

\paragraph{Handler}
The idea of a handler is to map the syntax tree onto its
meaning. 
We illustrate this on the \ensuremath{\Conid{LC}\;\Varid{v}\;\Varid{a}} type, where we use the
type \ensuremath{\Varid{v}\mathrel{=}\Conid{Closure}} for values.

\noindent
\begin{minipage}[t][][t]{0.6\textwidth}
\vspace{-\abovedisplayskip}
\begin{hscode}\SaveRestoreHook
\column{B}{@{}>{\hspre}l<{\hspost}@{}}%
\column{3}{@{}>{\hspre}l<{\hspost}@{}}%
\column{E}{@{}>{\hspre}l<{\hspost}@{}}%
\>[B]{}\mathbf{data}\;\Varid{Closure_{1}}\;\mathbf{where}{}\<[E]%
\\
\>[B]{}\hsindent{3}{}\<[3]%
\>[3]{}\Varid{Clos_{1}}\mathbin{::}\Varid{FunPtr_{1}}\to \Varid{Env_{1}}\to \Varid{Closure_{1}}{}\<[E]%
\\[\blanklineskip]%
\>[B]{}\mathbf{type}\;\Varid{Store_{1}}\mathrel{=}[\mskip1.5mu \Conid{LC}\;\Varid{Closure_{1}}\;\Varid{Closure_{1}}\mskip1.5mu]{}\<[E]%
\ColumnHook
\end{hscode}\resethooks
\end{minipage}%
\begin{minipage}[t][][t]{0.4\textwidth}
\vspace{-\abovedisplayskip}
\begin{hscode}\SaveRestoreHook
\column{B}{@{}>{\hspre}l<{\hspost}@{}}%
\column{E}{@{}>{\hspre}l<{\hspost}@{}}%
\>[B]{}\mathbf{type}\;\Varid{FunPtr_{1}}\mathrel{=}\Conid{Int}{}\<[E]%
\\[\blanklineskip]%
\>[B]{}\mathbf{type}\;\Varid{Env_{1}}\mathrel{=}[\mskip1.5mu \Varid{Closure_{1}}\mskip1.5mu]{}\<[E]%
\ColumnHook
\end{hscode}\resethooks
\end{minipage}
\\%
A closure contains a function pointer and an environment. 
The function pointer is an index into a list of deferred computations (i.e.,
function bodies) that we call the \emph{(resumption) store}.
The environment is a list that maps the closure's parameters (which are
indexes) onto their values.

Now we are ready to define the handler \ensuremath{\Varid{handleAbs}} as a function that, 
given an initial environment and store,
maps an \ensuremath{\Conid{LC}\;\Conid{Closure}\;\Varid{a}} computation onto its meaning, which is a
tuple of the result value of type \ensuremath{\Varid{a}} and the final store.

\begin{hscode}\SaveRestoreHook
\column{B}{@{}>{\hspre}l<{\hspost}@{}}%
\column{3}{@{}>{\hspre}l<{\hspost}@{}}%
\column{10}{@{}>{\hspre}l<{\hspost}@{}}%
\column{12}{@{}>{\hspre}l<{\hspost}@{}}%
\column{14}{@{}>{\hspre}l<{\hspost}@{}}%
\column{16}{@{}>{\hspre}l<{\hspost}@{}}%
\column{28}{@{}>{\hspre}l<{\hspost}@{}}%
\column{30}{@{}>{\hspre}l<{\hspost}@{}}%
\column{34}{@{}>{\hspre}c<{\hspost}@{}}%
\column{34E}{@{}l@{}}%
\column{35}{@{}>{\hspre}c<{\hspost}@{}}%
\column{35E}{@{}l@{}}%
\column{37}{@{}>{\hspre}l<{\hspost}@{}}%
\column{38}{@{}>{\hspre}l<{\hspost}@{}}%
\column{E}{@{}>{\hspre}l<{\hspost}@{}}%
\>[B]{}\Varid{handleAbs}{}\<[12]%
\>[12]{}\mathbin{::}\Varid{Env_{1}}\to \Varid{Store_{1}}\to \Conid{LC}\;\Varid{Closure_{1}}\;\Varid{Closure_{1}}\to (\Varid{Store_{1}},\Varid{Closure_{1}}){}\<[E]%
\\
\>[B]{}\Varid{handleAbs}\;\anonymous \;{}\<[16]%
\>[16]{}\Varid{r}\;(\Conid{Return}\;\Varid{x}){}\<[34]%
\>[34]{}\mathrel{=}{}\<[34E]%
\>[37]{}(\Varid{r},\Varid{x}){}\<[E]%
\\
\>[B]{}\Varid{handleAbs}\;\Varid{env}\;{}\<[16]%
\>[16]{}\Varid{r}\;(\Conid{Var}\;\Varid{n}\;{}\<[30]%
\>[30]{}\Varid{k}){}\<[34]%
\>[34]{}\mathrel{=}{}\<[34E]%
\>[37]{}\Varid{handleAbs}\;\Varid{env}\;\Varid{r}\;(\Varid{k}\;(\Varid{env}\mathbin{!!}\Varid{n})){}\<[E]%
\\
\>[B]{}\Varid{handleAbs}\;\Varid{env}\;{}\<[16]%
\>[16]{}\Varid{r}\;(\Conid{App}\;\Varid{v}_{1}\;\Varid{v}_{2}\;{}\<[30]%
\>[30]{}\Varid{k}){}\<[34]%
\>[34]{}\mathrel{=}{}\<[34E]%
\>[37]{}\Varid{handleAbs}\;\Varid{env}\;\Varid{r'}\;(\Varid{k}\;\Varid{v}){}\<[E]%
\\
\>[B]{}\hsindent{3}{}\<[3]%
\>[3]{}\mathbf{where}\;{}\<[10]%
\>[10]{}(\Varid{Clos_{1}}\;\Varid{fp}\;\Varid{env'}){}\<[28]%
\>[28]{}\mathrel{=}\Varid{v}_{1}{}\<[E]%
\\
\>[10]{}(\Varid{r'},\Varid{v}){}\<[28]%
\>[28]{}\mathrel{=}\Varid{handleAbs}\;(\Varid{v}_{2}\mathbin{:}\Varid{env'})\;\Varid{r}\;(\Varid{r}\mathbin{!!}\Varid{fp}){}\<[E]%
\\
\>[B]{}\Varid{handleAbs}\;\Varid{env}\;{}\<[16]%
\>[16]{}\Varid{r}\;(\Conid{Abs}\;\Varid{e}\;{}\<[30]%
\>[30]{}\Varid{k}){}\<[35]%
\>[35]{}\mathrel{=}{}\<[35E]%
\>[38]{}\Varid{handleAbs}\;\Varid{env}\;\Varid{r'}\;(\Varid{k}\;\Varid{v}){}\<[E]%
\\
\>[B]{}\hsindent{3}{}\<[3]%
\>[3]{}\mathbf{where}\;{}\<[10]%
\>[10]{}\Varid{v}{}\<[14]%
\>[14]{}\mathrel{=}\Varid{Clos_{1}}\;(\Varid{length}\;\Varid{r})\;\Varid{env}{}\<[E]%
\\
\>[10]{}\Varid{r'}{}\<[14]%
\>[14]{}\mathrel{=}\Varid{r}\plus [\mskip1.5mu \Varid{e}\mskip1.5mu]{}\<[E]%
\ColumnHook
\end{hscode}\resethooks

First, the leaf case of the handler returns the value in that leaf, supplemented with the resumption store.
Next, the variable case looks up the variable in the environment and passes it to the continuation.
Then, the application case unpacks the closure, 
retrieves the corresponding function body from the resumption store 
and applies it to the extended environment and store.
The resulting value is passed to the continuation.
Finally, the abstraction case adds the function body to the resumption store, 
creates a closure that indexes this new entry, 
and calls the continuation on this closure value.

\subsection{Modular Latent Effect Signatures and Trees, Version 1}
\label{a-signature-for-lambda}

We now modularize the definition of \ensuremath{\Conid{LC}} by separating  the recursive structure of
the monadic syntax from the node shapes of 
the \ensuremath{\Conid{Var}}, \ensuremath{\Conid{App}} and \ensuremath{\Conid{Abs}} operations.

\paragraph{Latent Effect Signature}
We call the node shapes the latent effect
signature. In this case, it is called \ensuremath{\Conid{Abstracting}\;\Varid{v}} with \ensuremath{\Varid{v}} the type of values.
\begin{hscode}\SaveRestoreHook
\column{B}{@{}>{\hspre}l<{\hspost}@{}}%
\column{3}{@{}>{\hspre}l<{\hspost}@{}}%
\column{22}{@{}>{\hspre}l<{\hspost}@{}}%
\column{E}{@{}>{\hspre}l<{\hspost}@{}}%
\>[B]{}\mathbf{data}\;\Conid{Abstracting}\;\Varid{v}\mathbin{::}\mathbin{*}\to (\mathbin{*}\to \mathbin{*})\to \mathbin{*}\mathbf{where}{}\<[E]%
\\
\>[B]{}\hsindent{3}{}\<[3]%
\>[3]{}\Conid{Var'}\mathbin{::}\Conid{Int}\to {}\<[22]%
\>[22]{}\Conid{Abstracting}\;\Varid{v}\;\Varid{v}\;\Conid{NoSub}{}\<[E]%
\\
\>[B]{}\hsindent{3}{}\<[3]%
\>[3]{}\Conid{App'}\mathbin{::}\Varid{v}\to \Varid{v}\to {}\<[22]%
\>[22]{}\Conid{Abstracting}\;\Varid{v}\;\Varid{v}\;\Conid{NoSub}{}\<[E]%
\\
\>[B]{}\hsindent{3}{}\<[3]%
\>[3]{}\Conid{Abs'}\mathbin{::}{}\<[22]%
\>[22]{}\Conid{Abstracting}\;\Varid{v}\;\Varid{v}\;(\Conid{OneSub}\;\Varid{v}){}\<[E]%
\ColumnHook
\end{hscode}\resethooks
Besides its first parameter \ensuremath{\Varid{v}}, the type \ensuremath{\Conid{Abstracting}\;\Varid{v}\;\Varid{p}\;\Varid{c}} is indexed by two
further parameters: The parameter \ensuremath{\Varid{p}} is the return type of the primitive
operations; this is the type of value they pass to their continuation. As all
three primitive operations return a value of type \ensuremath{\Varid{v}}, we have that \ensuremath{\Varid{p}\mathrel{=}\Varid{v}}.
The parameter \ensuremath{\Varid{c}\mathbin{::}\mathbin{*}\to \mathbin{*}} captures the number and result type of the subcomputations.
As \ensuremath{\Conid{Var'}} and \ensuremath{\Conid{App'}} have no subcomputations, they use the type \ensuremath{\Varid{c}\mathrel{=}\Conid{NoSub}} to indicate that.
However, \ensuremath{\Conid{Abs'}} has a subcomputation and it indicates this with \ensuremath{\Varid{c}\mathrel{=}\Conid{OneSub}\;\Varid{v}}. This subcomputation is the body of the function abstraction, whose return
type is \ensuremath{\Varid{v}}. Hence, \ensuremath{\Conid{OneSub}\;\Varid{v}} has one constructor \ensuremath{\Conid{One}\mathbin{::}\Conid{OneSub}\;\Varid{v}\;\Varid{v}}.
\noindent
\begin{minipage}[t][][t]{0.5\textwidth}
\vspace{-\abovedisplayskip}
\begin{hscode}\SaveRestoreHook
\column{B}{@{}>{\hspre}l<{\hspost}@{}}%
\column{E}{@{}>{\hspre}l<{\hspost}@{}}%
\>[B]{}\mathbf{data}\;\Conid{NoSub}\mathbin{::}\mathbin{*}\to \mathbin{*}\mathbf{where}{}\<[E]%
\ColumnHook
\end{hscode}\resethooks
\end{minipage}
\begin{minipage}[t][][t]{0.5\textwidth}
\vspace{-\abovedisplayskip}
\begin{hscode}\SaveRestoreHook
\column{B}{@{}>{\hspre}l<{\hspost}@{}}%
\column{3}{@{}>{\hspre}l<{\hspost}@{}}%
\column{E}{@{}>{\hspre}l<{\hspost}@{}}%
\>[B]{}\mathbf{data}\;\Conid{OneSub}\;\Varid{v}\mathbin{::}\mathbin{*}\to \mathbin{*}\mathbf{where}{}\<[E]%
\\
\>[B]{}\hsindent{3}{}\<[3]%
\>[3]{}\Conid{One}\mathbin{::}\Conid{OneSub}\;\Varid{v}\;\Varid{v}{}\<[E]%
\ColumnHook
\end{hscode}\resethooks
\end{minipage}
\vspace{-5mm}

\paragraph{Latent Effect Tree, Version 1}
The type \ensuremath{\Conid{Tree}_{1}\;\Varid{\sigma}\;\Varid{a}} extends a latent effect signature \ensuremath{\Varid{\sigma}} into a recursive
syntactic structure that is a monad in \ensuremath{\Varid{a}}.
\begin{hscode}\SaveRestoreHook
\column{B}{@{}>{\hspre}l<{\hspost}@{}}%
\column{3}{@{}>{\hspre}l<{\hspost}@{}}%
\column{10}{@{}>{\hspre}c<{\hspost}@{}}%
\column{10E}{@{}l@{}}%
\column{14}{@{}>{\hspre}l<{\hspost}@{}}%
\column{26}{@{}>{\hspre}l<{\hspost}@{}}%
\column{61}{@{}>{\hspre}l<{\hspost}@{}}%
\column{84}{@{}>{\hspre}l<{\hspost}@{}}%
\column{E}{@{}>{\hspre}l<{\hspost}@{}}%
\>[B]{}\mathbf{data}\;\Conid{Tree}_{1}\;(\Varid{\sigma}\mathbin{::}\mathbin{*}\to (\mathbin{*}\to \mathbin{*})\to \mathbin{*})\;\Varid{a}\;\mathbf{where}{}\<[E]%
\\
\>[B]{}\hsindent{3}{}\<[3]%
\>[3]{}\Conid{Leaf}_{1}{}\<[10]%
\>[10]{}\mathbin{::}{}\<[10E]%
\>[14]{}\Varid{a}\to \Conid{Tree}_{1}\;\Varid{\sigma}\;\Varid{a}{}\<[E]%
\\
\>[B]{}\hsindent{3}{}\<[3]%
\>[3]{}\Conid{Node}_{1}{}\<[10]%
\>[10]{}\mathbin{::}{}\<[10E]%
\>[14]{}\Varid{\sigma}\;\Varid{p}\;\Varid{c}\to {}\<[26]%
\>[26]{}(\forall \Varid{x}\hsforall \hsdot{\circ }{.}\Varid{c}\;\Varid{x}\to \Conid{Tree}_{1}\;\Varid{\sigma}\;\Varid{x})\to {}\<[61]%
\>[61]{}(\Varid{p}\to \Conid{Tree}_{1}\;\Varid{\sigma}\;\Varid{a})\to {}\<[84]%
\>[84]{}\Conid{Tree}_{1}\;\Varid{\sigma}\;\Varid{a}{}\<[E]%
\ColumnHook
\end{hscode}\resethooks
The \ensuremath{\Conid{Leaf}_{1}} constructor is trivial; \ensuremath{\Conid{Leaf}_{1}\;\Varid{x}} returns a pure computation with result \ensuremath{\Varid{x}}.  
The internal nodes are of the form \ensuremath{\Conid{Node}\;\Varid{op}\;\Varid{sub}\;\Varid{k}} where the fields have the following meaning.
The first, \ensuremath{\Varid{op}}, identifies what primitive operation the node represents. 
Next, \ensuremath{\Varid{sub}} is a function that, in case of a non-atomic primitive, selects
the subcomputations of the node. 
Finally, \ensuremath{\Varid{k}} is the continuation of further operations
to perform after the current one.

Some notable characteristics are as follows:
\begin{itemize}
\item
Every operation has a result type \ensuremath{\Varid{p}} that is made available to its
continuation, and a number of subcomputations \ensuremath{\Varid{c}}. To model these two, the
signature of an operation \ensuremath{\Varid{op}\mathbin{::}\Varid{\sigma}\;\Varid{p}\;\Varid{c}} is parameterized by \ensuremath{\Varid{p}} and \ensuremath{\Varid{c}}.

\item
The function \ensuremath{\Varid{sub}} has type \ensuremath{\forall \Varid{x}\hsforall \hsdot{\circ }{.}\Varid{c}\;\Varid{x}\to \Conid{Tree}_{1}\;\Varid{\sigma}\;\Varid{x}}. The
input of type \ensuremath{\Varid{c}\;\Varid{x}} determines what subcomputation to select; the parameter \ensuremath{\Varid{x}}
indicates the result type of that subcomputation. 

\item
Likewise, continuations take as input the operation's output value (\ensuremath{\Varid{p}}).

\item The \ensuremath{\Conid{Tree}_{1}} data type is monadic, with a similar notion of return and bind as the \emph{free monad}~\cite{Swierstra08}:

\begin{hscode}\SaveRestoreHook
\column{B}{@{}>{\hspre}l<{\hspost}@{}}%
\column{3}{@{}>{\hspre}l<{\hspost}@{}}%
\column{E}{@{}>{\hspre}l<{\hspost}@{}}%
\>[B]{}\mathbf{instance}\;\Conid{Monad}\;(\Conid{Tree}_{1}\;\Varid{\sigma})\;\mathbf{where}{}\<[E]%
\\
\>[B]{}\hsindent{3}{}\<[3]%
\>[3]{}\Varid{return}\mathrel{=}\Conid{Leaf}_{1}{}\<[E]%
\\
\>[B]{}\hsindent{3}{}\<[3]%
\>[3]{}(\Conid{Leaf}_{1}\;\Varid{x})\bind \Varid{f}\mathrel{=}\Varid{f}\;\Varid{x}{}\<[E]%
\\
\>[B]{}\hsindent{3}{}\<[3]%
\>[3]{}(\Conid{Node}_{1}\;\Varid{op}\;\Varid{sub}\;\Varid{k})\bind \Varid{f}\mathrel{=}\Conid{Node}_{1}\;\Varid{op}\;\Varid{sub}\;(\lambda \Varid{x}\to \Varid{k}\;\Varid{x}\bind \Varid{f}){}\<[E]%
\ColumnHook
\end{hscode}\resethooks

A monadic binding \ensuremath{\Varid{t}\bind \Varid{f}} thus ``concatenates'' the tree in \ensuremath{\Varid{f}} to the leaf
positions in the continuation (only) of \ensuremath{\Varid{t}}.
\end{itemize}

We can emulate the non-modular type \ensuremath{\Conid{LC}\;\Varid{v}\;\Varid{a}} with \ensuremath{\Conid{LC'}\;\Varid{v}\;\Varid{a}}.
\begin{hscode}\SaveRestoreHook
\column{B}{@{}>{\hspre}l<{\hspost}@{}}%
\column{E}{@{}>{\hspre}l<{\hspost}@{}}%
\>[B]{}\mathbf{type}\;\Conid{LC'}\;\Varid{v}\;\Varid{a}\mathrel{=}\Conid{Tree}_{1}\;(\Conid{Abstracting}\;\Varid{v})\;\Varid{a}{}\<[E]%
\ColumnHook
\end{hscode}\resethooks
The corresponding representation for \ensuremath{\Conid{LC'}}s \ensuremath{\Conid{Return}} constructor is \ensuremath{\Conid{Leaf}_{1}}.
The \ensuremath{\Conid{Var}} constructor is represented with a \ensuremath{\Conid{Node}_{1}}.
\begin{hscode}\SaveRestoreHook
\column{B}{@{}>{\hspre}l<{\hspost}@{}}%
\column{11}{@{}>{\hspre}l<{\hspost}@{}}%
\column{48}{@{}>{\hspre}l<{\hspost}@{}}%
\column{E}{@{}>{\hspre}l<{\hspost}@{}}%
\>[B]{}\Varid{var}_{1}{}\<[11]%
\>[11]{}\mathbin{::}\Conid{Int}\to (\Varid{v}\to \Conid{LC'}\;\Varid{v}\;\Varid{a})\to \Conid{LC'}\;\Varid{v}\;\Varid{a}{}\<[E]%
\\
\>[B]{}\Varid{var}_{1}\;\Varid{i}\;\Varid{k}{}\<[11]%
\>[11]{}\mathrel{=}\Conid{Node}_{1}\;(\Conid{Var'}\;\Varid{i})\;(\lambda \Varid{x}\to \mathbf{case}\;\Varid{x}\;\mathbf{of}\;)\;{}\<[48]%
\>[48]{}\Varid{k}{}\<[E]%
\ColumnHook
\end{hscode}\resethooks
This is a \ensuremath{\Conid{Var'}\;\Varid{i}} node. As there are no subcomputations, there are no branches in the
pattern match in the selection function on the value \ensuremath{\Varid{x}} of the empty type
\ensuremath{\Conid{NoSub}}. Lastly, the continuation \ensuremath{\Varid{k}} receives the value produced
by the operation.

The encodings of the two other operations are similar.
One notable aspect of \ensuremath{\Varid{abs}_{1}} is that it does have one subcomputation. Hence,
the selection function matches on the \ensuremath{\Conid{One}} constructor and returns
\ensuremath{\Varid{t}}:
\begin{hscode}\SaveRestoreHook
\column{B}{@{}>{\hspre}l<{\hspost}@{}}%
\column{44}{@{}>{\hspre}l<{\hspost}@{}}%
\column{E}{@{}>{\hspre}l<{\hspost}@{}}%
\>[B]{}\Varid{app}_{1}\mathbin{::}\Varid{v}\to \Varid{v}\to (\Varid{v}\to \Conid{LC'}\;\Varid{v}\;\Varid{a})\to \Conid{LC'}\;\Varid{v}\;{}\<[44]%
\>[44]{}\Varid{a}{}\<[E]%
\\
\>[B]{}\Varid{app}_{1}\;\Varid{v}_{1}\;\Varid{v}_{2}\;\Varid{k}\mathrel{=}\Conid{Node}_{1}\;(\Conid{App'}\;\Varid{v}_{1}\;\Varid{v}_{2})\;(\lambda \Varid{x}\to \mathbf{case}\;\Varid{x}\;\mathbf{of}\;)\;\Varid{k}{}\<[E]%
\\[\blanklineskip]%
\>[B]{}\Varid{abs}_{1}\mathbin{::}\Conid{LC'}\;\Varid{v}\;\Varid{v}\to (\Varid{v}\to \Conid{LC'}\;\Varid{v}\;\Varid{a})\to \Conid{LC'}\;\Varid{v}\;\Varid{a}{}\<[E]%
\\
\>[B]{}\Varid{abs}_{1}\;\Varid{t}\;\Varid{k}\mathrel{=}\Conid{Node}_{1}\;\Conid{Abs'}\;(\lambda \Conid{One}\to \Varid{t})\;\Varid{k}{}\<[E]%
\ColumnHook
\end{hscode}\resethooks

\paragraph{Modular Tree Constructors}
We can create modular constructors for latent effect operations, similarly to
how DTC admits modular syntax constructors~\cite{Swierstra08}.  To this end, we
use a co-product operator \ensuremath{\mathbin{+}} that combines latent effect signatures, and a
subtyping relation \ensuremath{\Varid{\sigma}_{1}\mathbin{<}\Varid{\sigma}_{2}} with an associated injection function,
\ensuremath{\Varid{injSig}\mathbin{::}\Varid{\sigma}_{1}\;\Varid{p}\;\Varid{c}\to \Varid{\sigma}_{2}\;\Varid{p}\;\Varid{c}}.
Using these, we can implement the modular constructor functions in Figure~\ref{fig:modular_abstracting}, that allow
combining \ensuremath{\Conid{Abstracting}} with other latent and algebraic effects.
The subtyping requirements in the type signatures are automatically inferrable by type class instance resolution in Haskell.
The implementation details of \ensuremath{\mathbin{+}} and \ensuremath{\mathbin{<}} are given in Section \ref{app:effect-lib}.

\begin{figure}[t]
\begin{hscode}\SaveRestoreHook
\column{B}{@{}>{\hspre}l<{\hspost}@{}}%
\column{8}{@{}>{\hspre}l<{\hspost}@{}}%
\column{12}{@{}>{\hspre}l<{\hspost}@{}}%
\column{44}{@{}>{\hspre}l<{\hspost}@{}}%
\column{54}{@{}>{\hspre}l<{\hspost}@{}}%
\column{62}{@{}>{\hspre}l<{\hspost}@{}}%
\column{E}{@{}>{\hspre}l<{\hspost}@{}}%
\>[B]{}\Varid{var}\mathbin{::}(\Conid{Abstracting}\;\Varid{v}\mathbin{<}\Varid{\sigma})\Rightarrow \Conid{Int}\to \Conid{Tree}_{1}\;\Varid{\sigma}\;\Varid{v}{}\<[E]%
\\
\>[B]{}\Varid{var}\;\Varid{n}{}\<[8]%
\>[8]{}\mathrel{=}\Conid{Node}_{1}\;(\Varid{injSig}\;(\Conid{Var'}\;\Varid{n}))\;(\lambda \Varid{x}\to \mathbf{case}\;\Varid{x}\;\mathbf{of}\;)\;{}\<[54]%
\>[54]{}\Conid{Leaf}_{1}{}\<[E]%
\\[\blanklineskip]%
\>[B]{}\Varid{app}\mathbin{::}(\Conid{Abstracting}\;\Varid{v}\mathbin{<}\Varid{\sigma})\Rightarrow \Varid{v}\to \Varid{v}\to \Conid{Tree}_{1}\;\Varid{\sigma}\;\Varid{v}{}\<[E]%
\\
\>[B]{}\Varid{app}\;\Varid{v}_{1}\;\Varid{v}_{2}{}\<[12]%
\>[12]{}\mathrel{=}\Conid{Node}_{1}\;(\Varid{injSig}\;(\Conid{App'}\;\Varid{v}_{1}\;\Varid{v}_{2}))\;(\lambda \Varid{x}\to \mathbf{case}\;\Varid{x}\;\mathbf{of}\;)\;{}\<[62]%
\>[62]{}\Conid{Leaf}_{1}{}\<[E]%
\\[\blanklineskip]%
\>[B]{}\Varid{abs}\mathbin{::}(\Conid{Abstracting}\;\Varid{v}\mathbin{<}\Varid{\sigma})\Rightarrow \Conid{Tree}_{1}\;\Varid{\sigma}\;\Varid{v}\to \Conid{Tree}_{1}\;\Varid{\sigma}\;\Varid{v}{}\<[E]%
\\
\>[B]{}\Varid{abs}\;\Varid{t}{}\<[8]%
\>[8]{}\mathrel{=}\Conid{Node}_{1}\;(\Varid{injSig}\;\Conid{Abs'})\;(\lambda \Conid{One}\to \Varid{t})\;{}\<[44]%
\>[44]{}\Conid{Leaf}_{1}{}\<[E]%
\ColumnHook
\end{hscode}\resethooks
\caption{The modular constructor functions of the  \ensuremath{\Conid{Abstracting}} effect. These functions
all fix the continuation to \ensuremath{\Conid{Leaf}_{1}}, which can easily be replaced by an arbitrary continuation \ensuremath{\Varid{k}}
using the \ensuremath{\bind } operator of \ensuremath{\Conid{Tree}_{1}}'s monad instance.}
\label{fig:modular_abstracting}
\end{figure}


We can now use these modular constructors to implement denotation function cases.
We can also write programs using the constructors directly.
For example, the following program with a lambda abstraction with an effectful body:
\label{code:prog}
\begin{hscode}\SaveRestoreHook
\column{B}{@{}>{\hspre}l<{\hspost}@{}}%
\column{3}{@{}>{\hspre}l<{\hspost}@{}}%
\column{E}{@{}>{\hspre}l<{\hspost}@{}}%
\>[B]{}\Varid{prog}\mathbin{::}\forall \Varid{v}\hsforall \hsdot{\circ }{.}\Conid{Num}\;\Varid{v}\Rightarrow \Conid{Tree}_{1}\;(\Conid{Mutating}\;\Varid{v}\mathbin{+}\Conid{Abstracting}\;\Varid{v}\mathbin{+}\Varid{Ending})\;\Varid{v}{}\<[E]%
\\
\>[B]{}\Varid{prog}\mathrel{=}\mathbf{do}{}\<[E]%
\\
\>[B]{}\hsindent{3}{}\<[3]%
\>[3]{}\Varid{put}\;(\mathrm{1}\mathbin{::}\Varid{v}){}\<[E]%
\\
\>[B]{}\hsindent{3}{}\<[3]%
\>[3]{}\Varid{f}\leftarrow \Varid{abs}\;(\mathbf{do}\;\Varid{m}\leftarrow \Varid{var}\;\mathrm{0};(\Varid{n}\mathbin{::}\Varid{v})\leftarrow \Varid{get};\Varid{return}\;(\Varid{m}\mathbin{+}\Varid{n})){}\<[E]%
\\
\>[B]{}\hsindent{3}{}\<[3]%
\>[3]{}\Varid{put}\;(\mathrm{2}\mathbin{::}\Varid{v}){}\<[E]%
\\
\>[B]{}\hsindent{3}{}\<[3]%
\>[3]{}\Varid{app}\;\Varid{f}\;(\mathrm{3}\mathbin{::}\Varid{v}){}\<[E]%
\ColumnHook
\end{hscode}\resethooks
The body of the function abstraction increments the function argument value (de Bruijn index 0) by the \ensuremath{\Varid{n}} yielded by the \ensuremath{\Varid{get}} operation.
The signature of the program tree is the co-product of three signatures:
(1) \ensuremath{\Conid{Mutating}\;\Conid{V}} for mutable state;
(2) function abstractions \ensuremath{\Conid{Abstracting}\;\Conid{V}};
and (3) the empty signature \ensuremath{\Varid{Ending}}, which provides no operations and serves as the base case.
The \ensuremath{\Conid{Mutating}\;\Conid{V}} effect recasts the traditional algebraic state effect as a latent effect, and has two operations, \ensuremath{\Varid{get}} and \ensuremath{\Varid{put}}.
Observe that \ensuremath{\Varid{prog}} is essentially an AST, with multiple possible interpretations.
If \ensuremath{\Conid{Mutating}\;\Conid{V}} is dynamic (runtime) state, then \ensuremath{\Varid{prog}} evaluates to $3+2=5$.
However, if it is for macro bindings that are expanded \emph{statically}, then the \ensuremath{\Varid{get}} in the body of the abstraction is evaluated under the state at the ``definition site'' of the lambda, and \ensuremath{\Varid{prog}} evaluates to \ensuremath{\mathrm{3}\mathbin{+}\mathrm{1}\mathrel{=}\mathrm{4}}.
Next, we show how handlers can map the \ensuremath{\Varid{prog}} syntax tree to different semantics.

\subsection{Trees with Support for Modular Handlers}

In the case of a modularly composed signature \ensuremath{\Varid{\sigma}\mathrel{=}\Varid{\sigma}_{1}\mathbin{+}\mathbin{...}\mathbin{+}\sigma_n\mathbin{+}\Varid{Ending}},
the idea is to compose the handler function from individual handlers for the different
components of the signature
$\ensuremath{\Varid{h}\mathrel{=}\Varid{hEnd}\hsdot{\circ }{.}\Varid{h}_n\hsdot{\circ }{.}\mathbin{...}\hsdot{\circ }{.}\Varid{h}_{1}}$.
The type of each handler would be \ensuremath{\Varid{h}_i\mathbin{::}\forall \Varid{\sigma}\hsforall \hsdot{\circ }{.}\Conid{Tree}_{1}\;(\sigma_i\mathbin{+}\Varid{\sigma})\;\Varid{a}\to \Conid{Tree}_{1}\;\Varid{\sigma}\;(\Varid{L}_i\;\Varid{a})}.  Hence, it is polymorphic in the remaining part of the
signature and preserves those nodes in the resulting tree. It  only replaces
the nodes of \ensuremath{\sigma_i} with their meaning, which is given in the form of a functor \ensuremath{\Varid{L}_i} that
decorates the result type \ensuremath{\Varid{a}}.

Unfortunately, our \ensuremath{\Conid{Tree}_{1}} type and, in particular, the type of its \ensuremath{\Conid{Node}_{1}} constructor,
needs some further refinement to  fully support this idea. Indeed, if the signature is
\ensuremath{\sigma_i\mathbin{+}\Varid{\sigma}}, and we apply \ensuremath{\Varid{h}_i} to all the recursive occurrences of \ensuremath{\Conid{Tree}\;(\sigma_i\mathbin{+}\Varid{\sigma})}
in a \ensuremath{\Varid{\sigma}}-node \ensuremath{\Conid{Node}_{1}\;(\Conid{Inr'}\;\Varid{op})\;\Varid{sub}\;\Varid{k}}, we get:\\
%
%
\setlength\mathindent{0.5em}%
\begin{minipage}{0.495\linewidth}%
\textbf{Before applying \ensuremath{\Varid{h}_i}}\vspace{-0.5em}\begin{hscode}\SaveRestoreHook
\column{B}{@{}>{\hspre}l<{\hspost}@{}}%
\column{3}{@{}>{\hspre}l<{\hspost}@{}}%
\column{12}{@{}>{\hspre}l<{\hspost}@{}}%
\column{E}{@{}>{\hspre}l<{\hspost}@{}}%
\>[3]{}\Conid{Inr'}\;\Varid{op}{}\<[12]%
\>[12]{}\mathbin{::}(\sigma_i\mathbin{+}\Varid{\sigma})\;\Varid{p}\;\Varid{c}{}\<[E]%
\\
\>[3]{}\Varid{sub}{}\<[12]%
\>[12]{}\mathbin{::}\forall \Varid{x}\hsforall \hsdot{\circ }{.}\Varid{c}\;\Varid{x}\to \Conid{Tree}_{1}\;(\sigma_i\mathbin{+}\Varid{\sigma})\;\Varid{x}{}\<[E]%
\\
\>[3]{}\Varid{k}{}\<[12]%
\>[12]{}\mathbin{::}\Varid{p}\to \Conid{Tree}_{1}\;(\sigma_i\mathbin{+}\Varid{\sigma})\;\Varid{a}{}\<[E]%
\ColumnHook
\end{hscode}\resethooks
\end{minipage}%
\ \vrule\ %
\begin{minipage}{0.495\linewidth}%
\textbf{After applying \ensuremath{\Varid{h}_i}}\vspace{-0.5em}\begin{hscode}\SaveRestoreHook
\column{B}{@{}>{\hspre}l<{\hspost}@{}}%
\column{3}{@{}>{\hspre}l<{\hspost}@{}}%
\column{13}{@{}>{\hspre}l<{\hspost}@{}}%
\column{22}{@{}>{\hspre}l<{\hspost}@{}}%
\column{29}{@{}>{\hspre}l<{\hspost}@{}}%
\column{E}{@{}>{\hspre}l<{\hspost}@{}}%
\>[3]{}\Varid{op}{}\<[13]%
\>[13]{}\mathbin{::}\Varid{\sigma}\;\Varid{p}\;\Varid{c}{}\<[E]%
\\
\>[3]{}\Varid{h}_i\hsdot{\circ }{.}\Varid{sub}{}\<[13]%
\>[13]{}\mathbin{::}\forall \Varid{x}\hsforall \hsdot{\circ }{.}\Varid{c}\;\Varid{x}\to \Conid{Tree}_{1}\;\Varid{\sigma}\;(\colorbox{black!10}{$\Varid{L}_i\;\Varid{x}$}){}\<[E]%
\\
\>[3]{}\Varid{h}_i\hsdot{\circ }{.}\Varid{k}{}\<[13]%
\>[13]{}\mathbin{::}\Varid{p}\to {}\<[22]%
\>[22]{}\Conid{Tree}_{1}\;{}\<[29]%
\>[29]{}\Varid{\sigma}\;(\Varid{L}_i\;\Varid{a}){}\<[E]%
\ColumnHook
\end{hscode}\resethooks
\end{minipage} \\
\setlength\mathindent{1em}%
The resulting fields do not together form a node of type \ensuremath{\Conid{Tree}_{1}\;\Varid{\sigma}\;(\Varid{L}_i\;\Varid{a})} because
the highlighted result type of the subcomputations is \ensuremath{(\Varid{L}_i\;\Varid{x})} rather than 
\ensuremath{\Varid{x}} which the \ensuremath{\Conid{Node}_{1}} constructor requires \ensuremath{\Varid{sub}} to have as return type.

The problem is that the \ensuremath{\Conid{Tree}_{1}} type is oblivious to the \emph{effect functor} that the return type of subcomputations in the tree are decorated by.
To solve this problem, we can expose the effect functor decoration in the tree type itself; e.g.,

\begin{hscode}\SaveRestoreHook
\column{B}{@{}>{\hspre}l<{\hspost}@{}}%
\column{3}{@{}>{\hspre}l<{\hspost}@{}}%
\column{4}{@{}>{\hspre}l<{\hspost}@{}}%
\column{11}{@{}>{\hspre}c<{\hspost}@{}}%
\column{11E}{@{}l@{}}%
\column{15}{@{}>{\hspre}l<{\hspost}@{}}%
\column{24}{@{}>{\hspre}c<{\hspost}@{}}%
\column{24E}{@{}l@{}}%
\column{28}{@{}>{\hspre}l<{\hspost}@{}}%
\column{60}{@{}>{\hspre}l<{\hspost}@{}}%
\column{E}{@{}>{\hspre}l<{\hspost}@{}}%
\>[3]{}\mathbf{data}\;\Conid{Tree}_{2}\;(\Varid{\sigma}\mathbin{::}\mathbin{*}\to (\mathbin{*}\to \mathbin{*})\to \mathbin{*})\;(\colorbox{black!10}{$\Varid{l}\mathbin{::}\mathbin{*}\to \mathbin{*}$})\;\Varid{a}\;\mathbf{where}{}\<[E]%
\\
\>[3]{}\hsindent{1}{}\<[4]%
\>[4]{}\Conid{Leaf}_{2}{}\<[11]%
\>[11]{}\mathbin{::}{}\<[11E]%
\>[15]{}\Varid{a}\to \Conid{Tree}_{2}\;\Varid{\sigma}\;\Varid{l}\;\Varid{a}{}\<[E]%
\\
\>[3]{}\hsindent{1}{}\<[4]%
\>[4]{}\Conid{Node}_{2}{}\<[11]%
\>[11]{}\mathbin{::}{}\<[11E]%
\>[15]{}\Varid{\sigma}\;\Varid{p}\;\Varid{c}{}\<[24]%
\>[24]{}\to {}\<[24E]%
\>[28]{}(\forall \Varid{x}\hsforall \hsdot{\circ }{.}\Varid{c}\;\Varid{x}\to \Conid{Tree}_{2}\;\Varid{\sigma}\;\Varid{l}\;(\colorbox{black!10}{$\Varid{l}$}\;\Varid{x})){}\<[E]%
\\
\>[24]{}\to {}\<[24E]%
\>[28]{}(\colorbox{black!10}{$\Varid{l}$}\;\Varid{p}\to \Conid{Tree}_{2}\;\Varid{\sigma}\;\Varid{l}\;\Varid{a})\to {}\<[60]%
\>[60]{}\Conid{Tree}_{2}\;\Varid{\sigma}\;\Varid{l}\;\Varid{a}{}\<[E]%
\ColumnHook
\end{hscode}\resethooks
But the \ensuremath{\Conid{Tree}_{2}} type requires effect handlers to be applied to subcomputations \emph{immediately}.
Motivated by modeling constructs that defer computation, we generalize the type further by parameterizing subcomputations by the effect functor state, and making each node ``remember'' the effect state (the \emph{latent effects}):
\begin{hscode}\SaveRestoreHook
\column{B}{@{}>{\hspre}l<{\hspost}@{}}%
\column{3}{@{}>{\hspre}l<{\hspost}@{}}%
\column{5}{@{}>{\hspre}l<{\hspost}@{}}%
\column{11}{@{}>{\hspre}c<{\hspost}@{}}%
\column{11E}{@{}l@{}}%
\column{15}{@{}>{\hspre}l<{\hspost}@{}}%
\column{27}{@{}>{\hspre}l<{\hspost}@{}}%
\column{40}{@{}>{\hspre}l<{\hspost}@{}}%
\column{E}{@{}>{\hspre}l<{\hspost}@{}}%
\>[3]{}\mathbf{data}\;\Conid{Tree}\;(\Varid{\sigma}\mathbin{::}\mathbin{*}\to (\mathbin{*}\to \mathbin{*})\to \mathbin{*})\;(\Varid{l}\mathbin{::}\mathbin{*}\to \mathbin{*})\;\Varid{a}\;\mathbf{where}{}\<[E]%
\\
\>[3]{}\hsindent{2}{}\<[5]%
\>[5]{}\Conid{Leaf}{}\<[11]%
\>[11]{}\mathbin{::}{}\<[11E]%
\>[15]{}\Varid{a}\to \Conid{Tree}\;\Varid{\sigma}\;\Varid{l}\;\Varid{a}{}\<[E]%
\\
\>[3]{}\hsindent{2}{}\<[5]%
\>[5]{}\Conid{Node}{}\<[11]%
\>[11]{}\mathbin{::}{}\<[11E]%
\>[15]{}\Varid{\sigma}\;\Varid{p}\;\Varid{c}\to {}\<[27]%
\>[27]{}\colorbox{black!10}{$\Varid{l}\;()$}{}\<[40]%
\>[40]{}\to (\forall \Varid{x}\hsforall \hsdot{\circ }{.}\Varid{c}\;\Varid{x}\to \colorbox{black!10}{$\Varid{l}\;()$}\to \Conid{Tree}\;\Varid{\sigma}\;\Varid{l}\;(\Varid{l}\;\Varid{x})){}\<[E]%
\\
\>[40]{}\to (\Varid{l}\;\Varid{p}\to \Conid{Tree}\;\Varid{\sigma}\;\Varid{l}\;\Varid{a})\to \Conid{Tree}\;\Varid{\sigma}\;\Varid{l}\;\Varid{a}{}\<[E]%
\ColumnHook
\end{hscode}\resethooks

In section \ref{two-modular-handlers-for-function-abstraction} we discuss how \ensuremath{\Conid{Tree}} supports deferring computation.
\ensuremath{\Conid{Tree}} is a monad with a return and bind defined similarly to the ones for \ensuremath{\Conid{Tree}_{1}} in section \ref{a-signature-for-lambda}.
We can also define modular tree constructors using similar techniques as in section \ref{a-signature-for-lambda}.
For instance, using \ensuremath{\Conid{Tree}} instead of \ensuremath{\Conid{Tree}_{1}}, the type of \ensuremath{\Varid{prog}} from Section \ref{code:prog} becomes:
\begin{hscode}\SaveRestoreHook
\column{B}{@{}>{\hspre}l<{\hspost}@{}}%
\column{3}{@{}>{\hspre}l<{\hspost}@{}}%
\column{E}{@{}>{\hspre}l<{\hspost}@{}}%
\>[3]{}\Varid{prog}\mathbin{::}\Conid{Tree}\;(\Conid{Mutating}\;\Conid{V}\mathbin{+}\Conid{Abstracting}\;\Conid{V}\mathbin{+}\Varid{Ending})\;\Conid{Id}\;\Conid{V}{}\<[E]%
\ColumnHook
\end{hscode}\resethooks
Here, the \ensuremath{\Conid{Id}} functor models the absence of latent effects in the tree.
The type \ensuremath{\Conid{V}} represents a concrete value type.

\paragraph{Example}
Figure~\ref{fig:example:types} shows how the type of the \ensuremath{\Varid{prog}} tree evolves
when applying successive handlers for the three parts of the signature:
\begin{hscode}\SaveRestoreHook
\column{B}{@{}>{\hspre}l<{\hspost}@{}}%
\column{3}{@{}>{\hspre}l<{\hspost}@{}}%
\column{E}{@{}>{\hspre}l<{\hspost}@{}}%
\>[3]{}(\Varid{hEnd}\hsdot{\circ }{.}\Varid{hAbs}\;[\mskip1.5mu \mskip1.5mu]\;[\mskip1.5mu \mskip1.5mu]\hsdot{\circ }{.}\Varid{hMut}\;\mathrm{0})\;\Varid{prog}{}\<[E]%
\ColumnHook
\end{hscode}\resethooks
First, we run the modular handler for mutable state \ensuremath{\Conid{Mutating}\;\Varid{s}}, which
has type:
\begin{equation*}
\ensuremath{\Varid{hMut}\mathbin{::}\Conid{Functor}\;\Varid{l}\Rightarrow \Varid{s}\to \Conid{Tree}\;(\Conid{Mutating}\;\Varid{s}\mathbin{+}\Varid{\sigma})\;\Varid{l}\;\Varid{a}\to \Conid{Tree}\;\Varid{\sigma}\;(\Conid{StateL}\;\Varid{s}\;\Varid{l})\;(\Varid{s},\Varid{a})}
\end{equation*}
Given an initial state of type \ensuremath{\Varid{s}}, this handler transforms a tree 
into another tree. The
signature of the tree evolves from \ensuremath{\Conid{Mutating}\;\Varid{s}\mathbin{+}\Varid{\sigma}} to \ensuremath{\Varid{\sigma}} because the handler
interprets the mutable state, but not the other effects. Also, the latent
effect functor evolves from \ensuremath{\Varid{l}} (the latent effects already present) to \ensuremath{\Conid{StateL}\;\Varid{s}\;\Varid{l}}, which augments \ensuremath{\Varid{l}} with the value of the intermediate state.

\begin{hscode}\SaveRestoreHook
\column{B}{@{}>{\hspre}l<{\hspost}@{}}%
\column{23}{@{}>{\hspre}l<{\hspost}@{}}%
\column{E}{@{}>{\hspre}l<{\hspost}@{}}%
\>[B]{}\mathbf{newtype}\;\Conid{StateL}\;\Varid{s}\;\Varid{l}\;\Varid{a}{}\<[23]%
\>[23]{}\mathrel{=}\Conid{StateL}\;\{\mskip1.5mu \Varid{unStateL}\mathbin{::}(\Varid{s},\Varid{l}\;\Varid{a})\mskip1.5mu\}{}\<[E]%
\ColumnHook
\end{hscode}\resethooks

\noindent
The result type evolves from \ensuremath{\Varid{a}} to \ensuremath{(\Varid{s},\Varid{a})}, which augments it with the final state. 

Second, the handler for \ensuremath{\Conid{Abstracting}\;\Conid{V}} behaves similarly to \ensuremath{\Varid{hMut}}, removing
itself from the signature and growing the latent effect functor.
Finally, \ensuremath{\Varid{hEnd}} handles the \ensuremath{\Varid{Ending}} base case. It takes a tree with an empty signature, which thus necessarily
only contains a leaf, and extracts the final result out of it.

\begin{hscode}\SaveRestoreHook
\column{B}{@{}>{\hspre}l<{\hspost}@{}}%
\column{E}{@{}>{\hspre}l<{\hspost}@{}}%
\>[B]{}\Varid{hEnd}\mathbin{::}\Conid{Tree}\;\Varid{Ending}\;\Varid{l}\;\Varid{a}\to \Varid{a}{}\<[E]%
\\
\>[B]{}\Varid{hEnd}\;(\Conid{Leaf}\;\Varid{x})\mathrel{=}\Varid{x}{}\<[E]%
\ColumnHook
\end{hscode}\resethooks

The remainder of this section illustrates how modular handlers are implemented.

\begin{figure}[t!]
\centering
{\scriptsize
\begin{tabular}{l}
\ensuremath{\Conid{Tree}\;($\underline{$\Conid{Mutating}\;\Conid{Int}$}$\mathbin{+}\Conid{Abstracting}\;\Conid{V}\mathbin{+}\Varid{Ending})\;\Conid{Id}\;\Conid{V}} \\
\\

\xhrulefill{black}{.2pt} $\downarrow$~\ensuremath{\Varid{hMut}\;\mathrm{0}} \hspace{1mm} \xhrulefill{black}{.2pt} \\ 
\\
\ensuremath{\Conid{Tree}\;($\underline{$\Conid{Abstracting}\;\Conid{V}$}$\mathbin{+}\Varid{Ending})\;(\colorbox{black!10}{$\Conid{StateL}\;\Conid{Int}$}\;\Conid{Id})\;(\Conid{Int},\Conid{V})} \\
\\
\xhrulefill{black}{.2pt} $\downarrow$~\ensuremath{\Varid{hAbs}\;[\mskip1.5mu \mskip1.5mu]\;[\mskip1.5mu \mskip1.5mu]} \hspace{1mm} \xhrulefill{black}{.2pt} \\ 
\\
\ensuremath{\Conid{Tree}\;($\underline{$\Varid{Ending}$}$)\;(\colorbox{black!10}{$\Conid{StateL}\;(\Conid{Store}\;\Varid{Ending}\;(\Conid{StateL}\;\Conid{Int}\;\Conid{Id})\;\Conid{V})$}\;(\Conid{StateL}\;\Conid{Int}\;\Conid{Id})} \\
\hspace{1,765cm} \ensuremath{(\colorbox{black!10}{$\Conid{Store}\;\Varid{Ending}\;(\Conid{StateL}\;\Conid{Int}\;\Conid{Id})\;\Conid{V}$},(\Conid{Int},\Conid{V}))} \\
\\
\xhrulefill{black}{.2pt} $\downarrow$~\ensuremath{\Varid{hEnd}} \hspace{1mm} \xhrulefill{black}{.2pt}\\ 
\\
\ensuremath{(\Conid{Store}\;\Varid{Ending}\;(\Conid{StateL}\;\Conid{Int}\;\Conid{Id})\;\Conid{V},(\Conid{Int},\Conid{V}))} \\
\end{tabular}
}
\caption{The type of \ensuremath{\Varid{prog}} after successive handling steps.}\label{fig:example:types}
\end{figure}

\subsection{Example: Two Modular Handlers for Function Abstractions}
\label{two-modular-handlers-for-function-abstraction}

We implement two different modular handlers for the operations in
\ensuremath{\Conid{Abstracting}}, which illustrate (1) how latent effects let us write handlers for
function abstraction, and (2) the kind of fine-grained control the handlers
provide.
The first handler we consider evaluates the body of a function abstraction using the latent effects
of its \emph{call site}. Hence, the evaluation of
side-effectful operations is postponed until the function is applied.
The second handler evaluates the body of the function abstraction using the
latent effects of its \emph{definition site}. This immediately enacts the
latent effects introduced by previously-applied handlers.

\subsubsection{Modular Closure Values}

A concern that arises when we step away from the earlier non-modular handler
for \ensuremath{\Conid{Abstracting}} is \emph{reuse}. Notably, in a modular setting we want to
allow reuse of both handlers with different notions of values. For that reason, they are
parameterized in the type of values \ensuremath{\Varid{v}}. This type may comprise various shapes
of values; all the function abstraction handlers require is that closures are
one possible shape of value.

To express this requirement, we introduce another type class \ensuremath{\Varid{v}_{1}\mathbin{<:}\Varid{v}_{2}} for subtyping, this
time at kind \ensuremath{\mathbin{*}}, which witnesses (1) that any \ensuremath{\Varid{v}_{1}} can be ``upcast'' to the
type \ensuremath{\Varid{v}_{2}}; and (2) that a \ensuremath{\Varid{v}_{2}} can be ``downcast'' to type \ensuremath{\Varid{v}_{1}}. The latter may
fail, but the former does not. The minimal requirement for lambda
abstractions is that the value type includes closure values.

\noindent
\begin{minipage}[t][][t]{0.4\textwidth}
\vspace{-\abovedisplayskip}
\begin{hscode}\SaveRestoreHook
\column{B}{@{}>{\hspre}l<{\hspost}@{}}%
\column{3}{@{}>{\hspre}l<{\hspost}@{}}%
\column{9}{@{}>{\hspre}l<{\hspost}@{}}%
\column{E}{@{}>{\hspre}l<{\hspost}@{}}%
\>[B]{}\mathbf{class}\;\Varid{v}_{1}\mathbin{<:}\Varid{v}_{2}\;\mathbf{where}{}\<[E]%
\\
\>[B]{}\hsindent{3}{}\<[3]%
\>[3]{}\Varid{inj}_{\mathrm{v}}{}\<[9]%
\>[9]{}\mathbin{::}\Varid{v}_{1}\to \Varid{v}_{2}{}\<[E]%
\\
\>[B]{}\hsindent{3}{}\<[3]%
\>[3]{}\Varid{proj}_{\mathrm{v}}\mathbin{::}\Varid{v}_{2}\to \Conid{Maybe}\;\Varid{v}_{1}{}\<[E]%
\ColumnHook
\end{hscode}\resethooks
\end{minipage}%
\begin{minipage}[t][][t]{0.6\textwidth}
\vspace{-\abovedisplayskip}
\begin{hscode}\SaveRestoreHook
\column{B}{@{}>{\hspre}l<{\hspost}@{}}%
\column{3}{@{}>{\hspre}l<{\hspost}@{}}%
\column{E}{@{}>{\hspre}l<{\hspost}@{}}%
\>[B]{}\mathbf{data}\;\Conid{Closure}\;\Varid{v}\;\mathbf{where}{}\<[E]%
\\
\>[B]{}\hsindent{3}{}\<[3]%
\>[3]{}\Conid{Clos}\mathbin{::}\Conid{FunPtr}\to \Conid{Env}\;\Varid{v}\to \Conid{Closure}\;\Varid{v}{}\<[E]%
\\[\blanklineskip]%
\>[B]{}\mathbf{type}\;\Conid{Env}\;\Varid{v}\mathrel{=}[\mskip1.5mu \Varid{v}\mskip1.5mu]{}\<[E]%
\ColumnHook
\end{hscode}\resethooks
\end{minipage}
\\%
In the modular setting, the types \ensuremath{\Conid{Closure}\;\Varid{v}} and \ensuremath{\Conid{Env}\;\Varid{v}}, of respectively closures and 
value environments, are 
parameterized in the type of values used.

%

\subsubsection{Modular Resumption Store}

Recall that the resumption store keeps track of the function bodies whose
execution has been deferred; i.e., it is a list of resumptions. 
In the modular setting, the type of resumptions is parametric in the specific
type of signature, latent effect functor and value type.
\begin{hscode}\SaveRestoreHook
\column{B}{@{}>{\hspre}l<{\hspost}@{}}%
\column{3}{@{}>{\hspre}l<{\hspost}@{}}%
\column{E}{@{}>{\hspre}l<{\hspost}@{}}%
\>[3]{}\mathbf{type}\;\Conid{Store}\;\Varid{\sigma}\;\Varid{l}\;\Varid{v}\mathrel{=}[\mskip1.5mu \Conid{R}\;\Varid{\sigma}\;\Varid{l}\;\Varid{v}\mskip1.5mu]{}\<[E]%
\ColumnHook
\end{hscode}\resethooks
Moreover, depending on whether we want to handle latent effects on the call site or
definition site, the definition of a resumption differs.

A resumption of a call-site effect is a function that takes an \ensuremath{\Varid{l}\;()} input,
which is the latent effect context of the call site where the
resumption is evaluated. 
\begin{hscode}\SaveRestoreHook
\column{B}{@{}>{\hspre}l<{\hspost}@{}}%
\column{23}{@{}>{\hspre}l<{\hspost}@{}}%
\column{E}{@{}>{\hspre}l<{\hspost}@{}}%
\>[B]{}\mathbf{type}\;\Varid{R}_{CS}\;\Varid{\sigma}\;\Varid{l}\;\Varid{v}{}\<[23]%
\>[23]{}\mathrel{=}\Varid{l}\;()\to \Conid{Tree}\;(\Conid{Abstracting}\;\Varid{v}\mathbin{+}\Varid{\sigma})\;\Varid{l}\;(\Varid{l}\;\Varid{v}){}\<[E]%
\ColumnHook
\end{hscode}\resethooks
The resumptions of a definition-site effect store are trees instead of functions that produce trees. 
Indeed, they have no dependency on the latent effects of the call site. 
Instead, they have been fully determined by the definition site.
\begin{hscode}\SaveRestoreHook
\column{B}{@{}>{\hspre}l<{\hspost}@{}}%
\column{24}{@{}>{\hspre}l<{\hspost}@{}}%
\column{E}{@{}>{\hspre}l<{\hspost}@{}}%
\>[B]{}\mathbf{type}\;\Varid{R}_{DS}\;\Varid{\sigma}\;\Varid{l}\;\Varid{v}{}\<[24]%
\>[24]{}\mathrel{=}\Conid{Tree}\;(\Conid{Abstracting}\;\Varid{v}\mathbin{+}\Varid{\sigma})\;\Varid{l}\;(\Varid{l}\;\Varid{v}){}\<[E]%
\ColumnHook
\end{hscode}\resethooks
Although the resumption store makes the handlers verbose, 
it is a more modular solution than storing \ensuremath{\Conid{Tree}}s in values.

\subsubsection{Modular Handlers}

Figure~\ref{fig:handleAbsCS} shows the modular handler \ensuremath{\Varid{hAbs}_{CS}} that uses
the call-site latent effects when executing a function body\footnote{The function \ensuremath{(\mathrel{{<\kern-1pt}{\$}})} is short for \ensuremath{\Varid{fmap}\hsdot{\circ }{.}\Varid{const}}}. 

Compared to the non-modular definition, there are several differences.
Firstly, the handler only interprets part of the work and thus returns the
remaining tree instead. Hence, the \ensuremath{\Conid{Leaf}} case now returns a new leaf, and the
other cases use monadic \ensuremath{\mathbf{do}}-notation to build a new tree.  Secondly, because
the signature is a composition and the resulting value type is too, the pattern
matching on operations involves the \ensuremath{\Conid{Inl'}} and \ensuremath{\Conid{Inr'}} tags of the \ensuremath{\mathbin{+}}
co-product.
The pattern matching and construction of values involves \ensuremath{\Varid{inj}_{\mathrm{v}}}
and \ensuremath{\Varid{proj}_{\mathrm{v}}} calls for the same reason.
Thirdly, the latent effects now matter and need to be properly threaded in all
the operation cases.
Finally, there is an additional operation case, to handle unknown operations 
from the remaining part of the signature by ``forwarding'' them, i.e., copying
them to the resulting tree for later handling.

\begin{figure}[t]
\small
\centering
\begin{hscode}\SaveRestoreHook
\column{B}{@{}>{\hspre}l<{\hspost}@{}}%
\column{3}{@{}>{\hspre}l<{\hspost}@{}}%
\column{5}{@{}>{\hspre}l<{\hspost}@{}}%
\column{8}{@{}>{\hspre}l<{\hspost}@{}}%
\column{14}{@{}>{\hspre}l<{\hspost}@{}}%
\column{17}{@{}>{\hspre}l<{\hspost}@{}}%
\column{25}{@{}>{\hspre}l<{\hspost}@{}}%
\column{29}{@{}>{\hspre}l<{\hspost}@{}}%
\column{32}{@{}>{\hspre}l<{\hspost}@{}}%
\column{40}{@{}>{\hspre}l<{\hspost}@{}}%
\column{42}{@{}>{\hspre}l<{\hspost}@{}}%
\column{44}{@{}>{\hspre}l<{\hspost}@{}}%
\column{45}{@{}>{\hspre}l<{\hspost}@{}}%
\column{46}{@{}>{\hspre}l<{\hspost}@{}}%
\column{48}{@{}>{\hspre}l<{\hspost}@{}}%
\column{49}{@{}>{\hspre}l<{\hspost}@{}}%
\column{50}{@{}>{\hspre}l<{\hspost}@{}}%
\column{52}{@{}>{\hspre}l<{\hspost}@{}}%
\column{54}{@{}>{\hspre}l<{\hspost}@{}}%
\column{56}{@{}>{\hspre}l<{\hspost}@{}}%
\column{63}{@{}>{\hspre}l<{\hspost}@{}}%
\column{67}{@{}>{\hspre}l<{\hspost}@{}}%
\column{E}{@{}>{\hspre}l<{\hspost}@{}}%
\>[B]{}\Varid{hAbs}_{CS}{}\<[14]%
\>[14]{}\mathbin{::}(\Conid{Closure}\;\Varid{v}\mathbin{<:}\Varid{v},\Conid{Functor}\;\Varid{l}){}\<[E]%
\\
\>[14]{}\Rightarrow \Conid{Env}\;\Varid{v}\to \Conid{Store}\;\Varid{\sigma}\;\Varid{l}\;\Varid{v}\to \Conid{Tree}\;(\Conid{Abstracting}\;\Varid{v}\mathbin{+}\Varid{\sigma})\;\Varid{l}\;\Varid{a}{}\<[E]%
\\
\>[14]{}\to \Conid{Tree}\;\Varid{\sigma}\;(\Conid{StateL}\;(\Conid{Store}\;\Varid{\sigma}\;\Varid{l}\;\Varid{v})\;\Varid{l})\;(\Conid{Store}\;\Varid{\sigma}\;\Varid{l}\;\Varid{v},\Varid{a}){}\<[E]%
\\
\>[B]{}\Varid{hAbs}_{CS}\;\anonymous \;{}\<[17]%
\>[17]{}\Varid{r}\;(\Conid{Leaf}\;\Varid{x}){}\<[48]%
\>[48]{}\mathrel{=}\Conid{Leaf}\;(\Varid{r},\Varid{x}){}\<[E]%
\\
\>[B]{}\Varid{hAbs}_{CS}\;\Varid{nv}\;{}\<[17]%
\>[17]{}\Varid{r}\;(\Conid{Node}\;(\Conid{Inl'}\;\Conid{Abs'})\;\Varid{l}\;{}\<[40]%
\>[40]{}\Varid{st}\;{}\<[44]%
\>[44]{}\Varid{k}){}\<[48]%
\>[48]{}\mathrel{=}\mathbf{do}\;{}\<[54]%
\>[54]{}\mathbf{let}\;\Varid{v}{}\<[63]%
\>[63]{}\mathrel{=}\Varid{inj}_{\mathrm{v}}\;(\Conid{Clos}\;(\Varid{length}\;\Varid{r})\;\Varid{nv}){}\<[E]%
\\
\>[54]{}\mathbf{let}\;\Varid{r'}{}\<[63]%
\>[63]{}\mathrel{=}\Varid{r}\plus [\mskip1.5mu \colorbox{black!10}{$\Varid{st}\;\Conid{One}$}\mskip1.5mu]{}\<[E]%
\\
\>[54]{}\Varid{hAbs}_{CS}\;{}\<[67]%
\>[67]{}\Varid{nv}\;\Varid{r'}\;(\Varid{k}\;(\Varid{v}\mathrel{{<\kern-1pt}{\$}}\Varid{l})){}\<[E]%
\\
\>[B]{}\Varid{hAbs}_{CS}\;\Varid{nv}\;{}\<[17]%
\>[17]{}\Varid{r}\;(\Conid{Node}\;(\Conid{Inl'}\;(\Conid{App'}\;\Varid{v}_{1}\;\Varid{v}_{2}))\;{}\<[46]%
\>[46]{}\Varid{l}\;{}\<[49]%
\>[49]{}\anonymous \;{}\<[52]%
\>[52]{}\Varid{k}){}\<[56]%
\>[56]{}\mathrel{=}\mathbf{case}\;\Varid{proj}_{\mathrm{v}}\;\Varid{v}_{1}\;\mathbf{of}{}\<[E]%
\\
\>[B]{}\hsindent{5}{}\<[5]%
\>[5]{}\Conid{Just}\;(\Conid{Clos}\;\Varid{fp}\;\Varid{nv'}){}\<[25]%
\>[25]{}\to \mathbf{do}\;{}\<[32]%
\>[32]{}(\Varid{r'},\Varid{v})\leftarrow \Varid{hAbs}_{CS}\;(\Varid{v}_{2}\mathbin{:}\Varid{nv'})\;\Varid{r}\;(\colorbox{black!10}{$(\Varid{r}\mathbin{!!}\Varid{fp})\;\Varid{l}$}){}\<[E]%
\\
\>[32]{}\Varid{hAbs}_{CS}\;\Varid{nv}\;\Varid{r'}\;(\Varid{k}\;\Varid{v}){}\<[E]%
\\
\>[B]{}\hsindent{5}{}\<[5]%
\>[5]{}\Conid{Nothing}{}\<[25]%
\>[25]{}\to \Varid{error}\;\text{\tt \char34 application~error\char34}{}\<[E]%
\\
\>[B]{}\Varid{hAbs}_{CS}\;\Varid{nv}\;{}\<[17]%
\>[17]{}\Varid{r}\;(\Conid{Node}\;(\Conid{Inl'}\;(\Conid{Var'}\;\Varid{n}))\;{}\<[42]%
\>[42]{}\Varid{l}\;{}\<[45]%
\>[45]{}\anonymous \;{}\<[50]%
\>[50]{}\Varid{k}){}\<[54]%
\>[54]{}\mathrel{=}\Varid{hAbs}_{CS}\;\Varid{nv}\;\Varid{r}\;(\Varid{k}\;((\Varid{nv}\mathbin{!!}\Varid{n})\mathrel{{<\kern-1pt}{\$}}\Varid{l})){}\<[E]%
\\
\>[B]{}\Varid{hAbs}_{CS}\;\Varid{nv}\;{}\<[17]%
\>[17]{}\Varid{r}\;(\Conid{Node}\;(\Conid{Inr'}\;\Varid{op})\;{}\<[42]%
\>[42]{}\Varid{l}\;{}\<[45]%
\>[45]{}\Varid{st}\;{}\<[50]%
\>[50]{}\Varid{k}){}\<[54]%
\>[54]{}\mathrel{=}\Conid{Node}\;\Varid{op}\;(\Conid{StateL}\;(\Varid{r},\Varid{l})){}\<[E]%
\\
\>[B]{}\hsindent{3}{}\<[3]%
\>[3]{}(\lambda \Varid{c}\;{}\<[8]%
\>[8]{}(\Conid{StateL}\;(\Varid{r'},\Varid{l}{}\<[25]%
\>[25]{})){}\<[29]%
\>[29]{}\to \Conid{StateL}\mathrel{{<\kern-1pt}{\$}{\kern-1pt>}}\Varid{hAbs}_{CS}\;\Varid{nv}\;\Varid{r'}\;(\Varid{st}\;\Varid{c}\;\Varid{l})){}\<[E]%
\\
\>[B]{}\hsindent{3}{}\<[3]%
\>[3]{}(\lambda {}\<[8]%
\>[8]{}(\Conid{StateL}\;(\Varid{r'},\Varid{lv}{}\<[25]%
\>[25]{})){}\<[29]%
\>[29]{}\to \Varid{hAbs}_{CS}\;\Varid{nv}\;\Varid{r'}\;(\Varid{k}\;\Varid{lv})){}\<[E]%
\ColumnHook
\end{hscode}\resethooks
\caption{Modular call-site abstraction handler. The gray highlights indicate the places
where it differs from a definition-site handler.}
\label{fig:handleAbsCS}
\end{figure}

As discussed, in a modular setting a second handler is possible:
one that uses the latent effects of the definition site
for function bodies rather than their call site.
This definition-site handler, \ensuremath{\Varid{hAbs}_{DS}}, looks much like its sibling.
The key difference is the type of resumptions, 
which affects the code in two places (highlighted in gray).
Firstly, the abstraction case applies the subtree function
to the latent effect of the definition site instead of deferring the application
(\ensuremath{\Varid{st}\;\Conid{One}\;\Varid{l}} instead of \ensuremath{\Varid{st}\;\Conid{One}}).
Dually, the application case does not have to apply the resumption to the
call-site latent effect (\ensuremath{\Varid{r}\mathbin{!!}\Varid{p}} instead of \ensuremath{(\Varid{r}\mathbin{!!}\Varid{p})\;\Varid{l}}).

\paragraph{Example}
\label{cmacro:solution}

With the abstraction handlers in place, let us revisit the \ensuremath{\Varid{prog}} example. 
We run the handlers with default initial values, i.e., 0 for the state, 
the empty variable environment and the empty resumption store. 
When using the call-site abstraction handler after the state handler,
the function body uses the value of the state that was written right
before its invocation.
\begin{hscode}\SaveRestoreHook
\column{B}{@{}>{\hspre}l<{\hspost}@{}}%
\column{3}{@{}>{\hspre}l<{\hspost}@{}}%
\column{E}{@{}>{\hspre}l<{\hspost}@{}}%
\>[3]{}\mathbin{>}\Varid{example}_{CS}\mathrel{=}\Varid{inspect}\mathbin{\$}\Varid{hEnd}\mathbin{\$}\Varid{hAbs}_{CS}\;[\mskip1.5mu \mskip1.5mu]\;[\mskip1.5mu \mskip1.5mu]\mathbin{\$}\Varid{hMut}\;\mathrm{0}\;\Varid{prog}{}\<[E]%
\\
\>[3]{}\mathrm{5}{}\<[E]%
\ColumnHook
\end{hscode}\resethooks
If we use the definition-site handler instead, the function body uses the
state value that was written right before the abstraction was created.
\begin{hscode}\SaveRestoreHook
\column{B}{@{}>{\hspre}l<{\hspost}@{}}%
\column{3}{@{}>{\hspre}l<{\hspost}@{}}%
\column{E}{@{}>{\hspre}l<{\hspost}@{}}%
\>[3]{}\mathbin{>}\Varid{example}_{DS}\mathrel{=}\Varid{inspect}\mathbin{\$}\Varid{hEnd}\mathbin{\$}\Varid{hAbs}_{DS}\;[\mskip1.5mu \mskip1.5mu]\;[\mskip1.5mu \mskip1.5mu]\mathbin{\$}\Varid{hMut}\;\mathrm{0}\;\Varid{prog}{}\<[E]%
\\
\>[3]{}\mathrm{4}{}\<[E]%
\ColumnHook
\end{hscode}\resethooks




\section{Case Study}
\label{case_studies}

This section reports on a case study implementation of a library with a
range of modular effects (Section \ref{app:effect-lib}), and two advanced control-flow features implemented using this library: 
\emph{call-by-need lambdas} (section \ref{call-by-need-evaluation}) and \emph{multi-staging} (section \ref{staging}).
For the source code of these case studies, we refer to the implementation available at
{\tt \url{https://github.com/birthevdb/Latent-Effect-and-Handlers.git}}.

\subsection{Call-by-Need Evaluation}
\label{call-by-need-evaluation}

We have implemented two different evaluation strategies, call-by-need (lazy) and
call-by-value (CBV), for lambdas by using different latent effect handlers. Our
approach is inspired by Levy's call-by-push-value~\cite{Levy06},
which can express both strategies. We summarize here; 
Appendix \ref{app:cbn} has all the details.


Call-by-need evaluation lazily delays the evaluation of argument expressions of
function applications, and uses \emph{memoization} to ensure that evaluation
only happens once for delayed expressions. We build a lazy semantics for
function abstractions out of three primitive effects: 
\begin{enumerate}
  \item The \ensuremath{\Conid{Reading}} effect corresponds to the well-known \emph{reader monad} 
  from the Haskell monad transformer library~\cite{liang95monad}.
  \item The \ensuremath{\Conid{Suspending}} effect delays the evaluation of function bodies, 
  without memoizing the result of the evaluation of the delayed subtrees.
  \item The \ensuremath{\Conid{Thunking}} effect delays the evaluation of argument expressions of 
  function applications, memoizing the result of forcing a thunked computation.
\end{enumerate}
The definition of these effects and their handlers can be found in the effect library of Appendix~\ref{app:effect-lib}.
Using these effects, Appendix~\ref{app:cbn-eval} defines three operations for lazy
evaluation (\ensuremath{\Varid{abs}_{lazy}}, \ensuremath{\Varid{var}_{lazy}}, and \ensuremath{\Varid{app}_{lazy}}).
Lambda abstraction suspends the body of a lambda, and pairs a pointer to the suspension with the environment that the thunk should be evaluated under.
The \ensuremath{\Varid{var}_{lazy}} and \ensuremath{\Varid{app}_{lazy}} functions memoize and recall argument values 
(possibly by forcing the evaluation of a thunked computation), and evaluate the body of a lambda.
Application evaluates the first argument to a function value, 
and memoizes the second argument, which is placed in the current environment.
Then, the function body is executed.
%


The following example program evaluates to 0 when using lazy evaluation:
\begin{hscode}\SaveRestoreHook
\column{B}{@{}>{\hspre}l<{\hspost}@{}}%
\column{3}{@{}>{\hspre}l<{\hspost}@{}}%
\column{22}{@{}>{\hspre}c<{\hspost}@{}}%
\column{22E}{@{}l@{}}%
\column{28}{@{}>{\hspre}l<{\hspost}@{}}%
\column{E}{@{}>{\hspre}l<{\hspost}@{}}%
\>[3]{}\Varid{prog}_{lazy}\mathbin{::}\Conid{Tree}\;{}\<[22]%
\>[22]{}({}\<[22E]%
\>[28]{}\Conid{Mutating}\;\Conid{V}\mathbin{+}\Conid{Reading}\;[\mskip1.5mu \Conid{V}\mskip1.5mu]\mathbin{+}\Conid{Suspending}\;\Conid{V}{}\<[E]%
\\
\>[22]{}\mathbin{+}{}\<[22E]%
\>[28]{}\Conid{Thunking}\;\Conid{V}\mathbin{+}\Varid{Ending})\;\Conid{Id}\;\Conid{V}{}\<[E]%
\\
\>[3]{}\Varid{prog}_{lazy}\mathrel{=}\Varid{app}_{lazy}\;(\Varid{abs}_{lazy}\;\Varid{get})\;(\mathbf{do}\;\Varid{put}\;\mathrm{42};\Varid{get}){}\<[E]%
\ColumnHook
\end{hscode}\resethooks
Function application delays the evaluation of \ensuremath{\Varid{put}} in the argument, 
and is never executed
because the function body does not reference its parameter.%
We can run the program with call-by-need by applying its handlers:
 \begin{hscode}\SaveRestoreHook
\column{B}{@{}>{\hspre}l<{\hspost}@{}}%
\column{3}{@{}>{\hspre}l<{\hspost}@{}}%
\column{E}{@{}>{\hspre}l<{\hspost}@{}}%
\>[3]{}\mathbin{>}\Varid{inspect}\mathbin{\$}\Varid{hEnd}\mathbin{\$}\Varid{hThunk}\;[\mskip1.5mu \mskip1.5mu]\mathbin{\$}\Varid{hSuspend}\;[\mskip1.5mu \mskip1.5mu]\mathbin{\$}\Varid{hRead}\;[\mskip1.5mu \mskip1.5mu]\mathbin{\$}\Varid{hMut}\;\mathrm{0}\;\Varid{prog}_{lazy}{}\<[E]%
\\
\>[3]{}\mathrm{0}{}\<[E]%
\ColumnHook
\end{hscode}\resethooks
The inspect function extracts the final value out of the result that is decorated with the latent
effect functor (in this case nested \ensuremath{\Conid{StateL}}'s).


We can also recover a CBV semantics from \ensuremath{\Varid{app}_{lazy}}, \ensuremath{\Varid{abs}_{lazy}}, and \ensuremath{\Varid{app}_{lazy}} 
by implementing an alternative handler for the \ensuremath{\Conid{Thunking}} effect.
This handler eagerly evaluates subtrees and stores their value in a store.
\begin{hscode}\SaveRestoreHook
\column{B}{@{}>{\hspre}l<{\hspost}@{}}%
\column{3}{@{}>{\hspre}l<{\hspost}@{}}%
\column{E}{@{}>{\hspre}l<{\hspost}@{}}%
\>[3]{}\mathbin{>}\Varid{inspect}\mathbin{\$}\Varid{hEnd}\mathbin{\$}\Varid{hEager}\;[\mskip1.5mu \mskip1.5mu]\mathbin{\$}\Varid{hSuspend}\;[\mskip1.5mu \mskip1.5mu]\mathbin{\$}\Varid{hRead}\;[\mskip1.5mu \mskip1.5mu]\mathbin{\$}\Varid{hMut}\;\mathrm{0}\;\Varid{prog}_{lazy}{}\<[E]%
\\
\>[3]{}\mathrm{42}{}\<[E]%
\ColumnHook
\end{hscode}\resethooks
This case study demonstrates that modular call-by-need can be implemented
by decomposing it into modular, primitive latent effects and handlers.
It also shows how \emph{overloading} the handler of the \ensuremath{\Conid{Thunking}} effect 
provides a means of changing the semantics of a program without touching the
program itself.

\subsection{Staging}
\label{staging}

Another advanced control-flow feature that we have implemented with
latent effects is multi-staging. 
By applying effect handlers \emph{before} the handler for the staging effect, 
we can control which effects should be staged, and which not.
The implementation of these staging constructs 
can be found in Appendix~\ref{app:latent-staging}.

Our inspiration are the three constructs of 
MetaML~\cite{TahaS2000}: (1) \emph{bracket} expressions (\lstinline[language=ml]{<_>}) delay execution to a later stage; (2) \emph{escape} expressions (\lstinline[language=ml]{~_}) splice a staged expression into another; and (3) \emph{run} expressions (\lstinline[language=ml]{run _}) run an expression that has been dynamically generated by bracket and escape expressions.

A key feature of MetaML is that staged code is \emph{statically typed} and \emph{lexically scoped}.
The staging constructs that we implement differ in two ways: our staging constructs are untyped, and we provide \emph{two} constructs for splicing code (\ensuremath{\Varid{push}} and \ensuremath{\Varid{splice}}) instead of the single \emph{escape} expression found in MetaML.

We use \ensuremath{\Varid{push}} for writing programs with escape expressions under binders in staged code.
The dynamic semantics of \ensuremath{\Varid{push}} creates an environment with ``holes'' that represent unknown bindings, and \ensuremath{\Varid{splice}} automatically fills in these holes with bindings from the dynamic context of the \ensuremath{\Varid{splice}} expression.

The four staging constructs we implement are thus: (1) \ensuremath{\Varid{quote}}, corresponding to brackets in MetaML; (2) \ensuremath{\Varid{unquote}}, corresponding to \lstinline[language=ml]{run _} in MetaML; and (3+4) \ensuremath{\Varid{splice}} and \ensuremath{\Varid{push}} for code splicing.
The programs in section \ref{fig:metaml-basic} illustrate the difference in how splicing works.
The MetaML program on the left prints the string \lstinline{"foobar"} and returns the value \lstinline[language=ml]{3}.
The program on the right desugars into latent effects.
With the appropriate handlers, it gives the same output.

Yet, by switching the order of handlers for the \ensuremath{\Varid{print}} effect and staging, we obtain a different semantics that eagerly handles \ensuremath{\Varid{print}} operations inside \ensuremath{\Varid{quote}}d code.
This makes the program on the right print \ensuremath{\text{\tt \char34 barfoo\char34}} instead.

\begin{figure}[t]
\begin{tabular}{c|c}
\begin{minipage}{0.35\linewidth}\scriptsize
\begin{lstlisting}[language=ml,basicstyle=\scriptsize\ttfamily]
run <print "bar";
  1 + ~(print "foo"; <2>)>
\end{lstlisting}
\end{minipage}
&
\vspace{-5pt}
\begin{minipage}{0.72\linewidth}\scriptsize\begin{hscode}\SaveRestoreHook
\column{B}{@{}>{\hspre}l<{\hspost}@{}}%
\column{3}{@{}>{\hspre}l<{\hspost}@{}}%
\column{5}{@{}>{\hspre}l<{\hspost}@{}}%
\column{27}{@{}>{\hspre}l<{\hspost}@{}}%
\column{E}{@{}>{\hspre}l<{\hspost}@{}}%
\>[3]{}\Varid{letbind}\;(\Varid{seq}\;(\Varid{print}\;\text{\tt \char34 foo\char34})\;(\Varid{quote}\;(\Varid{num}\;\mathrm{2})))\;{}\<[E]%
\\
\>[3]{}\hsindent{2}{}\<[5]%
\>[5]{}(\Varid{unquote}\;(\Varid{quote}\;(\Varid{seq}\;{}\<[27]%
\>[27]{}(\Varid{print}\;\text{\tt \char34 bar\char34})\;{}\<[E]%
\\
\>[27]{}(\Varid{add}\;(\Varid{num}\;\mathrm{1})\;(\Varid{splice}\;(\Varid{var}\;\mathrm{0})))))){}\<[E]%
\ColumnHook
\end{hscode}\resethooks
\end{minipage}
\vspace{-5pt}
\end{tabular}
\caption{A MetaML program (left) and its latent effects implementation (right).}
\label{fig:metaml-basic}
\end{figure}

\subsection{Library Summary}
\label{library-of-effects}

We have given two examples where latent 
effects can be modularly composed to form language features. 
Figure \ref{fig:stats} gives an overview of our effect library in Appendix~\ref{app:effect-lib} and how the primitive 
effects are combined into language features.

\begin{figure}[t]
\centering
\includegraphics[width=\textwidth]{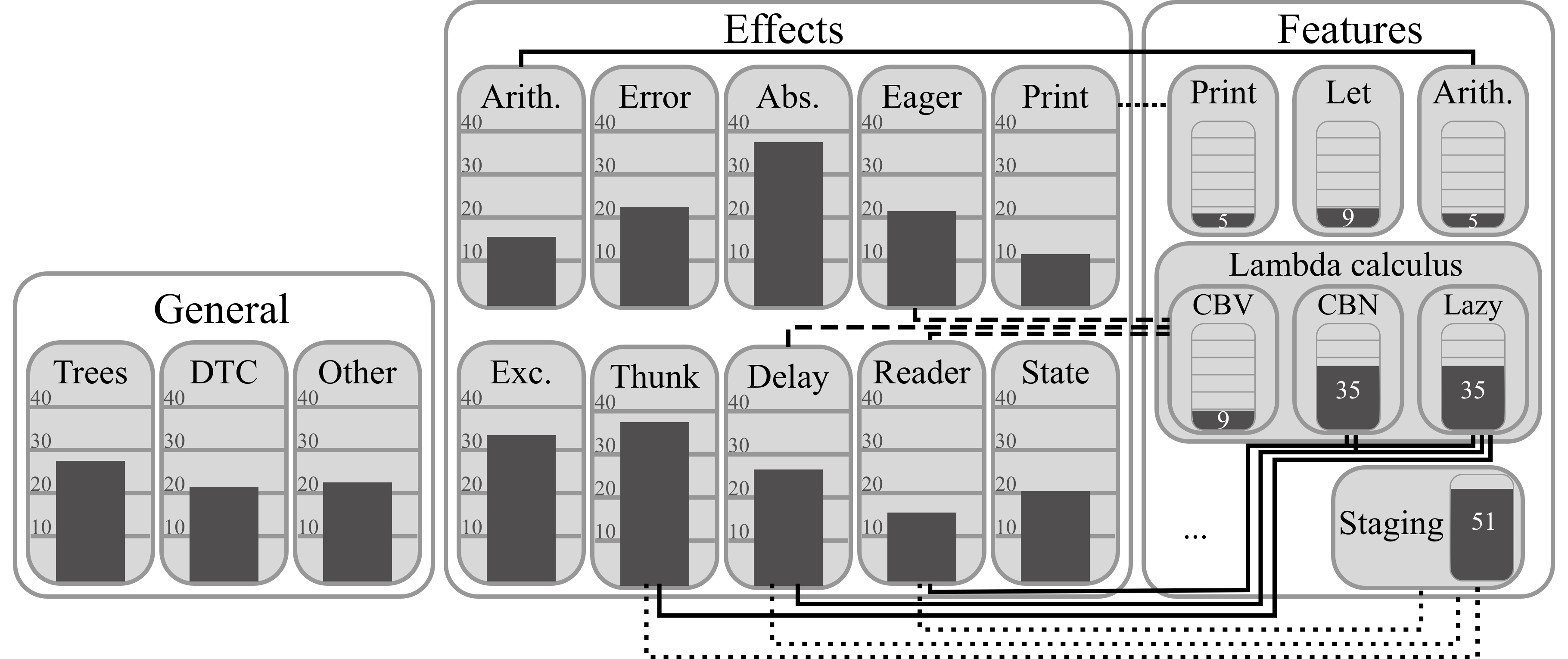}
\caption{Effect Library with Lines of Code (LoC) per Effect.}
\label{fig:stats}
\end{figure}

The left part shows the general framework code for implementing latent effects,
consisting of \ensuremath{\Conid{Tree}}s, 
the DTC approach and 
helper definitions (e.g. \ensuremath{\Conid{Id}}, \ensuremath{\Conid{Void}});
the figure also indicates the associated lines of code (LoC).
The middle part shows ten different effects and their LoC. 
Each effect comes with an effect signature, a handler, and smart constructors
for their operations.
For the detailed implementation of these effects, 
we refer to the effect library (Section \ref{app:effect-lib}).
The right part contains several language features that can be 
implemented using these effects, with their associated LoC.
Each feature comes with its object language syntax and a mapping onto the
effects. 
Each language requires an additional two LoC, to indicate the effects and handlers used
and their order. A different order of effects and handlers may give different
semantics.
%

Figure~\ref{fig:stats} only includes a few language features covered in the paper.
However, as we provide ten effects and handlers, they can be modularly composed
in different order, using different combinations.
In theory, when algebras are fixed, we can define ($10! + 9! + \ldots + 2! + 1!$) = 4,037,913 semantics, 
although some compositions may result in the same behaviour.
Even more variations are possible, varying the algebra that maps the object 
language syntax to the effects.

\section{Related Work}
\label{related_work}

\paragraph{Modular Semantics and Effects}

%
Modularity has received much attention both at the level of language
definitions and of the effects used by those languages.  
A landmark is the formulation of the expression problem~\cite{wadler98expression}, 
the challenge to
modularly extend languages with new features and new interpretation functions.
%
As different language features use different effects, this also requires the
modular composition of those effects. 
Monad transformers \cite{liang95monad} are the most prominent approach, 
including alternative implementations 
such as
Filinksi's layered monads~\cite{Filinski:LayeredMonads} and Jaskelioff's
Monatron~\cite{Monatron}. 

\paragraph{Algebraic Effects}

Algebraic effects~\cite{fossacs/PlotkinP02} have
been proposed as a more structured approach to monads that can
also be composed~\cite{HylandPP:2006}. The subsequent introduction of
handlers~\cite{PlotkinP09} to deal with exceptions has enabled practical
language and library implementations, e.g.,~\cite{KiselyovI15,LindleyMM17,Koka17}.
Schrijvers et al.~\cite{SchrijversPWJ19} identified when algebraic effect handlers are modular
and related this to a subclass of monad transformers, using the notion of
modules~\cite{PirogWG15}. 
%
Wu et al.~\cite{WuSH14} have identified a class of what they call \emph{scoped} effects,
which cannot be expressed as algebraic operations.
To remedy the
situation, they have proposed a practical generalization of algebraic effects.
Pir\'og et al.~\cite{PirogSWJ18} have put this ad-hoc approach for scoped
effects on formal footing in terms of a free monad on a level-indexed category.


\paragraph{Latent Effects}

There are many works on specific types of latent effects. For instance, staging
is a widely studied
area~\cite{RompfO10,TahaS97,SheardJ02}.  Some
works have also combined algebraic effects with staging
mechanisms~\cite{Yallop17,SchusterBO20,WeiBTR20}. Yet, we are, to the best of
our knowledge, the first to consider staging using effect
handlers.

The call-by-push-value calculus of Levy~\cite{Levy06} provides primitives for
expressing both call-by-name and call-by-value. These have been an
inspiration for our modular thunking handler.
A more generic work is that of Atkey and Johann~\cite{AtkeyJ15} on interleaving data and effects
to model the incremental production of the data, and on interpreting these with
$f$-and-$m$ algebras.

Various forms of delimited control have been used in the literature to realize
sophisticated control mechanisms, such as the simulation of call-by-need by
Garcia et al.~\cite{GarciaLS10}. 
Moreover, several
works~\cite{ForsterKLP19}
show the
interdefinability of conventional algebraic effects and delimited control. A
further investigation into the relative expressiveness of latent effects would
be interesting.

In future work we would like to demonstrate the performance of latent effects, 
using the techniques of fusion by Wu and Schrijvers~\cite{mpc2015}.


\section{Conclusion}
\label{conclusion}

This paper has introduced the notion of latent effects. These extend algebraic
effects with the ability to modularly model advanced control-flow mechanisms that can
postpone the execution of certain computations and require fine-grained control
over the effects inside them. Lambda abstraction, lazy evaluation,
and staging were shown to be three prominent instances.

\subsubsection*{Acknowledgments}

This work has been supported by EPSRC grant number EP/S028129/1 on ‘Scoped
Contextual Operations and Effects’, by the NWO VENI project on `Composable and
Safe-by-Construction Programming Language Definitions' (VI.Veni.192.259), 
by FWO project G095917N, and KU Leuven project C14/20/079.

%
%
%
\bibliographystyle{splncs04}
\bibliography{bibliography}

\clearpage
\appendix

\section{An Effect Library}
\label{app:effect-lib}

This Appendix contains a library of effects that can be modularly composed. 
Each effect can be implemented with a limited number of lines of code.

\subsection{General}

\subsubsection{Trees and Their Instances}
The \ensuremath{\Conid{Tree}} datatype for representing latent effects with its corresponding
instances.

\begin{hscode}\SaveRestoreHook
\column{B}{@{}>{\hspre}l<{\hspost}@{}}%
\column{3}{@{}>{\hspre}l<{\hspost}@{}}%
\column{9}{@{}>{\hspre}c<{\hspost}@{}}%
\column{9E}{@{}l@{}}%
\column{10}{@{}>{\hspre}l<{\hspost}@{}}%
\column{13}{@{}>{\hspre}l<{\hspost}@{}}%
\column{18}{@{}>{\hspre}c<{\hspost}@{}}%
\column{18E}{@{}l@{}}%
\column{23}{@{}>{\hspre}l<{\hspost}@{}}%
\column{26}{@{}>{\hspre}l<{\hspost}@{}}%
\column{E}{@{}>{\hspre}l<{\hspost}@{}}%
\>[B]{}\mathbf{data}\;\Conid{Tree}\;(\Varid{\sigma}\mathbin{::}\mathbin{*}\to (\mathbin{*}\to \mathbin{*})\to \mathbin{*})\;(\Varid{l}\mathbin{::}\mathbin{*}\to \mathbin{*})\;\Varid{a}\;\mathbf{where}{}\<[E]%
\\
\>[B]{}\hsindent{3}{}\<[3]%
\>[3]{}\Conid{Leaf}{}\<[9]%
\>[9]{}\mathbin{::}{}\<[9E]%
\>[13]{}\Varid{a}\to \Conid{Tree}\;\Varid{\sigma}\;\Varid{l}\;\Varid{a}{}\<[E]%
\\
\>[B]{}\hsindent{3}{}\<[3]%
\>[3]{}\Conid{Node}{}\<[9]%
\>[9]{}\mathbin{::}{}\<[9E]%
\>[13]{}\Varid{\sigma}\;\Varid{p}\;\Varid{c}{}\<[E]%
\\
\>[9]{}\to {}\<[9E]%
\>[13]{}\Varid{l}\;(){}\<[E]%
\\
\>[9]{}\to {}\<[9E]%
\>[13]{}(\forall \Varid{x}\hsforall \hsdot{\circ }{.}\Varid{c}\;\Varid{x}\to \Varid{l}\;()\to \Conid{Tree}\;\Varid{\sigma}\;\Varid{l}\;(\Varid{l}\;\Varid{x})){}\<[E]%
\\
\>[9]{}\to {}\<[9E]%
\>[13]{}(\Varid{l}\;\Varid{p}\to \Conid{Tree}\;\Varid{\sigma}\;\Varid{l}\;\Varid{a}){}\<[E]%
\\
\>[9]{}\to {}\<[9E]%
\>[13]{}\Conid{Tree}\;\Varid{\sigma}\;\Varid{l}\;\Varid{a}{}\<[E]%
\\[\blanklineskip]%
\>[B]{}\mathbf{instance}\;\Conid{Functor}\;(\Conid{Tree}\;\Varid{\sigma}\;\Varid{l})\;\mathbf{where}{}\<[E]%
\\
\>[B]{}\hsindent{3}{}\<[3]%
\>[3]{}\Varid{fmap}\mathrel{=}\Varid{liftM}{}\<[E]%
\\[\blanklineskip]%
\>[B]{}\mathbf{instance}\;\Conid{Applicative}\;(\Conid{Tree}\;\Varid{\sigma}\;\Varid{l})\;\mathbf{where}{}\<[E]%
\\
\>[B]{}\hsindent{3}{}\<[3]%
\>[3]{}\Varid{pure}{}\<[10]%
\>[10]{}\mathrel{=}\Varid{return}{}\<[E]%
\\
\>[B]{}\hsindent{3}{}\<[3]%
\>[3]{}(\mathrel{{<\kern-1pt}{\ast}{\kern-1pt>}}){}\<[10]%
\>[10]{}\mathrel{=}\Varid{ap}{}\<[E]%
\\[\blanklineskip]%
\>[B]{}\mathbf{instance}\;\Conid{Monad}\;(\Conid{Tree}\;\Varid{\sigma}\;\Varid{l})\;\mathbf{where}{}\<[E]%
\\
\>[B]{}\hsindent{3}{}\<[3]%
\>[3]{}\Varid{return}{}\<[26]%
\>[26]{}\mathrel{=}\Conid{Leaf}{}\<[E]%
\\
\>[B]{}\hsindent{3}{}\<[3]%
\>[3]{}\Conid{Leaf}\;\Varid{x}{}\<[18]%
\>[18]{}\bind {}\<[18E]%
\>[23]{}\Varid{f}{}\<[26]%
\>[26]{}\mathrel{=}\Varid{f}\;\Varid{x}{}\<[E]%
\\
\>[B]{}\hsindent{3}{}\<[3]%
\>[3]{}\Conid{Node}\;\Varid{c}\;\Varid{l}\;\Varid{st}\;\Varid{k}{}\<[18]%
\>[18]{}\bind {}\<[18E]%
\>[23]{}\Varid{f}{}\<[26]%
\>[26]{}\mathrel{=}\Conid{Node}\;\Varid{c}\;\Varid{l}\;\Varid{st}\;(\lambda \Varid{x}\to \Varid{k}\;\Varid{x}\bind \Varid{f}){}\<[E]%
\ColumnHook
\end{hscode}\resethooks

\subsubsection{Datatypes \`a la Carte}
We use a DTC approach to compose and project values and signatures. 

\begin{hscode}\SaveRestoreHook
\column{B}{@{}>{\hspre}l<{\hspost}@{}}%
\column{3}{@{}>{\hspre}l<{\hspost}@{}}%
\column{9}{@{}>{\hspre}l<{\hspost}@{}}%
\column{10}{@{}>{\hspre}l<{\hspost}@{}}%
\column{22}{@{}>{\hspre}l<{\hspost}@{}}%
\column{E}{@{}>{\hspre}l<{\hspost}@{}}%
\>[B]{}\mathbf{class}\;\Varid{v}_{1}\mathbin{<:}\Varid{v}_{2}\;\mathbf{where}{}\<[E]%
\\
\>[B]{}\hsindent{3}{}\<[3]%
\>[3]{}\Varid{inj}_{\mathrm{v}}{}\<[10]%
\>[10]{}\mathbin{::}\Varid{v}_{1}\to \Varid{v}_{2}{}\<[E]%
\\
\>[B]{}\hsindent{3}{}\<[3]%
\>[3]{}\Varid{proj}_{\mathrm{v}}{}\<[10]%
\>[10]{}\mathbin{::}\Varid{v}_{2}\to \Conid{Maybe}\;\Varid{v}_{1}{}\<[E]%
\\[\blanklineskip]%
\>[B]{}\mathbf{infixr}\mathbin{+}{}\<[E]%
\\[\blanklineskip]%
\>[B]{}\mathbf{data}\;(\Varid{\sigma}_{1}\mathbin{+}\Varid{\sigma}_{2})\mathbin{::}\mathbin{*}\to (\mathbin{*}\to \mathbin{*})\to \mathbin{*}\mathbf{where}{}\<[E]%
\\
\>[B]{}\hsindent{3}{}\<[3]%
\>[3]{}\Conid{Inl'}{}\<[9]%
\>[9]{}\mathbin{::}\Varid{\sigma}_{1}\;\Varid{p}\;\Varid{c}{}\<[22]%
\>[22]{}\to (\Varid{\sigma}_{1}\mathbin{+}\Varid{\sigma}_{2})\;\Varid{p}\;\Varid{c}{}\<[E]%
\\
\>[B]{}\hsindent{3}{}\<[3]%
\>[3]{}\Conid{Inr'}{}\<[9]%
\>[9]{}\mathbin{::}\Varid{\sigma}_{2}\;\Varid{p}\;\Varid{c}{}\<[22]%
\>[22]{}\to (\Varid{\sigma}_{1}\mathbin{+}\Varid{\sigma}_{2})\;\Varid{p}\;\Varid{c}{}\<[E]%
\\[\blanklineskip]%
\>[B]{}\mathbf{class}\;(\Varid{\sigma}_{1}\mathbin{::}\mathbin{*}\to (\mathbin{*}\to \mathbin{*})\to \mathbin{*})\mathbin{<}\Varid{\sigma}_{2}\;\mathbf{where}{}\<[E]%
\\
\>[B]{}\hsindent{3}{}\<[3]%
\>[3]{}\Varid{injSig}\mathbin{::}\Varid{\sigma}_{1}\;\Varid{p}\;\Varid{c}\to \Varid{\sigma}_{2}\;\Varid{p}\;\Varid{c}{}\<[E]%
\\[\blanklineskip]%
\>[B]{}\mathbf{instance}\;\Varid{\sigma}\mathbin{<}\Varid{\sigma}\;\mathbf{where}{}\<[E]%
\\
\>[B]{}\hsindent{3}{}\<[3]%
\>[3]{}\Varid{injSig}\mathrel{=}\Varid{id}{}\<[E]%
\\[\blanklineskip]%
\>[B]{}\mathbf{instance}\;\Varid{\sigma}_{1}\mathbin{<}(\Varid{\sigma}_{1}\mathbin{+}\Varid{\sigma}_{2})\;\mathbf{where}{}\<[E]%
\\
\>[B]{}\hsindent{3}{}\<[3]%
\>[3]{}\Varid{injSig}\mathrel{=}\Conid{Inl'}{}\<[E]%
\\[\blanklineskip]%
\>[B]{}\mathbf{instance}\;(\Varid{\sigma}_{1}\mathbin{<}\Varid{\sigma}_{3})\Rightarrow \Varid{\sigma}_{1}\mathbin{<}(\Varid{\sigma}_{2}\mathbin{+}\Varid{\sigma}_{3})\;\mathbf{where}{}\<[E]%
\\
\>[B]{}\hsindent{3}{}\<[3]%
\>[3]{}\Varid{injSig}\mathrel{=}\Conid{Inr'}\hsdot{\circ }{.}\Varid{injSig}{}\<[E]%
\ColumnHook
\end{hscode}\resethooks

\subsubsection{Other Datatypes and Helper Functions}
Other straightforward datatypes such as the identity functor and \ensuremath{\Conid{Void}}.

\begin{hscode}\SaveRestoreHook
\column{B}{@{}>{\hspre}l<{\hspost}@{}}%
\column{5}{@{}>{\hspre}l<{\hspost}@{}}%
\column{20}{@{}>{\hspre}l<{\hspost}@{}}%
\column{E}{@{}>{\hspre}l<{\hspost}@{}}%
\>[B]{}\mathbf{data}\;\Conid{Void}{}\<[E]%
\\[\blanklineskip]%
\>[B]{}\mathbf{newtype}\;\Conid{Id}\;\Varid{a}\mathrel{=}\Conid{Id}\;\{\mskip1.5mu \Varid{unId}\mathbin{::}\Varid{a}\mskip1.5mu\}\;\mathbf{deriving}\;(\Conid{Functor}){}\<[E]%
\\[\blanklineskip]%
\>[B]{}\mathbf{instance}\;\Conid{Applicative}\;\Conid{Id}\;\mathbf{where}{}\<[E]%
\\
\>[B]{}\hsindent{5}{}\<[5]%
\>[5]{}\Varid{pure}\;\Varid{a}{}\<[20]%
\>[20]{}\mathrel{=}\Conid{Id}\;\Varid{a}{}\<[E]%
\\
\>[B]{}\hsindent{5}{}\<[5]%
\>[5]{}\Conid{Id}\;\Varid{f}\mathrel{{<\kern-1pt}{\ast}{\kern-1pt>}}\Conid{Id}\;\Varid{x}{}\<[20]%
\>[20]{}\mathrel{=}\Conid{Id}\;(\Varid{f}\;\Varid{x}){}\<[E]%
\ColumnHook
\end{hscode}\resethooks

The \ensuremath{\Conid{StateL}\;\Varid{s}\;\Varid{l}\;\Varid{a}} type is a wrapper for the state type \ensuremath{(\Varid{s},\Varid{a})}, where the
value is wrapped in a latent effect functor \ensuremath{\Varid{l}}.
Similarly, \ensuremath{\Conid{EitherL}\;\Varid{left}\;\Varid{l}\;\Varid{a}} is a wrapper for the \ensuremath{\Conid{Either}\;\Varid{left}\;\Varid{a}} type, where
the right value is wrapped in a latent effect functor \ensuremath{\Varid{l}}.

\begin{hscode}\SaveRestoreHook
\column{B}{@{}>{\hspre}l<{\hspost}@{}}%
\column{3}{@{}>{\hspre}l<{\hspost}@{}}%
\column{23}{@{}>{\hspre}l<{\hspost}@{}}%
\column{E}{@{}>{\hspre}l<{\hspost}@{}}%
\>[B]{}\mathbf{newtype}\;\Conid{StateL}\;\Varid{s}\;\Varid{l}\;\Varid{a}{}\<[23]%
\>[23]{}\mathrel{=}\Conid{StateL}\;\{\mskip1.5mu \Varid{unStateL}\mathbin{::}(\Varid{s},\Varid{l}\;\Varid{a})\mskip1.5mu\}{}\<[E]%
\\
\>[B]{}\hsindent{3}{}\<[3]%
\>[3]{}\mathbf{deriving}\;\Conid{Show}{}\<[E]%
\\[\blanklineskip]%
\>[B]{}\mathbf{instance}\;\Conid{Functor}\;\Varid{l}\Rightarrow \Conid{Functor}\;(\Conid{StateL}\;\Varid{s}\;\Varid{l})\;\mathbf{where}{}\<[E]%
\\
\>[B]{}\hsindent{3}{}\<[3]%
\>[3]{}\Varid{fmap}\;\Varid{f}\;(\Conid{StateL}\;(\Varid{s},\Varid{la}))\mathrel{=}\Conid{StateL}\;(\Varid{s},\Varid{fmap}\;\Varid{f}\;\Varid{la}){}\<[E]%
\\[\blanklineskip]%
\>[B]{}\mathbf{newtype}\;\Conid{EitherL}\;\Varid{left}\;\Varid{l}\;\Varid{a}\mathrel{=}\Conid{EitherL}\;\{\mskip1.5mu \Varid{unEitherL}\mathbin{::}\Conid{Either}\;\Varid{left}\;(\Varid{l}\;\Varid{a})\mskip1.5mu\}{}\<[E]%
\ColumnHook
\end{hscode}\resethooks

\subsubsection{Extracting the Value from the Effect Tree}
\ensuremath{\Varid{Ending}} has an empty signature, so the tree can only consist of
  a leave. 
The handler extracts the value from that leave, possibly decorated with latent
effects.

\begin{hscode}\SaveRestoreHook
\column{B}{@{}>{\hspre}l<{\hspost}@{}}%
\column{E}{@{}>{\hspre}l<{\hspost}@{}}%
\>[B]{}\mathbf{data}\;\Varid{Ending}\mathbin{::}\mathbin{*}\to (\mathbin{*}\to \mathbin{*})\to \mathbin{*}{}\<[E]%
\\[\blanklineskip]%
\>[B]{}\Varid{hEnd}\mathbin{::}\Conid{Tree}\;\Varid{Ending}\;\Varid{l}\;\Varid{a}\to \Varid{a}{}\<[E]%
\\
\>[B]{}\Varid{hEnd}\;(\Conid{Leaf}\;\Varid{x})\mathrel{=}\Varid{x}{}\<[E]%
\ColumnHook
\end{hscode}\resethooks

\subsubsection{Subtrees of the \ensuremath{\Conid{Tree}} type}

\begin{hscode}\SaveRestoreHook
\column{B}{@{}>{\hspre}l<{\hspost}@{}}%
\column{3}{@{}>{\hspre}l<{\hspost}@{}}%
\column{13}{@{}>{\hspre}l<{\hspost}@{}}%
\column{16}{@{}>{\hspre}l<{\hspost}@{}}%
\column{18}{@{}>{\hspre}l<{\hspost}@{}}%
\column{26}{@{}>{\hspre}l<{\hspost}@{}}%
\column{38}{@{}>{\hspre}l<{\hspost}@{}}%
\column{40}{@{}>{\hspre}l<{\hspost}@{}}%
\column{47}{@{}>{\hspre}l<{\hspost}@{}}%
\column{70}{@{}>{\hspre}l<{\hspost}@{}}%
\column{E}{@{}>{\hspre}l<{\hspost}@{}}%
\>[B]{}\mathbf{data}\;\Conid{NoSub}{}\<[16]%
\>[16]{}\mathbin{::}\mathbin{*}\to \mathbin{*}\mathbf{where}{}\<[E]%
\\[\blanklineskip]%
\>[B]{}\mathbf{data}\;\Conid{OneSub}\;\Varid{v}{}\<[16]%
\>[16]{}\mathbin{::}\mathbin{*}\to \mathbin{*}\mathbf{where}{}\<[E]%
\\
\>[B]{}\hsindent{3}{}\<[3]%
\>[3]{}\Conid{One}\mathbin{::}\Conid{OneSub}\;\Varid{v}\;\Varid{v}{}\<[E]%
\\[\blanklineskip]%
\>[B]{}\Varid{noSubNode}{}\<[18]%
\>[18]{}\mathbin{::}\forall \Varid{v}\hsforall \;\Varid{\sigma}_{1}\;\Varid{\sigma}_{2}\hsdot{\circ }{.}\Varid{\sigma}_{1}\mathbin{<}\Varid{\sigma}_{2}{}\<[E]%
\\
\>[18]{}\Rightarrow \Varid{\sigma}_{1}\;\Varid{v}\;\Conid{NoSub}\to \Conid{Tree}\;\Varid{\sigma}_{2}\;\Conid{Id}\;\Varid{v}{}\<[E]%
\\
\>[B]{}\Varid{noSubNode}\;{}\<[13]%
\>[13]{}\Varid{n}{}\<[18]%
\>[18]{}\mathrel{=}\Conid{Node}\;{}\<[26]%
\>[26]{}(\Varid{injSig}\;\Varid{n})\;{}\<[38]%
\>[38]{}(\Conid{Id}\;())\;{}\<[47]%
\>[47]{}(\lambda \Varid{x}\to \mathbf{case}\;\Varid{x}\;\mathbf{of}\;)\;{}\<[70]%
\>[70]{}(\Conid{Leaf}\hsdot{\circ }{.}\Varid{unId}){}\<[E]%
\\[\blanklineskip]%
\>[B]{}\Varid{oneSubNode}{}\<[18]%
\>[18]{}\mathbin{::}\forall \Varid{v}\hsforall \;\Varid{w}\;\Varid{\sigma}_{1}\;\Varid{\sigma}_{2}\hsdot{\circ }{.}\Varid{\sigma}_{1}\mathbin{<}\Varid{\sigma}_{2}{}\<[E]%
\\
\>[18]{}\Rightarrow \Varid{\sigma}_{1}\;\Varid{w}\;(\Conid{OneSub}\;\Varid{v}){}\<[40]%
\>[40]{}\to \Conid{Tree}\;\Varid{\sigma}_{2}\;\Conid{Id}\;\Varid{v}\to \Conid{Tree}\;\Varid{\sigma}_{2}\;\Conid{Id}\;\Varid{w}{}\<[E]%
\\
\>[B]{}\Varid{oneSubNode}\;{}\<[13]%
\>[13]{}\Varid{n}\;\Varid{t}{}\<[18]%
\>[18]{}\mathrel{=}\Conid{Node}\;{}\<[26]%
\>[26]{}(\Varid{injSig}\;\Varid{n})\;{}\<[38]%
\>[38]{}(\Conid{Id}\;())\;{}\<[47]%
\>[47]{}(\lambda \Conid{One}\;\anonymous \to \Conid{Id}\mathrel{{<\kern-1pt}{\$}{\kern-1pt>}}\Varid{t})\;{}\<[70]%
\>[70]{}(\Conid{Leaf}\hsdot{\circ }{.}\Varid{unId}){}\<[E]%
\ColumnHook
\end{hscode}\resethooks

\subsection{Effects}

In this section, we present small standalone implementations of a single effect, 
that can be combined in a modular fashion into languages with different effect
features.
These implementations come with a definition of the effect, 
a handler for interpreting the effect, 
and smart constructors.

\subsubsection{Reader}
The \ensuremath{\Conid{Reading}} effect provides two operations: \ensuremath{\Varid{ask}} retrieves the current environment, and \ensuremath{\Varid{local}} passes down an updated environment to a subtree.

\begin{hscode}\SaveRestoreHook
\column{B}{@{}>{\hspre}l<{\hspost}@{}}%
\column{3}{@{}>{\hspre}l<{\hspost}@{}}%
\column{8}{@{}>{\hspre}l<{\hspost}@{}}%
\column{10}{@{}>{\hspre}l<{\hspost}@{}}%
\column{12}{@{}>{\hspre}l<{\hspost}@{}}%
\column{14}{@{}>{\hspre}l<{\hspost}@{}}%
\column{20}{@{}>{\hspre}l<{\hspost}@{}}%
\column{26}{@{}>{\hspre}l<{\hspost}@{}}%
\column{33}{@{}>{\hspre}l<{\hspost}@{}}%
\column{36}{@{}>{\hspre}l<{\hspost}@{}}%
\column{40}{@{}>{\hspre}l<{\hspost}@{}}%
\column{44}{@{}>{\hspre}l<{\hspost}@{}}%
\column{E}{@{}>{\hspre}l<{\hspost}@{}}%
\>[B]{}\mathbf{data}\;\Conid{Reading}\;\Varid{r}\mathbin{::}{}\<[20]%
\>[20]{}\mathbin{*}\to (\mathbin{*}\to \mathbin{*})\to \mathbin{*}\mathbf{where}{}\<[E]%
\\
\>[B]{}\hsindent{3}{}\<[3]%
\>[3]{}\Conid{Local}{}\<[10]%
\>[10]{}\mathbin{::}(\Varid{r}\to \Varid{r})\to {}\<[26]%
\>[26]{}\Conid{Reading}\;\Varid{r}\;\Varid{a}\;(\Conid{OneSub}\;\Varid{a}){}\<[E]%
\\
\>[B]{}\hsindent{3}{}\<[3]%
\>[3]{}\Conid{Ask}{}\<[10]%
\>[10]{}\mathbin{::}{}\<[26]%
\>[26]{}\Conid{Reading}\;\Varid{r}\;\Varid{r}\;\Conid{NoSub}{}\<[E]%
\\[\blanklineskip]%
\>[B]{}\Varid{hRead}{}\<[8]%
\>[8]{}\mathbin{::}\Conid{Functor}\;\Varid{l}{}\<[E]%
\\
\>[8]{}\Rightarrow \Varid{r}{}\<[E]%
\\
\>[8]{}\to \Conid{Tree}\;(\Conid{Reading}\;\Varid{r}\mathbin{+}\sigma)\;\Varid{l}\;(\Varid{l}\;\Varid{a}){}\<[E]%
\\
\>[8]{}\to \Conid{Tree}\;\sigma\;\Varid{l}\;(\Varid{l}\;\Varid{a}){}\<[E]%
\\
\>[B]{}\Varid{hRead}\;\Varid{r}\;(\Conid{Leaf}\;\Varid{x}){}\<[44]%
\>[44]{}\mathrel{=}\Conid{Leaf}\;\Varid{x}{}\<[E]%
\\
\>[B]{}\Varid{hRead}\;\Varid{r}\;(\Conid{Node}\;(\Conid{Inl'}\;(\Conid{Local}\;\Varid{f}))\;{}\<[33]%
\>[33]{}\Varid{l}\;{}\<[36]%
\>[36]{}\Varid{st}\;{}\<[40]%
\>[40]{}\Varid{k}){}\<[44]%
\>[44]{}\mathrel{=}\Varid{hRead}\;(\Varid{f}\;\Varid{r})\;(\Varid{st}\;\Conid{One}\;\Varid{l})\bind \Varid{hRead}\;\Varid{r}\hsdot{\circ }{.}\Varid{k}{}\<[E]%
\\
\>[B]{}\Varid{hRead}\;\Varid{r}\;(\Conid{Node}\;(\Conid{Inl'}\;\Conid{Ask})\;{}\<[33]%
\>[33]{}\Varid{l}\;{}\<[36]%
\>[36]{}\anonymous \;{}\<[40]%
\>[40]{}\Varid{k}){}\<[44]%
\>[44]{}\mathrel{=}\Varid{hRead}\;\Varid{r}\;(\Varid{k}\;(\Varid{r}\mathrel{{<\kern-1pt}{\$}}\Varid{l})){}\<[E]%
\\
\>[B]{}\Varid{hRead}\;\Varid{r}\;(\Conid{Node}\;(\Conid{Inr'}\;\Varid{op})\;{}\<[33]%
\>[33]{}\Varid{l}\;{}\<[36]%
\>[36]{}\Varid{st}\;{}\<[40]%
\>[40]{}\Varid{k}){}\<[44]%
\>[44]{}\mathrel{=}{}\<[E]%
\\
\>[B]{}\hsindent{3}{}\<[3]%
\>[3]{}\Conid{Node}\;\Varid{op}\;\Varid{l}\;{}\<[14]%
\>[14]{}(\lambda \Varid{c}_{2}\to \Varid{hRead}\;\Varid{r}\hsdot{\circ }{.}\Varid{st}\;\Varid{c}_{2})\;(\Varid{hRead}\;\Varid{r}\hsdot{\circ }{.}\Varid{k}){}\<[E]%
\\[\blanklineskip]%
\>[B]{}\Varid{local}{}\<[12]%
\>[12]{}\mathbin{::}\forall \Varid{r}\hsforall \;\sigma\;\Varid{a}\hsdot{\circ }{.}(\Conid{Reading}\;\Varid{r}\mathbin{<}\sigma)\Rightarrow (\Varid{r}\to \Varid{r})\to \Conid{Tree}\;\sigma\;\Conid{Id}\;\Varid{a}\to \Conid{Tree}\;\sigma\;\Conid{Id}\;\Varid{a}{}\<[E]%
\\
\>[B]{}\Varid{local}\;\Varid{f}\;\Varid{t}{}\<[12]%
\>[12]{}\mathrel{=}\Varid{oneSubNode}\;(\Conid{Local}\;\Varid{f})\;\Varid{t}{}\<[E]%
\\[\blanklineskip]%
\>[B]{}\Varid{ask}{}\<[12]%
\>[12]{}\mathbin{::}(\Conid{Reading}\;\Varid{r}\mathbin{<}\sigma)\Rightarrow \Conid{Tree}\;\sigma\;\Conid{Id}\;\Varid{r}{}\<[E]%
\\
\>[B]{}\Varid{ask}{}\<[12]%
\>[12]{}\mathrel{=}\Varid{noSubNode}\;\Conid{Ask}{}\<[E]%
\ColumnHook
\end{hscode}\resethooks

\subsubsection{State}
The \ensuremath{\Conid{Mutating}} effect provides the two standard operations for state: 
\ensuremath{\Varid{get}} and \ensuremath{\Varid{put}}.

\begin{hscode}\SaveRestoreHook
\column{B}{@{}>{\hspre}l<{\hspost}@{}}%
\column{3}{@{}>{\hspre}l<{\hspost}@{}}%
\column{8}{@{}>{\hspre}l<{\hspost}@{}}%
\column{9}{@{}>{\hspre}l<{\hspost}@{}}%
\column{11}{@{}>{\hspre}l<{\hspost}@{}}%
\column{13}{@{}>{\hspre}l<{\hspost}@{}}%
\column{17}{@{}>{\hspre}l<{\hspost}@{}}%
\column{27}{@{}>{\hspre}l<{\hspost}@{}}%
\column{32}{@{}>{\hspre}l<{\hspost}@{}}%
\column{33}{@{}>{\hspre}l<{\hspost}@{}}%
\column{36}{@{}>{\hspre}l<{\hspost}@{}}%
\column{40}{@{}>{\hspre}l<{\hspost}@{}}%
\column{44}{@{}>{\hspre}l<{\hspost}@{}}%
\column{53}{@{}>{\hspre}l<{\hspost}@{}}%
\column{E}{@{}>{\hspre}l<{\hspost}@{}}%
\>[B]{}\mathbf{data}\;\Conid{Mutating}\;\Varid{v}\mathbin{::}\mathbin{*}\to (\mathbin{*}\to \mathbin{*})\to \mathbin{*}\mathbf{where}{}\<[E]%
\\
\>[B]{}\hsindent{3}{}\<[3]%
\>[3]{}\Conid{Get}{}\<[8]%
\>[8]{}\mathbin{::}{}\<[17]%
\>[17]{}\Conid{Mutating}\;\Varid{v}\;\Varid{v}\;{}\<[32]%
\>[32]{}\Conid{NoSub}{}\<[E]%
\\
\>[B]{}\hsindent{3}{}\<[3]%
\>[3]{}\Conid{Put}{}\<[8]%
\>[8]{}\mathbin{::}\Varid{v}\to {}\<[17]%
\>[17]{}\Conid{Mutating}\;\Varid{v}\;()\;{}\<[32]%
\>[32]{}\Conid{NoSub}{}\<[E]%
\\[\blanklineskip]%
\>[B]{}\Varid{hMut}{}\<[9]%
\>[9]{}\mathbin{::}{}\<[13]%
\>[13]{}\Conid{Functor}\;\Varid{l}{}\<[E]%
\\
\>[9]{}\Rightarrow \Varid{s}{}\<[E]%
\\
\>[9]{}\to \Conid{Tree}\;(\Conid{Mutating}\;\Varid{s}\mathbin{+}\Varid{\sigma})\;\Varid{l}\;\Varid{a}{}\<[E]%
\\
\>[9]{}\to \Conid{Tree}\;\Varid{\sigma}\;(\Conid{StateL}\;\Varid{s}\;\Varid{l})\;(\Conid{StateL}\;\Varid{s}\;\Conid{Id}\;\Varid{a}){}\<[E]%
\\
\>[B]{}\Varid{hMut}\;\Varid{s}\;{}\<[11]%
\>[11]{}(\Conid{Leaf}\;\Varid{x}){}\<[44]%
\>[44]{}\mathrel{=}\Conid{Leaf}\mathbin{\$}\Conid{StateL}\;(\Varid{s},\Conid{Id}\;\Varid{x}){}\<[E]%
\\
\>[B]{}\Varid{hMut}\;\Varid{s}\;{}\<[11]%
\>[11]{}(\Conid{Node}\;(\Conid{Inl'}\;\Conid{Get})\;{}\<[33]%
\>[33]{}\Varid{l}\;{}\<[36]%
\>[36]{}\anonymous \;{}\<[40]%
\>[40]{}\Varid{k}){}\<[44]%
\>[44]{}\mathrel{=}\Varid{hMut}\;\Varid{s}\;(\Varid{k}\;(\Varid{fmap}\;(\lambda \anonymous \to \Varid{s})\;\Varid{l})){}\<[E]%
\\
\>[B]{}\Varid{hMut}\;\anonymous \;{}\<[11]%
\>[11]{}(\Conid{Node}\;(\Conid{Inl'}\;(\Conid{Put}\;\Varid{s}))\;{}\<[33]%
\>[33]{}\Varid{l}\;{}\<[36]%
\>[36]{}\anonymous \;{}\<[40]%
\>[40]{}\Varid{k}){}\<[44]%
\>[44]{}\mathrel{=}\Varid{hMut}\;\Varid{s}\;(\Varid{k}\;\Varid{l}){}\<[E]%
\\
\>[B]{}\Varid{hMut}\;\Varid{s}\;{}\<[11]%
\>[11]{}(\Conid{Node}\;(\Conid{Inr'}\;\Varid{c})\;{}\<[33]%
\>[33]{}\Varid{l}\;{}\<[36]%
\>[36]{}\Varid{st}\;{}\<[40]%
\>[40]{}\Varid{k}){}\<[44]%
\>[44]{}\mathrel{=}{}\<[E]%
\\
\>[B]{}\hsindent{3}{}\<[3]%
\>[3]{}\Conid{Node}\;\Varid{c}\;(\Conid{StateL}\;(\Varid{s},\Varid{l}))\;{}\<[27]%
\>[27]{}(\lambda \Varid{c}\;{}\<[33]%
\>[33]{}(\Conid{StateL}\;(\Varid{s'},\Varid{l'})){}\<[53]%
\>[53]{}\to \Varid{transS}\mathrel{{<\kern-1pt}{\$}{\kern-1pt>}}\Varid{hMut}\;\Varid{s'}\;(\Varid{st}\;\Varid{c}\;\Varid{l'}))\;{}\<[E]%
\\
\>[27]{}(\lambda {}\<[33]%
\>[33]{}(\Conid{StateL}\;(\Varid{s'},\Varid{lv'})){}\<[53]%
\>[53]{}\to \Varid{hMut}\;\Varid{s'}\;(\Varid{k}\;\Varid{lv'})){}\<[E]%
\\[\blanklineskip]%
\>[B]{}\Varid{transS}\mathbin{::}\Conid{StateL}\;\Varid{s}\;\Conid{Id}\;(\Varid{l}\;\Varid{a})\to \Conid{StateL}\;\Varid{s}\;\Varid{l}\;\Varid{a}{}\<[E]%
\\
\>[B]{}\Varid{transS}\;(\Conid{StateL}\;(\Varid{s},\Varid{x}))\mathrel{=}\Conid{StateL}\;(\Varid{s},\Varid{unId}\;\Varid{x}){}\<[E]%
\\[\blanklineskip]%
\>[B]{}\Varid{get}{}\<[8]%
\>[8]{}\mathbin{::}(\Conid{Mutating}\;\Varid{s}\mathbin{<}\Varid{\sigma})\Rightarrow \Conid{Tree}\;\Varid{\sigma}\;\Conid{Id}\;\Varid{s}{}\<[E]%
\\
\>[B]{}\Varid{get}{}\<[8]%
\>[8]{}\mathrel{=}\Varid{noSubNode}\;\Conid{Get}{}\<[E]%
\\[\blanklineskip]%
\>[B]{}\Varid{put}{}\<[8]%
\>[8]{}\mathbin{::}(\Conid{Mutating}\;\Varid{s}\mathbin{<}\Varid{\sigma})\Rightarrow \Varid{s}\to \Conid{Tree}\;\Varid{\sigma}\;\Conid{Id}\;(){}\<[E]%
\\
\>[B]{}\Varid{put}\;\Varid{s}{}\<[8]%
\>[8]{}\mathrel{=}\Varid{noSubNode}\;(\Conid{Put}\;\Varid{s}){}\<[E]%
\ColumnHook
\end{hscode}\resethooks

\subsubsection{Exceptions}
The \ensuremath{\Conid{Throwing}} effect provides the two well-known operations for exceptions: \ensuremath{\Varid{throw}} and \ensuremath{\Varid{catch}}.

\begin{hscode}\SaveRestoreHook
\column{B}{@{}>{\hspre}l<{\hspost}@{}}%
\column{3}{@{}>{\hspre}l<{\hspost}@{}}%
\column{7}{@{}>{\hspre}l<{\hspost}@{}}%
\column{10}{@{}>{\hspre}l<{\hspost}@{}}%
\column{11}{@{}>{\hspre}l<{\hspost}@{}}%
\column{12}{@{}>{\hspre}l<{\hspost}@{}}%
\column{15}{@{}>{\hspre}l<{\hspost}@{}}%
\column{19}{@{}>{\hspre}l<{\hspost}@{}}%
\column{20}{@{}>{\hspre}l<{\hspost}@{}}%
\column{21}{@{}>{\hspre}l<{\hspost}@{}}%
\column{26}{@{}>{\hspre}l<{\hspost}@{}}%
\column{27}{@{}>{\hspre}l<{\hspost}@{}}%
\column{30}{@{}>{\hspre}l<{\hspost}@{}}%
\column{31}{@{}>{\hspre}l<{\hspost}@{}}%
\column{33}{@{}>{\hspre}l<{\hspost}@{}}%
\column{37}{@{}>{\hspre}l<{\hspost}@{}}%
\column{38}{@{}>{\hspre}l<{\hspost}@{}}%
\column{41}{@{}>{\hspre}l<{\hspost}@{}}%
\column{E}{@{}>{\hspre}l<{\hspost}@{}}%
\>[B]{}\mathbf{data}\;\Conid{Throwing}\;\Varid{x}\;\Varid{v}\mathbin{::}\mathbin{*}\to (\mathbin{*}\to \mathbin{*})\to \mathbin{*}\mathbf{where}{}\<[E]%
\\
\>[B]{}\hsindent{3}{}\<[3]%
\>[3]{}\Conid{Throw}{}\<[10]%
\>[10]{}\mathbin{::}\Varid{x}\to {}\<[19]%
\>[19]{}\Conid{Throwing}\;\Varid{x}\;\Varid{v}\;\Conid{Void}\;{}\<[38]%
\>[38]{}\Conid{NoSub}{}\<[E]%
\\
\>[B]{}\hsindent{3}{}\<[3]%
\>[3]{}\Conid{Catch}{}\<[10]%
\>[10]{}\mathbin{::}{}\<[19]%
\>[19]{}\Conid{Throwing}\;\Varid{x}\;\Varid{v}\;\Varid{v}\;{}\<[38]%
\>[38]{}(\Conid{MaybeSub}\;\Varid{x}\;\Varid{v}){}\<[E]%
\\[\blanklineskip]%
\>[B]{}\Varid{hExc}{}\<[7]%
\>[7]{}\mathbin{::}(\Conid{Functor}\;\Varid{l}){}\<[E]%
\\
\>[7]{}\Rightarrow \Conid{Tree}\;(\Conid{Throwing}\;\Varid{x}\;\Varid{v}\mathbin{+}\Varid{\sigma})\;\Varid{l}\;\Varid{a}{}\<[E]%
\\
\>[7]{}\to \Conid{Tree}\;\Varid{\sigma}\;(\Conid{EitherL}\;(\Varid{l}\;(),\Varid{x})\;\Varid{l})\;(\Conid{EitherL}\;(\Varid{l}\;(),\Varid{x})\;\Conid{Id}\;\Varid{a}){}\<[E]%
\\
\>[B]{}\Varid{hExc}\;(\Conid{Leaf}\;\Varid{x}){}\<[41]%
\>[41]{}\mathrel{=}\Conid{Leaf}\mathbin{\$}\Conid{EitherL}\mathbin{\$}\Conid{Right}\mathbin{\$}\Conid{Id}\;\Varid{x}{}\<[E]%
\\
\>[B]{}\Varid{hExc}\;(\Conid{Node}\;(\Conid{Inl'}\;(\Conid{Throw}\;\Varid{x}))\;{}\<[30]%
\>[30]{}\Varid{l}\;{}\<[33]%
\>[33]{}\anonymous \;{}\<[37]%
\>[37]{}\anonymous ){}\<[41]%
\>[41]{}\mathrel{=}\Conid{Leaf}\mathbin{\$}\Conid{EitherL}\mathbin{\$}\Conid{Left}\;(\Varid{l},\Varid{x}){}\<[E]%
\\
\>[B]{}\Varid{hExc}\;(\Conid{Node}\;(\Conid{Inl'}\;\Conid{Catch})\;{}\<[30]%
\>[30]{}\Varid{l}\;{}\<[33]%
\>[33]{}\Varid{st}\;{}\<[37]%
\>[37]{}\Varid{k}){}\<[41]%
\>[41]{}\mathrel{=}\Varid{hExc}\;(\Varid{st}\;\Conid{NothingS}\;\Varid{l}){}\<[E]%
\\
\>[B]{}\hsindent{3}{}\<[3]%
\>[3]{}\bind \Varid{either}\;{}\<[15]%
\>[15]{}(\lambda (\Varid{l'},\Varid{x})\to \Varid{hExc}\;(\Varid{st}\;(\Conid{JustS}\;\Varid{x})\;\Varid{l'}){}\<[E]%
\\
\>[15]{}\hsindent{12}{}\<[27]%
\>[27]{}\bind \Varid{either}\;(\Conid{Leaf}\hsdot{\circ }{.}\Conid{EitherL}\hsdot{\circ }{.}\Conid{Left})\;(\Varid{hExc}\hsdot{\circ }{.}\Varid{k}\hsdot{\circ }{.}\Varid{unId}))\;{}\<[E]%
\\
\>[15]{}(\Varid{hExc}\hsdot{\circ }{.}\Varid{k}\hsdot{\circ }{.}\Varid{unId}){}\<[E]%
\\
\>[B]{}\Varid{hExc}\;(\Conid{Node}\;(\Conid{Inr'}\;\Varid{c})\;{}\<[30]%
\>[30]{}\Varid{l}\;{}\<[33]%
\>[33]{}\Varid{st}\;{}\<[37]%
\>[37]{}\Varid{k}){}\<[41]%
\>[41]{}\mathrel{=}{}\<[E]%
\\
\>[B]{}\hsindent{3}{}\<[3]%
\>[3]{}\Conid{Node}\;\Varid{c}\;{}\<[11]%
\>[11]{}(\Conid{EitherL}\mathbin{\$}\Conid{Right}\;\Varid{l})\;{}\<[E]%
\\
\>[11]{}(\lambda \Varid{z}\to \Varid{either}\;{}\<[26]%
\>[26]{}(\Conid{Leaf}\hsdot{\circ }{.}\Conid{EitherL}\hsdot{\circ }{.}\Conid{Left})\;{}\<[E]%
\\
\>[26]{}(\lambda \Varid{lv}\to \Varid{transE}\mathrel{{<\kern-1pt}{\$}{\kern-1pt>}}\Varid{hExc}\;(\Varid{st}\;\Varid{z}\;\Varid{lv})))\;{}\<[E]%
\\
\>[11]{}(\Varid{either}\;(\Conid{Leaf}\hsdot{\circ }{.}\Conid{EitherL}\hsdot{\circ }{.}\Conid{Left})\;(\Varid{hExc}\hsdot{\circ }{.}\Varid{k})){}\<[E]%
\\[\blanklineskip]%
\>[B]{}\Varid{throw}{}\<[12]%
\>[12]{}\mathbin{::}\forall \Varid{x}\hsforall \;\Varid{v}\;\Varid{\sigma}\hsdot{\circ }{.}(\Conid{Throwing}\;\Varid{x}\;\Varid{v}\mathbin{<}\Varid{\sigma})\Rightarrow \Varid{x}\to \Conid{Tree}\;\Varid{\sigma}\;\Conid{Id}\;\Varid{v}{}\<[E]%
\\
\>[B]{}\Varid{throw}\;\Varid{e}{}\<[12]%
\>[12]{}\mathrel{=}\Conid{Node}\;{}\<[20]%
\>[20]{}(\Varid{injSig}\;(\Conid{Throw}\;\Varid{e}\mathbin{::}\Conid{Throwing}\;\Varid{x}\;\Varid{v}\;\Conid{Void}\;\Conid{NoSub}))\;{}\<[E]%
\\
\>[20]{}(\Conid{Id}\;())\;{}\<[E]%
\\
\>[20]{}(\lambda \Varid{x}\;\anonymous \to \mathbf{case}\;\Varid{x}\;\mathbf{of}\;)\;{}\<[E]%
\\
\>[20]{}(\lambda \Varid{x}\to \mathbf{case}\;\Varid{x}\;\mathbf{of}\;){}\<[E]%
\\[\blanklineskip]%
\>[B]{}\Varid{catch}{}\<[12]%
\>[12]{}\mathbin{::}\forall \Varid{x}\hsforall \;\Varid{v}\;\Varid{\sigma}\hsdot{\circ }{.}(\Conid{Throwing}\;\Varid{x}\;\Varid{v}\mathbin{<}\Varid{\sigma}){}\<[E]%
\\
\>[12]{}\Rightarrow \Conid{Tree}\;\Varid{\sigma}\;\Conid{Id}\;\Varid{v}\to (\Varid{x}\to \Conid{Tree}\;\Varid{\sigma}\;\Conid{Id}\;\Varid{v})\to \Conid{Tree}\;\Varid{\sigma}\;\Conid{Id}\;\Varid{v}{}\<[E]%
\\
\>[B]{}\Varid{catch}\;\Varid{p}\;\Varid{h}{}\<[12]%
\>[12]{}\mathrel{=}\Conid{Node}\;{}\<[20]%
\>[20]{}(\Varid{injSig}\;(\Conid{Catch}\mathbin{::}\Conid{Throwing}\;\Varid{x}\;\Varid{v}\;\Varid{v}\;(\Conid{MaybeSub}\;\Varid{x}\;\Varid{v})))\;{}\<[E]%
\\
\>[20]{}(\Conid{Id}\;())\;{}\<[E]%
\\
\>[20]{}(\lambda \Varid{z}\;\anonymous \to \mathbf{case}\;\Varid{z}\;\mathbf{of}{}\<[E]%
\\
\>[20]{}\hsindent{1}{}\<[21]%
\>[21]{}\Conid{JustS}\;\Varid{x}{}\<[31]%
\>[31]{}\to \Conid{Id}\mathrel{{<\kern-1pt}{\$}{\kern-1pt>}}(\Varid{h}\;\Varid{x}){}\<[E]%
\\
\>[20]{}\hsindent{1}{}\<[21]%
\>[21]{}\Conid{NothingS}{}\<[31]%
\>[31]{}\to \Conid{Id}\mathrel{{<\kern-1pt}{\$}{\kern-1pt>}}\Varid{p})\;{}\<[E]%
\\
\>[20]{}(\Conid{Leaf}\hsdot{\circ }{.}\Varid{unId}){}\<[E]%
\ColumnHook
\end{hscode}\resethooks

\subsubsection{Errors}
The \ensuremath{\Conid{Failing}} effect has a single operation \ensuremath{\Varid{err}} for constructing an error. 

\begin{hscode}\SaveRestoreHook
\column{B}{@{}>{\hspre}l<{\hspost}@{}}%
\column{3}{@{}>{\hspre}l<{\hspost}@{}}%
\column{7}{@{}>{\hspre}l<{\hspost}@{}}%
\column{8}{@{}>{\hspre}l<{\hspost}@{}}%
\column{11}{@{}>{\hspre}l<{\hspost}@{}}%
\column{12}{@{}>{\hspre}l<{\hspost}@{}}%
\column{14}{@{}>{\hspre}l<{\hspost}@{}}%
\column{25}{@{}>{\hspre}l<{\hspost}@{}}%
\column{28}{@{}>{\hspre}l<{\hspost}@{}}%
\column{30}{@{}>{\hspre}l<{\hspost}@{}}%
\column{31}{@{}>{\hspre}l<{\hspost}@{}}%
\column{35}{@{}>{\hspre}l<{\hspost}@{}}%
\column{36}{@{}>{\hspre}c<{\hspost}@{}}%
\column{36E}{@{}l@{}}%
\column{39}{@{}>{\hspre}l<{\hspost}@{}}%
\column{51}{@{}>{\hspre}l<{\hspost}@{}}%
\column{E}{@{}>{\hspre}l<{\hspost}@{}}%
\>[B]{}\mathbf{data}\;\Conid{Failing}\;\Varid{v}\mathbin{::}\mathbin{*}\to (\mathbin{*}\to \mathbin{*})\to \mathbin{*}\mathbf{where}{}\<[E]%
\\
\>[B]{}\hsindent{3}{}\<[3]%
\>[3]{}\Conid{Err}\mathbin{::}\Varid{v}\to \Conid{Failing}\;\Varid{v}\;\Conid{Void}\;\Conid{NoSub}{}\<[E]%
\\[\blanklineskip]%
\>[B]{}\Varid{hErr}{}\<[7]%
\>[7]{}\mathbin{::}\Conid{Functor}\;\Varid{l}{}\<[E]%
\\
\>[7]{}\Rightarrow \Conid{Tree}\;(\Conid{Failing}\;\Varid{v}\mathbin{+}\Varid{\sigma})\;\Varid{l}\;\Varid{a}{}\<[E]%
\\
\>[7]{}\to \Conid{Tree}\;\Varid{\sigma}\;(\Conid{EitherL}\;\Varid{v}\;\Varid{l})\;(\Conid{EitherL}\;\Varid{v}\;\Conid{Id}\;\Varid{a}){}\<[E]%
\\
\>[B]{}\Varid{hErr}\;(\Conid{Leaf}\;\Varid{x}){}\<[39]%
\>[39]{}\mathrel{=}\Conid{Leaf}\mathbin{\$}\Conid{EitherL}\mathbin{\$}\Conid{Right}\mathbin{\$}\Conid{Id}\;\Varid{x}{}\<[E]%
\\
\>[B]{}\Varid{hErr}\;(\Conid{Node}\;(\Conid{Inl'}\;(\Conid{Err}\;\Varid{x}))\;{}\<[28]%
\>[28]{}\anonymous \;{}\<[31]%
\>[31]{}\anonymous \;{}\<[35]%
\>[35]{}\anonymous ){}\<[39]%
\>[39]{}\mathrel{=}\Conid{Leaf}\mathbin{\$}\Conid{EitherL}\mathbin{\$}\Conid{Left}\;\Varid{x}{}\<[E]%
\\
\>[B]{}\Varid{hErr}\;(\Conid{Node}\;(\Conid{Inr'}\;\Varid{c})\;{}\<[28]%
\>[28]{}\Varid{l}\;{}\<[31]%
\>[31]{}\Varid{st}\;{}\<[35]%
\>[35]{}\Varid{k}){}\<[39]%
\>[39]{}\mathrel{=}{}\<[E]%
\\
\>[B]{}\hsindent{3}{}\<[3]%
\>[3]{}\Conid{Node}\;\Varid{c}\;{}\<[12]%
\>[12]{}(\Conid{EitherL}\mathbin{\$}\Conid{Right}\;\Varid{l})\;{}\<[E]%
\\
\>[12]{}(\lambda \Varid{z}\to \Varid{either}\;(\Conid{Leaf}\hsdot{\circ }{.}\Conid{EitherL}\hsdot{\circ }{.}\Conid{Left})\;{}\<[51]%
\>[51]{}(\lambda \Varid{lv}\to \Varid{transE}\mathrel{{<\kern-1pt}{\$}{\kern-1pt>}}\Varid{hErr}\;(\Varid{st}\;\Varid{z}\;\Varid{lv})))\;{}\<[E]%
\\
\>[12]{}(\Varid{either}\;(\Conid{Leaf}\hsdot{\circ }{.}\Conid{EitherL}\hsdot{\circ }{.}\Conid{Left})\;{}\<[51]%
\>[51]{}(\Varid{hErr}\hsdot{\circ }{.}\Varid{k})){}\<[E]%
\\[\blanklineskip]%
\>[B]{}\Varid{either}\mathbin{::}(\Varid{v}\to \Varid{c})\to (\Varid{l}\;\Varid{a}\to \Varid{c})\to \Conid{EitherL}\;\Varid{v}\;\Varid{l}\;\Varid{a}\to \Varid{c}{}\<[E]%
\\
\>[B]{}\Varid{either}\;\Varid{f}\;{}\<[11]%
\>[11]{}\anonymous \;{}\<[14]%
\>[14]{}(\Conid{EitherL}\;(\Conid{Left}\;{}\<[31]%
\>[31]{}\Varid{x})){}\<[36]%
\>[36]{}\mathrel{=}{}\<[36E]%
\>[39]{}\Varid{f}\;\Varid{x}{}\<[E]%
\\
\>[B]{}\Varid{either}\;\anonymous \;{}\<[11]%
\>[11]{}\Varid{g}\;{}\<[14]%
\>[14]{}(\Conid{EitherL}\;(\Conid{Right}\;{}\<[31]%
\>[31]{}\Varid{y})){}\<[36]%
\>[36]{}\mathrel{=}{}\<[36E]%
\>[39]{}\Varid{g}\;\Varid{y}{}\<[E]%
\\[\blanklineskip]%
\>[B]{}\Varid{transE}\mathbin{::}\Conid{EitherL}\;\Varid{v}\;\Conid{Id}\;(\Varid{l}\;\Varid{a})\to \Conid{EitherL}\;\Varid{v}\;\Varid{l}\;\Varid{a}{}\<[E]%
\\
\>[B]{}\Varid{transE}\;(\Conid{EitherL}\;(\Conid{Left}\;{}\<[25]%
\>[25]{}\Varid{x})){}\<[30]%
\>[30]{}\mathrel{=}\Conid{EitherL}\mathbin{\$}\Conid{Left}\;\Varid{x}{}\<[E]%
\\
\>[B]{}\Varid{transE}\;(\Conid{EitherL}\;(\Conid{Right}\;{}\<[25]%
\>[25]{}\Varid{y})){}\<[30]%
\>[30]{}\mathrel{=}\Conid{EitherL}\mathbin{\$}\Conid{Right}\mathbin{\$}\Varid{unId}\;\Varid{y}{}\<[E]%
\\[\blanklineskip]%
\>[B]{}\Varid{err}{}\<[8]%
\>[8]{}\mathbin{::}(\Conid{Failing}\;\Varid{v}\mathbin{<}\Varid{\sigma})\Rightarrow \Varid{v}\to \Conid{Tree}\;\Varid{\sigma}\;\Conid{Id}\;\Varid{a}{}\<[E]%
\\
\>[B]{}\Varid{err}\;\Varid{x}{}\<[8]%
\>[8]{}\mathrel{=}\Conid{Node}\;(\Varid{injSig}\;(\Conid{Err}\;\Varid{x}))\;(\Conid{Id}\;())\;(\lambda \Varid{x}\;\anonymous \to \mathbf{case}\;\Varid{x}\;\mathbf{of}\;)\;(\lambda \Varid{x}\to \mathbf{case}\;\Varid{x}\;\mathbf{of}\;){}\<[E]%
\ColumnHook
\end{hscode}\resethooks

\subsubsection{Arithmetics}
The \ensuremath{\Conid{Adding}} effect provides operations for adding natural numbers: 
\ensuremath{\Varid{plus}} and \ensuremath{\Varid{nat}}.

\begin{hscode}\SaveRestoreHook
\column{B}{@{}>{\hspre}l<{\hspost}@{}}%
\column{3}{@{}>{\hspre}l<{\hspost}@{}}%
\column{8}{@{}>{\hspre}l<{\hspost}@{}}%
\column{9}{@{}>{\hspre}l<{\hspost}@{}}%
\column{13}{@{}>{\hspre}l<{\hspost}@{}}%
\column{21}{@{}>{\hspre}l<{\hspost}@{}}%
\column{34}{@{}>{\hspre}l<{\hspost}@{}}%
\column{37}{@{}>{\hspre}l<{\hspost}@{}}%
\column{41}{@{}>{\hspre}l<{\hspost}@{}}%
\column{45}{@{}>{\hspre}l<{\hspost}@{}}%
\column{E}{@{}>{\hspre}l<{\hspost}@{}}%
\>[B]{}\mathbf{data}\;\Conid{Adding}\;\Varid{v}\mathbin{::}\mathbin{*}\to (\mathbin{*}\to \mathbin{*})\to \mathbin{*}\mathbf{where}{}\<[E]%
\\
\>[B]{}\hsindent{3}{}\<[3]%
\>[3]{}\Conid{Nat}{}\<[9]%
\>[9]{}\mathbin{::}\Conid{Integer}{}\<[21]%
\>[21]{}\to \Conid{Adding}\;\Varid{v}\;\Varid{v}\;\Conid{NoSub}{}\<[E]%
\\
\>[B]{}\hsindent{3}{}\<[3]%
\>[3]{}\Conid{Plus}{}\<[9]%
\>[9]{}\mathbin{::}\Varid{v}\to \Varid{v}{}\<[21]%
\>[21]{}\to \Conid{Adding}\;\Varid{v}\;\Varid{v}\;\Conid{NoSub}{}\<[E]%
\\[\blanklineskip]%
\>[B]{}\Varid{hPlus}{}\<[8]%
\>[8]{}\mathbin{::}(\Conid{Functor}\;\Varid{l},\Conid{Integer}\mathbin{<:}\Varid{v}){}\<[E]%
\\
\>[8]{}\Rightarrow \Conid{Tree}\;(\Conid{Adding}\;\Varid{v}\mathbin{+}\Varid{\sigma})\;\Varid{l}\;\Varid{a}{}\<[E]%
\\
\>[8]{}\to \Conid{Tree}\;\Varid{\sigma}\;\Varid{l}\;\Varid{a}{}\<[E]%
\\
\>[B]{}\Varid{hPlus}\;(\Conid{Leaf}\;\Varid{x}){}\<[45]%
\>[45]{}\mathrel{=}\Conid{Leaf}\;\Varid{x}{}\<[E]%
\\
\>[B]{}\Varid{hPlus}\;(\Conid{Node}\;(\Conid{Inl'}\;(\Conid{Nat}\;\Varid{n}))\;{}\<[34]%
\>[34]{}\Varid{l}\;{}\<[37]%
\>[37]{}\anonymous \;{}\<[41]%
\>[41]{}\Varid{k}){}\<[45]%
\>[45]{}\mathrel{=}\Varid{hPlus}\;(\Varid{k}\;(\Varid{const}\;(\Varid{inj}_{\mathrm{v}}\;\Varid{n})\mathrel{{<\kern-1pt}{\$}{\kern-1pt>}}\Varid{l})){}\<[E]%
\\
\>[B]{}\Varid{hPlus}\;(\Conid{Node}\;(\Conid{Inl'}\;(\Conid{Plus}\;\Varid{v}_{1}\;\Varid{v}_{2}))\;{}\<[34]%
\>[34]{}\Varid{l}\;{}\<[37]%
\>[37]{}\anonymous \;{}\<[41]%
\>[41]{}\Varid{k}){}\<[45]%
\>[45]{}\mathrel{=}\mathbf{case}\;(\Varid{proj}_{\mathrm{v}}\;\Varid{v}_{1},\Varid{proj}_{\mathrm{v}}\;\Varid{v}_{2})\;\mathbf{of}{}\<[E]%
\\
\>[B]{}\hsindent{3}{}\<[3]%
\>[3]{}(\Conid{Just}\;\Varid{n1},\Conid{Just}\;\Varid{n2})\to \Varid{hPlus}\;(\Varid{k}\;(\Varid{const}\;(\Varid{inj}_{\mathrm{v}}\;(\Varid{n1}\mathbin{+}\Varid{n2}))\mathrel{{<\kern-1pt}{\$}{\kern-1pt>}}\Varid{l})){}\<[E]%
\\
\>[B]{}\Varid{hPlus}\;(\Conid{Node}\;(\Conid{Inr'}\;\Varid{c})\;{}\<[34]%
\>[34]{}\Varid{l}\;{}\<[37]%
\>[37]{}\Varid{st}\;{}\<[41]%
\>[41]{}\Varid{k}){}\<[45]%
\>[45]{}\mathrel{=}{}\<[E]%
\\
\>[B]{}\hsindent{3}{}\<[3]%
\>[3]{}\Conid{Node}\;\Varid{c}\;\Varid{l}\;(\lambda \Varid{s}\to \Varid{hPlus}\hsdot{\circ }{.}\Varid{st}\;\Varid{s})\;(\Varid{hPlus}\hsdot{\circ }{.}\Varid{k}){}\<[E]%
\\[\blanklineskip]%
\>[B]{}\Varid{plus}{}\<[13]%
\>[13]{}\mathbin{::}(\Conid{Adding}\;\Varid{v}\mathbin{<}\sigma)\Rightarrow \Varid{v}\to \Varid{v}\to \Conid{Tree}\;\sigma\;\Conid{Id}\;\Varid{v}{}\<[E]%
\\
\>[B]{}\Varid{plus}\;\Varid{n1}\;\Varid{n2}{}\<[13]%
\>[13]{}\mathrel{=}\Varid{noSubNode}\;(\Conid{Plus}\;\Varid{n1}\;\Varid{n2}){}\<[E]%
\\[\blanklineskip]%
\>[B]{}\Varid{nat}{}\<[13]%
\>[13]{}\mathbin{::}(\Conid{Adding}\;\Varid{v}\mathbin{<}\Varid{\sigma})\Rightarrow \Conid{Integer}\to \Conid{Tree}\;\Varid{\sigma}\;\Conid{Id}\;\Varid{v}{}\<[E]%
\\
\>[B]{}\Varid{nat}\;\Varid{n}{}\<[13]%
\>[13]{}\mathrel{=}\Varid{noSubNode}\;(\Conid{Nat}\;\Varid{n}){}\<[E]%
\ColumnHook
\end{hscode}\resethooks

\subsubsection{Print}
The \ensuremath{\Conid{Printing}} effect has a single operation for printing a string: \ensuremath{\Varid{print}}.

\begin{hscode}\SaveRestoreHook
\column{B}{@{}>{\hspre}l<{\hspost}@{}}%
\column{3}{@{}>{\hspre}l<{\hspost}@{}}%
\column{9}{@{}>{\hspre}l<{\hspost}@{}}%
\column{31}{@{}>{\hspre}l<{\hspost}@{}}%
\column{37}{@{}>{\hspre}l<{\hspost}@{}}%
\column{41}{@{}>{\hspre}l<{\hspost}@{}}%
\column{E}{@{}>{\hspre}l<{\hspost}@{}}%
\>[B]{}\mathbf{data}\;\Conid{Printing}\;\Varid{v}\mathbin{::}\mathbin{*}\to (\mathbin{*}\to \mathbin{*})\to \mathbin{*}\mathbf{where}{}\<[E]%
\\
\>[B]{}\hsindent{3}{}\<[3]%
\>[3]{}\Conid{Print}\mathbin{::}\Conid{String}\to \Conid{Printing}\;\Varid{v}\;\Varid{v}\;\Conid{NoSub}{}\<[E]%
\\[\blanklineskip]%
\>[B]{}\Varid{hPrint}{}\<[9]%
\>[9]{}\mathbin{::}(\Conid{Functor}\;\Varid{l},\Conid{String}\mathbin{<:}\Varid{v}){}\<[E]%
\\
\>[9]{}\Rightarrow \Conid{Tree}\;(\Conid{Printing}\;\Varid{v}\mathbin{+}\Varid{\sigma})\;\Varid{l}\;\Varid{a}\to \Conid{Tree}\;\Varid{\sigma}\;\Varid{l}\;\Varid{a}{}\<[E]%
\\
\>[B]{}\Varid{hPrint}\;(\Conid{Leaf}\;\Varid{x}){}\<[41]%
\>[41]{}\mathrel{=}\Conid{Leaf}\;\Varid{x}{}\<[E]%
\\
\>[B]{}\Varid{hPrint}\;(\Conid{Node}\;(\Conid{Inl'}\;(\Conid{Print}\;\Varid{s}))\;\Varid{l}\;\anonymous \;{}\<[37]%
\>[37]{}\Varid{k}){}\<[41]%
\>[41]{}\mathrel{=}\Varid{hPrint}\;(\Varid{k}\;(\Varid{const}\;(\Varid{inj}_{\mathrm{v}}\;\Varid{s})\mathrel{{<\kern-1pt}{\$}{\kern-1pt>}}\Varid{l})){}\<[E]%
\\
\>[B]{}\Varid{hPrint}\;(\Conid{Node}\;(\Conid{Inr'}\;\Varid{c})\;{}\<[31]%
\>[31]{}\Varid{l}\;\Varid{st}\;{}\<[37]%
\>[37]{}\Varid{k}){}\<[41]%
\>[41]{}\mathrel{=}\Conid{Node}\;\Varid{c}\;\Varid{l}\;(\lambda \Varid{s}\to \Varid{hPrint}\hsdot{\circ }{.}\Varid{st}\;\Varid{s})\;(\Varid{hPrint}\hsdot{\circ }{.}\Varid{k}){}\<[E]%
\\[\blanklineskip]%
\>[B]{}\Varid{print}\mathbin{::}(\Conid{Printing}\;\Varid{v}\mathbin{<}\Varid{\sigma})\Rightarrow \Conid{String}\to \Conid{Tree}\;\Varid{\sigma}\;\Conid{Id}\;\Varid{v}{}\<[E]%
\\
\>[B]{}\Varid{print}\;\Varid{s}\mathrel{=}\Varid{noSubNode}\;(\Conid{Print}\;\Varid{s}){}\<[E]%
\ColumnHook
\end{hscode}\resethooks

\subsubsection{Delay Evaluation}
The \ensuremath{\Conid{Suspending}} effect provides two operations: 
\ensuremath{\Varid{suspend}} delays the evaluation of a recursive subtree, and 
\ensuremath{\Varid{enact}} forces the evaluation of the delayed subtree 
\emph{without memoizing the result}.

\begin{hscode}\SaveRestoreHook
\column{B}{@{}>{\hspre}l<{\hspost}@{}}%
\column{3}{@{}>{\hspre}l<{\hspost}@{}}%
\column{11}{@{}>{\hspre}l<{\hspost}@{}}%
\column{12}{@{}>{\hspre}l<{\hspost}@{}}%
\column{23}{@{}>{\hspre}l<{\hspost}@{}}%
\column{25}{@{}>{\hspre}l<{\hspost}@{}}%
\column{29}{@{}>{\hspre}l<{\hspost}@{}}%
\column{36}{@{}>{\hspre}l<{\hspost}@{}}%
\column{37}{@{}>{\hspre}l<{\hspost}@{}}%
\column{41}{@{}>{\hspre}l<{\hspost}@{}}%
\column{43}{@{}>{\hspre}l<{\hspost}@{}}%
\column{47}{@{}>{\hspre}l<{\hspost}@{}}%
\column{51}{@{}>{\hspre}l<{\hspost}@{}}%
\column{57}{@{}>{\hspre}l<{\hspost}@{}}%
\column{E}{@{}>{\hspre}l<{\hspost}@{}}%
\>[B]{}\mathbf{data}\;\Conid{Suspending}\;\Varid{v}\mathbin{::}\mathbin{*}\to (\mathbin{*}\to \mathbin{*})\to \mathbin{*}\mathbf{where}{}\<[E]%
\\
\>[B]{}\hsindent{3}{}\<[3]%
\>[3]{}\Conid{Suspend}{}\<[12]%
\>[12]{}\mathbin{::}{}\<[23]%
\>[23]{}\Conid{Suspending}\;\Varid{v}\;\Conid{Ptr}\;{}\<[41]%
\>[41]{}(\Conid{OneSub}\;\Varid{v}){}\<[E]%
\\
\>[B]{}\hsindent{3}{}\<[3]%
\>[3]{}\Conid{Enact}{}\<[12]%
\>[12]{}\mathbin{::}\Conid{Ptr}\to {}\<[23]%
\>[23]{}\Conid{Suspending}\;\Varid{v}\;\Varid{v}\;{}\<[41]%
\>[41]{}\Conid{NoSub}{}\<[E]%
\\[\blanklineskip]%
\>[B]{}\mathbf{type}\;\Conid{Ptr}{}\<[25]%
\>[25]{}\mathrel{=}\Conid{Int}{}\<[E]%
\\
\>[B]{}\mathbf{type}\;\Conid{Suspension}\;\sigma\;\Varid{l}\;\Varid{v}{}\<[25]%
\>[25]{}\mathrel{=}\Varid{l}\;()\to \Conid{Tree}\;(\Conid{Suspending}\;\Varid{v}\mathbin{+}\sigma)\;\Varid{l}\;(\Varid{l}\;\Varid{v}){}\<[E]%
\\[\blanklineskip]%
\>[B]{}\mathbf{data}\;\Conid{Suspended}\;\Varid{r}{}\<[25]%
\>[25]{}\mathrel{=}\Conid{Suspended}\;\Conid{Ptr}\;\Varid{r}\;\mathbf{deriving}\;(\Conid{Functor}){}\<[E]%
\\[\blanklineskip]%
\>[B]{}\Varid{hSuspend}{}\<[11]%
\>[11]{}\mathbin{::}\Conid{Functor}\;\Varid{l}{}\<[E]%
\\
\>[11]{}\Rightarrow [\mskip1.5mu \Conid{Suspension}\;\sigma\;\Varid{l}\;\Varid{v}\mskip1.5mu]{}\<[E]%
\\
\>[11]{}\to \Conid{Tree}\;(\Conid{Suspending}\;\Varid{v}\mathbin{+}\sigma)\;\Varid{l}\;(\Varid{l}\;\Varid{a}){}\<[E]%
\\
\>[11]{}\to \Conid{Tree}\;\sigma\;(\Conid{StateL}\;[\mskip1.5mu \Conid{Suspension}\;\sigma\;\Varid{l}\;\Varid{v}\mskip1.5mu]\;\Varid{l})\;(\Conid{StateL}\;[\mskip1.5mu \Conid{Suspension}\;\sigma\;\Varid{l}\;\Varid{v}\mskip1.5mu]\;\Varid{l}\;\Varid{a}){}\<[E]%
\\
\>[B]{}\Varid{hSuspend}\;\Varid{rs}\;(\Conid{Leaf}\;\Varid{x}){}\<[47]%
\>[47]{}\mathrel{=}\Conid{Leaf}\;(\Conid{StateL}\;(\Varid{rs},\Varid{x})){}\<[E]%
\\
\>[B]{}\Varid{hSuspend}\;\Varid{rs}\;(\Conid{Node}\;(\Conid{Inl'}\;\Conid{Suspend})\;{}\<[37]%
\>[37]{}\Varid{l}\;\Varid{st}\;{}\<[43]%
\>[43]{}\Varid{k}){}\<[47]%
\>[47]{}\mathrel{=}{}\<[E]%
\\
\>[B]{}\hsindent{3}{}\<[3]%
\>[3]{}\Varid{hSuspend}\;(\Varid{rs}\plus [\mskip1.5mu \Varid{st}\;\Conid{One}\mskip1.5mu])\;(\Varid{k}\;(\Varid{length}\;\Varid{rs}\mathrel{{<\kern-1pt}{\$}}\Varid{l})){}\<[E]%
\\
\>[B]{}\Varid{hSuspend}\;\Varid{rs}\;(\Conid{Node}\;(\Conid{Inl'}\;(\Conid{Enact}\;\Varid{p}))\;{}\<[37]%
\>[37]{}\Varid{l}\;\anonymous \;{}\<[43]%
\>[43]{}\Varid{k}){}\<[47]%
\>[47]{}\mathrel{=}\mathbf{do}{}\<[E]%
\\
\>[B]{}\hsindent{3}{}\<[3]%
\>[3]{}\Conid{StateL}\;(\Varid{rs'},\Varid{lv})\leftarrow \Varid{hSuspend}\;\Varid{rs}\;((\Varid{rs}\mathbin{!!}\Varid{p})\;\Varid{l});{}\<[51]%
\>[51]{}\Varid{hSuspend}\;\Varid{rs'}\;(\Varid{k}\;\Varid{lv}){}\<[E]%
\\
\>[B]{}\Varid{hSuspend}\;\Varid{rs}\;(\Conid{Node}\;(\Conid{Inr'}\;\Varid{op})\;{}\<[37]%
\>[37]{}\Varid{l}\;\Varid{st}\;{}\<[43]%
\>[43]{}\Varid{k}){}\<[47]%
\>[47]{}\mathrel{=}{}\<[E]%
\\
\>[B]{}\hsindent{3}{}\<[3]%
\>[3]{}\Conid{Node}\;\Varid{op}\;(\Conid{StateL}\;(\Varid{rs},\Varid{l}))\;{}\<[29]%
\>[29]{}(\lambda \Varid{c}_{2}\;{}\<[36]%
\>[36]{}(\Conid{StateL}\;(\Varid{rs'},\Varid{l'})){}\<[57]%
\>[57]{}\to \Varid{hSuspend}\;\Varid{rs'}\;(\Varid{st}\;\Varid{c}_{2}\;\Varid{l'}))\;{}\<[E]%
\\
\>[29]{}(\lambda {}\<[36]%
\>[36]{}(\Conid{StateL}\;(\Varid{rs'},\Varid{lv'})){}\<[57]%
\>[57]{}\to \Varid{hSuspend}\;\Varid{rs'}\;(\Varid{k}\;\Varid{lv'})){}\<[E]%
\\[\blanklineskip]%
\>[B]{}\Varid{suspend}{}\<[12]%
\>[12]{}\mathbin{::}\forall \Varid{v}\hsforall \;\sigma\hsdot{\circ }{.}(\Conid{Suspending}\;\Varid{v}\mathbin{<}\sigma)\Rightarrow \Conid{Tree}\;\sigma\;\Conid{Id}\;\Varid{v}\to \Conid{Tree}\;\sigma\;\Conid{Id}\;\Conid{Ptr}{}\<[E]%
\\
\>[B]{}\Varid{suspend}\;\Varid{t}{}\<[12]%
\>[12]{}\mathrel{=}\Varid{oneSubNode}\;(\Conid{Suspend})\;\Varid{t}{}\<[E]%
\\[\blanklineskip]%
\>[B]{}\Varid{enact}{}\<[12]%
\>[12]{}\mathbin{::}(\Conid{Suspending}\;\Varid{v}\mathbin{<}\sigma)\Rightarrow \Conid{Ptr}\to \Conid{Tree}\;\sigma\;\Conid{Id}\;\Varid{v}{}\<[E]%
\\
\>[B]{}\Varid{enact}\;\Varid{p}{}\<[12]%
\>[12]{}\mathrel{=}\Varid{noSubNode}\;(\Conid{Enact}\;\Varid{p}){}\<[E]%
\ColumnHook
\end{hscode}\resethooks

\subsubsection{Memoization}
The \ensuremath{\Conid{Thunking}} effect is analogous to \ensuremath{\Conid{Suspending}}, 
but its handler \emph{memoizes} the result of forcing a thunked computation.

\begin{hscode}\SaveRestoreHook
\column{B}{@{}>{\hspre}l<{\hspost}@{}}%
\column{3}{@{}>{\hspre}l<{\hspost}@{}}%
\column{10}{@{}>{\hspre}l<{\hspost}@{}}%
\column{21}{@{}>{\hspre}l<{\hspost}@{}}%
\column{36}{@{}>{\hspre}l<{\hspost}@{}}%
\column{E}{@{}>{\hspre}l<{\hspost}@{}}%
\>[B]{}\mathbf{data}\;\Conid{Thunking}\;\Varid{v}\mathbin{::}\mathbin{*}\to (\mathbin{*}\to \mathbin{*})\to \mathbin{*}\mathbf{where}{}\<[E]%
\\
\>[B]{}\hsindent{3}{}\<[3]%
\>[3]{}\Conid{Thunk}{}\<[10]%
\>[10]{}\mathbin{::}{}\<[21]%
\>[21]{}\Conid{Thunking}\;\Varid{v}\;\Conid{Ptr}\;(\Conid{OneSub}\;\Varid{v}){}\<[E]%
\\
\>[B]{}\hsindent{3}{}\<[3]%
\>[3]{}\Conid{Force}{}\<[10]%
\>[10]{}\mathbin{::}\Conid{Ptr}\to {}\<[21]%
\>[21]{}\Conid{Thunking}\;\Varid{v}\;\Varid{v}\;{}\<[36]%
\>[36]{}\Conid{NoSub}{}\<[E]%
\ColumnHook
\end{hscode}\resethooks
\begin{hscode}\SaveRestoreHook
\column{B}{@{}>{\hspre}l<{\hspost}@{}}%
\column{3}{@{}>{\hspre}l<{\hspost}@{}}%
\column{13}{@{}>{\hspre}l<{\hspost}@{}}%
\column{E}{@{}>{\hspre}l<{\hspost}@{}}%
\>[3]{}\mathbf{type}\;\Conid{Ptr}{}\<[13]%
\>[13]{}\mathrel{=}\Conid{Int}{}\<[E]%
\ColumnHook
\end{hscode}\resethooks
\begin{hscode}\SaveRestoreHook
\column{B}{@{}>{\hspre}l<{\hspost}@{}}%
\column{3}{@{}>{\hspre}l<{\hspost}@{}}%
\column{17}{@{}>{\hspre}l<{\hspost}@{}}%
\column{20}{@{}>{\hspre}l<{\hspost}@{}}%
\column{31}{@{}>{\hspre}l<{\hspost}@{}}%
\column{E}{@{}>{\hspre}l<{\hspost}@{}}%
\>[B]{}\mathbf{type}\;\Conid{Thunk}\;\sigma\;\Varid{l}\;\Varid{v}{}\<[20]%
\>[20]{}\mathrel{=}\Conid{Either}\;(\Varid{l}\;()\to \Conid{Tree}\;(\Conid{Thunking}\;\Varid{v}\mathbin{+}\sigma)\;\Varid{l}\;(\Varid{l}\;\Varid{v}))\;\Varid{v}{}\<[E]%
\\[\blanklineskip]%
\>[B]{}\mathbf{data}\;\Conid{TreeWith}\;\Varid{f}\;\sigma\;(\Varid{l}\mathbin{::}\mathbin{*}\to \mathbin{*})\;\Varid{a}\mathrel{=}{}\<[E]%
\\
\>[B]{}\hsindent{3}{}\<[3]%
\>[3]{}\Conid{TreeWith}\;\{\mskip1.5mu \Varid{unTreeWith}\mathbin{::}\Conid{Tree}\;\sigma\;(\Varid{f}\;\Varid{l})\;(\Varid{f}\;\Varid{l}\;\Varid{a})\mskip1.5mu\}{}\<[E]%
\\
\>[B]{}\mathbf{data}\;\Conid{Thunked}\;\Varid{r}{}\<[17]%
\>[17]{}\mathrel{=}\Conid{Thunked}\;\Conid{Ptr}\;\Varid{r}{}\<[E]%
\\[\blanklineskip]%
\>[B]{}\mathbf{class}\;\Conid{Modular}\;(\Varid{m}_{1}\mathbin{::}\mathbin{*}\to \mathbin{*})\;(\Varid{m}_{2}\mathbin{::}(\mathbin{*}\to \mathbin{*})\to \mathbin{*}\to \mathbin{*})\;\mathbf{where}{}\<[E]%
\\
\>[B]{}\hsindent{3}{}\<[3]%
\>[3]{}\Varid{fwd}\mathbin{::}\Varid{m}_{1}\;(\Varid{m}_{2}\;\Varid{m}_{1}\;\Varid{a})\to \Varid{m}_{2}\;\Varid{m}_{1}\;\Varid{a}{}\<[E]%
\\[\blanklineskip]%
\>[B]{}\Varid{replace}\mathbin{::}\Conid{Int}\to \Varid{a}\to [\mskip1.5mu \Varid{a}\mskip1.5mu]\to [\mskip1.5mu \Varid{a}\mskip1.5mu]{}\<[E]%
\\
\>[B]{}\Varid{replace}\;\mathrm{0}\;\Varid{x}\;(\anonymous \mathbin{:}\Varid{xs}){}\<[31]%
\>[31]{}\mathrel{=}\Varid{x}\mathbin{:}\Varid{xs}{}\<[E]%
\\
\>[B]{}\Varid{replace}\;\Varid{i}\;\Varid{x}\;(\Varid{y}\mathbin{:}\Varid{xs})\mid \Varid{i}\mathbin{>}\mathrm{0}{}\<[31]%
\>[31]{}\mathrel{=}\Varid{y}\mathbin{:}\Varid{replace}\;(\Varid{i}\mathbin{-}\mathrm{1})\;\Varid{x}\;\Varid{xs}{}\<[E]%
\\
\>[B]{}\Varid{replace}\;\anonymous \;\anonymous \;\anonymous {}\<[31]%
\>[31]{}\mathrel{=}\Varid{error}\;\text{\tt \char34 bad~index\char34}{}\<[E]%
\ColumnHook
\end{hscode}\resethooks

For this effect, we can provide two handlers, one for lazy evaluation and another
one for eager evaluation.

\begin{hscode}\SaveRestoreHook
\column{B}{@{}>{\hspre}l<{\hspost}@{}}%
\column{3}{@{}>{\hspre}l<{\hspost}@{}}%
\column{5}{@{}>{\hspre}l<{\hspost}@{}}%
\column{7}{@{}>{\hspre}l<{\hspost}@{}}%
\column{9}{@{}>{\hspre}l<{\hspost}@{}}%
\column{29}{@{}>{\hspre}l<{\hspost}@{}}%
\column{35}{@{}>{\hspre}l<{\hspost}@{}}%
\column{36}{@{}>{\hspre}l<{\hspost}@{}}%
\column{41}{@{}>{\hspre}l<{\hspost}@{}}%
\column{45}{@{}>{\hspre}l<{\hspost}@{}}%
\column{57}{@{}>{\hspre}l<{\hspost}@{}}%
\column{E}{@{}>{\hspre}l<{\hspost}@{}}%
\>[B]{}\Varid{hThunk}{}\<[9]%
\>[9]{}\mathbin{::}(\Conid{Functor}\;\Varid{l},\Conid{Modular}\;\Varid{l}\;(\Conid{TreeWith}\;(\Conid{StateL}\;([\mskip1.5mu \Conid{Thunk}\;\sigma\;\Varid{l}\;\Varid{v}\mskip1.5mu]))\;\sigma)){}\<[E]%
\\
\>[9]{}\Rightarrow [\mskip1.5mu \Conid{Thunk}\;\sigma\;\Varid{l}\;\Varid{v}\mskip1.5mu]{}\<[E]%
\\
\>[9]{}\to \Conid{Tree}\;(\Conid{Thunking}\;\Varid{v}\mathbin{+}\sigma)\;\Varid{l}\;(\Varid{l}\;\Varid{a}){}\<[E]%
\\
\>[9]{}\to \Conid{Tree}\;\sigma\;(\Conid{StateL}\;[\mskip1.5mu \Conid{Thunk}\;\sigma\;\Varid{l}\;\Varid{v}\mskip1.5mu]\;\Varid{l})\;(\Conid{StateL}\;[\mskip1.5mu \Conid{Thunk}\;\sigma\;\Varid{l}\;\Varid{v}\mskip1.5mu]\;\Varid{l}\;\Varid{a}){}\<[E]%
\\
\>[B]{}\Varid{hThunk}\;\Varid{rs}\;(\Conid{Leaf}\;\Varid{x}){}\<[45]%
\>[45]{}\mathrel{=}\Conid{Leaf}\;(\Conid{StateL}\;(\Varid{rs},\Varid{x})){}\<[E]%
\\
\>[B]{}\Varid{hThunk}\;\Varid{rs}\;(\Conid{Node}\;(\Conid{Inl'}\;\Conid{Thunk})\;{}\<[35]%
\>[35]{}\Varid{l}\;\Varid{st}\;{}\<[41]%
\>[41]{}\Varid{k}){}\<[45]%
\>[45]{}\mathrel{=}{}\<[E]%
\\
\>[B]{}\hsindent{3}{}\<[3]%
\>[3]{}\Varid{hThunk}\;(\Varid{rs}\plus [\mskip1.5mu \Conid{Left}\;(\Varid{st}\;\Conid{One})\mskip1.5mu])\;(\Varid{k}\;(\Varid{length}\;\Varid{rs}\mathrel{{<\kern-1pt}{\$}}\Varid{l})){}\<[E]%
\\
\>[B]{}\Varid{hThunk}\;\Varid{rs}\;(\Conid{Node}\;(\Conid{Inl'}\;(\Conid{Force}\;\Varid{p}))\;{}\<[35]%
\>[35]{}\Varid{l}\;\anonymous \;{}\<[41]%
\>[41]{}\Varid{k}){}\<[45]%
\>[45]{}\mathrel{=}\mathbf{case}\;(\Varid{rs}\mathbin{!!}\Varid{p})\;\mathbf{of}{}\<[E]%
\\
\>[B]{}\hsindent{3}{}\<[3]%
\>[3]{}\Conid{Left}\;\Varid{t}\to \mathbf{do}{}\<[E]%
\\
\>[3]{}\hsindent{2}{}\<[5]%
\>[5]{}\Conid{StateL}\;(\Varid{rs'},\Varid{lv})\leftarrow \Varid{hThunk}\;\Varid{rs}\;(\Varid{t}\;\Varid{l}){}\<[E]%
\\
\>[3]{}\hsindent{2}{}\<[5]%
\>[5]{}\Varid{unTreeWith}\mathbin{\$}\Varid{fwd}\mathbin{\$}{}\<[E]%
\\
\>[5]{}\hsindent{2}{}\<[7]%
\>[7]{}(\lambda \Varid{v}\to \Conid{TreeWith}\mathbin{\$}\Varid{hThunk}\;(\Varid{replace}\;\Varid{p}\;(\Conid{Right}\;\Varid{v})\;\Varid{rs'})\;(\Varid{k}\;\Varid{lv}))\mathrel{{<\kern-1pt}{\$}{\kern-1pt>}}\Varid{lv}{}\<[E]%
\\
\>[B]{}\hsindent{3}{}\<[3]%
\>[3]{}\Conid{Right}\;\Varid{v}\to \Varid{hThunk}\;\Varid{rs}\;(\Varid{k}\;(\Varid{v}\mathrel{{<\kern-1pt}{\$}}\Varid{l})){}\<[E]%
\\
\>[B]{}\Varid{hThunk}\;\Varid{rs}\;(\Conid{Node}\;(\Conid{Inr'}\;\Varid{op})\;{}\<[35]%
\>[35]{}\Varid{l}\;\Varid{st}\;{}\<[41]%
\>[41]{}\Varid{k}){}\<[45]%
\>[45]{}\mathrel{=}{}\<[E]%
\\
\>[B]{}\hsindent{3}{}\<[3]%
\>[3]{}\Conid{Node}\;\Varid{op}\;(\Conid{StateL}\;(\Varid{rs},\Varid{l}))\;{}\<[29]%
\>[29]{}(\lambda \Varid{c}_{2}\;{}\<[36]%
\>[36]{}(\Conid{StateL}\;(\Varid{rs'},\Varid{l'})){}\<[57]%
\>[57]{}\to \Varid{hThunk}\;\Varid{rs'}\;(\Varid{st}\;\Varid{c}_{2}\;\Varid{l'}))\;{}\<[E]%
\\
\>[29]{}(\lambda {}\<[36]%
\>[36]{}(\Conid{StateL}\;(\Varid{rs'},\Varid{lv'})){}\<[57]%
\>[57]{}\to \Varid{hThunk}\;\Varid{rs'}\;(\Varid{k}\;\Varid{lv'})){}\<[E]%
\\[\blanklineskip]%
\>[B]{}\Varid{hEager}{}\<[9]%
\>[9]{}\mathbin{::}(\Conid{Functor}\;\Varid{l},\Conid{Modular}\;\Varid{l}\;(\Conid{TreeWith}\;(\Conid{StateL}\;[\mskip1.5mu \Varid{v}\mskip1.5mu])\;\sigma)){}\<[E]%
\\
\>[9]{}\Rightarrow [\mskip1.5mu \Varid{v}\mskip1.5mu]{}\<[E]%
\\
\>[9]{}\to \Conid{Tree}\;(\Conid{Thunking}\;\Varid{v}\mathbin{+}\sigma)\;\Varid{l}\;(\Varid{l}\;\Varid{a}){}\<[E]%
\\
\>[9]{}\to \Conid{Tree}\;\sigma\;(\Conid{StateL}\;[\mskip1.5mu \Varid{v}\mskip1.5mu]\;\Varid{l})\;(\Conid{StateL}\;[\mskip1.5mu \Varid{v}\mskip1.5mu]\;\Varid{l}\;\Varid{a}){}\<[E]%
\\
\>[B]{}\Varid{hEager}\;\Varid{vs}\;(\Conid{Leaf}\;\Varid{x}){}\<[45]%
\>[45]{}\mathrel{=}\Conid{Leaf}\;(\Conid{StateL}\;(\Varid{vs},\Varid{x})){}\<[E]%
\\
\>[B]{}\Varid{hEager}\;\Varid{vs}\;(\Conid{Node}\;(\Conid{Inl'}\;\Conid{Thunk})\;{}\<[35]%
\>[35]{}\Varid{l}\;\Varid{st}\;{}\<[41]%
\>[41]{}\Varid{k}){}\<[45]%
\>[45]{}\mathrel{=}\mathbf{do}{}\<[E]%
\\
\>[B]{}\hsindent{3}{}\<[3]%
\>[3]{}\Conid{StateL}\;(\Varid{vs'},\Varid{lv})\leftarrow \Varid{hEager}\;\Varid{vs}\;(\Varid{st}\;\Conid{One}\;\Varid{l}){}\<[E]%
\\
\>[B]{}\hsindent{3}{}\<[3]%
\>[3]{}\Varid{unTreeWith}\mathbin{\$}\Varid{fwd}\mathbin{\$}{}\<[E]%
\\
\>[3]{}\hsindent{2}{}\<[5]%
\>[5]{}(\lambda \Varid{v}\to \Conid{TreeWith}\mathbin{\$}\Varid{hEager}\;(\Varid{vs'}\plus [\mskip1.5mu \Varid{v}\mskip1.5mu])\;(\Varid{k}\;(\Varid{length}\;\Varid{vs'}\mathrel{{<\kern-1pt}{\$}}\Varid{l})))\mathrel{{<\kern-1pt}{\$}{\kern-1pt>}}\Varid{lv}{}\<[E]%
\\
\>[B]{}\Varid{hEager}\;\Varid{vs}\;(\Conid{Node}\;(\Conid{Inl'}\;(\Conid{Force}\;\Varid{p}))\;{}\<[35]%
\>[35]{}\Varid{l}\;\anonymous \;{}\<[41]%
\>[41]{}\Varid{k}){}\<[45]%
\>[45]{}\mathrel{=}\mathbf{do}{}\<[E]%
\\
\>[B]{}\hsindent{3}{}\<[3]%
\>[3]{}\Varid{hEager}\;\Varid{vs}\;(\Varid{k}\;((\Varid{vs}\mathbin{!!}\Varid{p})\mathrel{{<\kern-1pt}{\$}}\Varid{l})){}\<[E]%
\\
\>[B]{}\Varid{hEager}\;\Varid{vs}\;(\Conid{Node}\;(\Conid{Inr'}\;\Varid{op})\;{}\<[35]%
\>[35]{}\Varid{l}\;\Varid{st}\;{}\<[41]%
\>[41]{}\Varid{k}){}\<[45]%
\>[45]{}\mathrel{=}{}\<[E]%
\\
\>[B]{}\hsindent{3}{}\<[3]%
\>[3]{}\Conid{Node}\;\Varid{op}\;(\Conid{StateL}\;(\Varid{vs},\Varid{l}))\;{}\<[29]%
\>[29]{}(\lambda \Varid{c}_{2}\;{}\<[36]%
\>[36]{}(\Conid{StateL}\;(\Varid{vs'},\Varid{l'})){}\<[57]%
\>[57]{}\to \Varid{hEager}\;\Varid{vs'}\;(\Varid{st}\;\Varid{c}_{2}\;\Varid{l'}))\;{}\<[E]%
\\
\>[29]{}(\lambda {}\<[36]%
\>[36]{}(\Conid{StateL}\;(\Varid{vs'},\Varid{lv'})){}\<[57]%
\>[57]{}\to \Varid{hEager}\;\Varid{vs'}\;(\Varid{k}\;\Varid{lv'})){}\<[E]%
\ColumnHook
\end{hscode}\resethooks

\begin{hscode}\SaveRestoreHook
\column{B}{@{}>{\hspre}l<{\hspost}@{}}%
\column{10}{@{}>{\hspre}l<{\hspost}@{}}%
\column{E}{@{}>{\hspre}l<{\hspost}@{}}%
\>[B]{}\Varid{thunk}{}\<[10]%
\>[10]{}\mathbin{::}\forall \Varid{v}\hsforall \;\sigma\hsdot{\circ }{.}(\Conid{Thunking}\;\Varid{v}\mathbin{<}\sigma)\Rightarrow \Conid{Tree}\;\sigma\;\Conid{Id}\;\Varid{v}\to \Conid{Tree}\;\sigma\;\Conid{Id}\;\Conid{Ptr}{}\<[E]%
\\
\>[B]{}\Varid{thunk}\;\Varid{t}\mathrel{=}\Varid{oneSubNode}\;\Conid{Thunk}\;\Varid{t}{}\<[E]%
\\[\blanklineskip]%
\>[B]{}\Varid{force}{}\<[10]%
\>[10]{}\mathbin{::}(\Conid{Thunking}\;\Varid{v}\mathbin{<}\sigma)\Rightarrow \Conid{Ptr}\to \Conid{Tree}\;\sigma\;\Conid{Id}\;\Varid{v}{}\<[E]%
\\
\>[B]{}\Varid{force}\;\Varid{p}{}\<[10]%
\>[10]{}\mathrel{=}\Varid{noSubNode}\;(\Conid{Force}\;\Varid{p}){}\<[E]%
\ColumnHook
\end{hscode}\resethooks

\subsubsection{Function Abstraction}
The \ensuremath{\Conid{Abstracting}} effect has three operators: 
\ensuremath{\Varid{var}} for representing a variable, 
\ensuremath{\Varid{abs}} for function abstraction, and
\ensuremath{\Varid{app}} for function application.

\begin{hscode}\SaveRestoreHook
\column{B}{@{}>{\hspre}l<{\hspost}@{}}%
\column{3}{@{}>{\hspre}l<{\hspost}@{}}%
\column{21}{@{}>{\hspre}l<{\hspost}@{}}%
\column{E}{@{}>{\hspre}l<{\hspost}@{}}%
\>[B]{}\mathbf{data}\;\Conid{Abstracting}\;\Varid{v}\mathbin{::}\mathbin{*}\to (\mathbin{*}\to \mathbin{*})\to \mathbin{*}\mathbf{where}{}\<[E]%
\\
\>[B]{}\hsindent{3}{}\<[3]%
\>[3]{}\Conid{Var}\mathbin{::}\Conid{Int}\to {}\<[21]%
\>[21]{}\Conid{Abstracting}\;\Varid{v}\;\Varid{v}\;\Conid{NoSub}{}\<[E]%
\\
\>[B]{}\hsindent{3}{}\<[3]%
\>[3]{}\Conid{App}\mathbin{::}\Varid{v}\to \Varid{v}\to {}\<[21]%
\>[21]{}\Conid{Abstracting}\;\Varid{v}\;\Varid{v}\;\Conid{NoSub}{}\<[E]%
\\
\>[B]{}\hsindent{3}{}\<[3]%
\>[3]{}\Conid{Abs}\mathbin{::}{}\<[21]%
\>[21]{}\Conid{Abstracting}\;\Varid{v}\;\Varid{v}\;(\Conid{OneSub}\;\Varid{v}){}\<[E]%
\\[\blanklineskip]%
\>[B]{}\mathbf{data}\;\Conid{Closure}\;\Varid{r}\;\mathbf{where}{}\<[E]%
\\
\>[B]{}\hsindent{3}{}\<[3]%
\>[3]{}\Conid{Clos}\mathbin{::}\Conid{Ptr}\to \Conid{Env}\;\Varid{r}\to \Conid{Closure}\;\Varid{r}{}\<[E]%
\\[\blanklineskip]%
\>[B]{}\mathbf{type}\;\Conid{Env}\;\Varid{v}\mathrel{=}[\mskip1.5mu \Varid{v}\mskip1.5mu]{}\<[E]%
\ColumnHook
\end{hscode}\resethooks
\begin{hscode}\SaveRestoreHook
\column{B}{@{}>{\hspre}l<{\hspost}@{}}%
\column{3}{@{}>{\hspre}l<{\hspost}@{}}%
\column{27}{@{}>{\hspre}l<{\hspost}@{}}%
\column{E}{@{}>{\hspre}l<{\hspost}@{}}%
\>[3]{}\mathbf{type}\;\Conid{Ptr}{}\<[27]%
\>[27]{}\mathrel{=}\Conid{Int}{}\<[E]%
\ColumnHook
\end{hscode}\resethooks
For this effect, we can define two handlers: one for call site handling of
  effects an another for definition site handling of effects

\begin{hscode}\SaveRestoreHook
\column{B}{@{}>{\hspre}l<{\hspost}@{}}%
\column{3}{@{}>{\hspre}l<{\hspost}@{}}%
\column{5}{@{}>{\hspre}l<{\hspost}@{}}%
\column{7}{@{}>{\hspre}l<{\hspost}@{}}%
\column{8}{@{}>{\hspre}l<{\hspost}@{}}%
\column{10}{@{}>{\hspre}l<{\hspost}@{}}%
\column{11}{@{}>{\hspre}l<{\hspost}@{}}%
\column{12}{@{}>{\hspre}l<{\hspost}@{}}%
\column{13}{@{}>{\hspre}l<{\hspost}@{}}%
\column{14}{@{}>{\hspre}l<{\hspost}@{}}%
\column{16}{@{}>{\hspre}l<{\hspost}@{}}%
\column{17}{@{}>{\hspre}l<{\hspost}@{}}%
\column{25}{@{}>{\hspre}l<{\hspost}@{}}%
\column{26}{@{}>{\hspre}l<{\hspost}@{}}%
\column{29}{@{}>{\hspre}l<{\hspost}@{}}%
\column{43}{@{}>{\hspre}l<{\hspost}@{}}%
\column{44}{@{}>{\hspre}l<{\hspost}@{}}%
\column{49}{@{}>{\hspre}l<{\hspost}@{}}%
\column{52}{@{}>{\hspre}l<{\hspost}@{}}%
\column{E}{@{}>{\hspre}l<{\hspost}@{}}%
\>[B]{}\mathbf{type}\;\Varid{Store}_{CS}\;{}\<[16]%
\>[16]{}\Varid{\sigma}\;\Varid{l}\;\Varid{v}{}\<[25]%
\>[25]{}\mathrel{=}[\mskip1.5mu \Varid{R}_{CS}\;\Varid{\sigma}\;\Varid{l}\;\Varid{v}\mskip1.5mu]{}\<[E]%
\\
\>[B]{}\mathbf{type}\;\Varid{R}_{CS}\;{}\<[16]%
\>[16]{}\Varid{\sigma}\;\Varid{l}\;\Varid{v}{}\<[25]%
\>[25]{}\mathrel{=}\Varid{l}\;()\to \Conid{Tree}\;(\Conid{Abstracting}\;\Varid{v}\mathbin{+}\Varid{\sigma})\;\Varid{l}\;(\Varid{l}\;\Varid{v}){}\<[E]%
\\
\>[B]{}\mathbf{type}\;\Varid{Store}_{DS}\;{}\<[16]%
\>[16]{}\Varid{\sigma}\;\Varid{l}\;\Varid{v}{}\<[25]%
\>[25]{}\mathrel{=}[\mskip1.5mu \Varid{R}_{DS}\;\Varid{\sigma}\;\Varid{l}\;\Varid{v}\mskip1.5mu]{}\<[E]%
\\
\>[B]{}\mathbf{type}\;\Varid{R}_{DS}\;{}\<[16]%
\>[16]{}\Varid{\sigma}\;\Varid{l}\;\Varid{v}{}\<[25]%
\>[25]{}\mathrel{=}\Conid{Tree}\;(\Conid{Abstracting}\;\Varid{v}\mathbin{+}\Varid{\sigma})\;\Varid{l}\;(\Varid{l}\;\Varid{v}){}\<[E]%
\\[\blanklineskip]%
\>[B]{}\Varid{hAbs}_{CS}{}\<[14]%
\>[14]{}\mathbin{::}(\Conid{Closure}\;\Varid{v}\mathbin{<:}\Varid{v},\Conid{Functor}\;\Varid{l}){}\<[E]%
\\
\>[14]{}\Rightarrow [\mskip1.5mu \Varid{v}\mskip1.5mu]{}\<[E]%
\\
\>[14]{}\to \Varid{Store}_{CS}\;\Varid{\sigma}\;\Varid{l}\;\Varid{v}{}\<[E]%
\\
\>[14]{}\to \Conid{Tree}\;(\Conid{Abstracting}\;\Varid{v}\mathbin{+}\Varid{\sigma})\;\Varid{l}\;\Varid{a}{}\<[E]%
\\
\>[14]{}\to \Conid{Tree}\;\Varid{\sigma}\;(\Conid{StateL}\;(\Varid{Store}_{CS}\;\Varid{\sigma}\;\Varid{l}\;\Varid{v})\;\Varid{l})\;(\Varid{Store}_{CS}\;\Varid{\sigma}\;\Varid{l}\;\Varid{v},\Varid{a}){}\<[E]%
\\
\>[B]{}\Varid{hAbs}_{CS}\;\anonymous \;{}\<[17]%
\>[17]{}\Varid{r}\;(\Conid{Leaf}\;\Varid{x})\mathrel{=}\Conid{Leaf}\;(\Varid{r},\Varid{x}){}\<[E]%
\\
\>[B]{}\Varid{hAbs}_{CS}\;\Varid{nv}\;{}\<[17]%
\>[17]{}\Varid{r}\;(\Conid{Node}\;(\Conid{Inl'}\;\Conid{Abs})\;\Varid{l}\;\Varid{st}\;\Varid{k})\mathrel{=}\mathbf{do}{}\<[E]%
\\
\>[B]{}\hsindent{3}{}\<[3]%
\>[3]{}\mathbf{let}\;\Varid{v}{}\<[12]%
\>[12]{}\mathrel{=}\Varid{inj}_{\mathrm{v}}\;(\Conid{Clos}\;(\Varid{length}\;\Varid{r})\;\Varid{nv}){}\<[E]%
\\
\>[B]{}\hsindent{3}{}\<[3]%
\>[3]{}\mathbf{let}\;\Varid{r'}{}\<[12]%
\>[12]{}\mathrel{=}\Varid{r}\plus [\mskip1.5mu \Varid{st}\;\Conid{One}\mskip1.5mu]{}\<[E]%
\\
\>[B]{}\hsindent{3}{}\<[3]%
\>[3]{}\Varid{hAbs}_{CS}\;{}\<[16]%
\>[16]{}\Varid{nv}\;\Varid{r'}\;(\Varid{k}\;(\Varid{v}\mathrel{{<\kern-1pt}{\$}}\Varid{l})){}\<[E]%
\\
\>[B]{}\Varid{hAbs}_{CS}\;\Varid{nv}\;{}\<[17]%
\>[17]{}\Varid{r}\;(\Conid{Node}\;(\Conid{Inl'}\;(\Conid{App}\;\Varid{v}_{1}\;\Varid{v}_{2}))\;\Varid{l}\;\anonymous \;\Varid{k}){}\<[52]%
\>[52]{}\mathrel{=}\mathbf{case}\;\Varid{proj}_{\mathrm{v}}\;\Varid{v}_{1}\;\mathbf{of}{}\<[E]%
\\
\>[B]{}\hsindent{5}{}\<[5]%
\>[5]{}\Conid{Just}\;(\Conid{Clos}\;\Varid{fp}\;\Varid{nv'}){}\<[25]%
\>[25]{}\to \mathbf{do}{}\<[E]%
\\
\>[5]{}\hsindent{2}{}\<[7]%
\>[7]{}(\Varid{r'},\Varid{v})\leftarrow \Varid{hAbs}_{CS}\;(\Varid{v}_{2}\mathbin{:}\Varid{nv'})\;\Varid{r}\;((\Varid{r}\mathbin{!!}\Varid{fp})\;\Varid{l}){}\<[E]%
\\
\>[5]{}\hsindent{2}{}\<[7]%
\>[7]{}\Varid{hAbs}_{CS}\;\Varid{nv}\;\Varid{r'}\;(\Varid{k}\;\Varid{v}){}\<[E]%
\\
\>[B]{}\hsindent{5}{}\<[5]%
\>[5]{}\Conid{Nothing}{}\<[25]%
\>[25]{}\to \Varid{error}\;\text{\tt \char34 application~error\char34}{}\<[E]%
\\
\>[B]{}\Varid{hAbs}_{CS}\;\Varid{nv}\;{}\<[17]%
\>[17]{}\Varid{r}\;(\Conid{Node}\;(\Conid{Inl'}\;(\Conid{Var}\;\Varid{n}))\;\Varid{l}\;{}\<[43]%
\>[43]{}\anonymous \;\Varid{k}){}\<[49]%
\>[49]{}\mathrel{=}\Varid{hAbs}_{CS}\;\Varid{nv}\;\Varid{r}\;(\Varid{k}\;((\Varid{nv}\mathbin{!!}\Varid{n})\mathrel{{<\kern-1pt}{\$}}\Varid{l})){}\<[E]%
\\
\>[B]{}\Varid{hAbs}_{CS}\;\Varid{nv}\;{}\<[17]%
\>[17]{}\Varid{r}\;(\Conid{Node}\;(\Conid{Inr'}\;\Varid{op})\;\Varid{l}\;\Varid{st}\;\Varid{k}){}\<[44]%
\>[44]{}\mathrel{=}\Conid{Node}\;\Varid{op}\;(\Conid{StateL}\;(\Varid{r},\Varid{l})){}\<[E]%
\\
\>[B]{}\hsindent{3}{}\<[3]%
\>[3]{}(\lambda \Varid{c}\;{}\<[8]%
\>[8]{}(\Conid{StateL}\;(\Varid{r'},\Varid{l}{}\<[25]%
\>[25]{})){}\<[29]%
\>[29]{}\to \Conid{StateL}\mathrel{{<\kern-1pt}{\$}{\kern-1pt>}}\Varid{hAbs}_{CS}\;\Varid{nv}\;\Varid{r'}\;(\Varid{st}\;\Varid{c}\;\Varid{l})){}\<[E]%
\\
\>[B]{}\hsindent{3}{}\<[3]%
\>[3]{}(\lambda {}\<[8]%
\>[8]{}(\Conid{StateL}\;(\Varid{r'},\Varid{lv}{}\<[25]%
\>[25]{})){}\<[29]%
\>[29]{}\to \Varid{hAbs}_{CS}\;\Varid{nv}\;\Varid{r'}\;(\Varid{k}\;\Varid{lv})){}\<[E]%
\\[\blanklineskip]%
\>[B]{}\Varid{hAbs}_{DS}\mathbin{::}(\Conid{Closure}\;\Varid{v}\mathbin{<:}\Varid{v},\Conid{Functor}\;\Varid{l}){}\<[E]%
\\
\>[B]{}\hsindent{13}{}\<[13]%
\>[13]{}\Rightarrow [\mskip1.5mu \Varid{v}\mskip1.5mu]{}\<[E]%
\\
\>[B]{}\hsindent{13}{}\<[13]%
\>[13]{}\to \Varid{Store}_{DS}\;\Varid{\sigma}\;\Varid{l}\;\Varid{v}{}\<[E]%
\\
\>[B]{}\hsindent{13}{}\<[13]%
\>[13]{}\to \Conid{Tree}\;(\Conid{Abstracting}\;\Varid{v}\mathbin{+}\Varid{\sigma})\;\Varid{l}\;\Varid{a}{}\<[E]%
\\
\>[B]{}\hsindent{13}{}\<[13]%
\>[13]{}\to \Conid{Tree}\;\Varid{\sigma}\;{}\<[26]%
\>[26]{}(\Conid{StateL}\;(\Varid{Store}_{DS}\;\Varid{\sigma}\;\Varid{l}\;\Varid{v})\;\Varid{l})\;{}\<[E]%
\\
\>[26]{}(\Varid{Store}_{DS}\;\Varid{\sigma}\;\Varid{l}\;\Varid{v},\Varid{a}){}\<[E]%
\\
\>[B]{}\Varid{hAbs}_{DS}\;\anonymous \;{}\<[17]%
\>[17]{}\Varid{r}\;(\Conid{Leaf}\;\Varid{x})\mathrel{=}\Conid{Leaf}\;(\Varid{r},\Varid{x}){}\<[E]%
\\
\>[B]{}\Varid{hAbs}_{DS}\;\Varid{nv}\;{}\<[17]%
\>[17]{}\Varid{r}\;(\Conid{Node}\;(\Conid{Inl'}\;\Conid{Abs})\;\Varid{l}\;\Varid{st}\;\Varid{k})\mathrel{=}\mathbf{do}{}\<[E]%
\\
\>[B]{}\hsindent{3}{}\<[3]%
\>[3]{}\mathbf{let}\;\Varid{v}{}\<[10]%
\>[10]{}\mathrel{=}\Varid{inj}_{\mathrm{v}}\;(\Conid{Clos}\;(\Varid{length}\;\Varid{r})\;\Varid{nv}){}\<[E]%
\\
\>[B]{}\hsindent{3}{}\<[3]%
\>[3]{}\mathbf{let}\;\Varid{r'}\mathrel{=}\Varid{r}\plus [\mskip1.5mu \Varid{st}\;\Conid{One}\;\Varid{l}\mskip1.5mu]{}\<[E]%
\\
\>[B]{}\hsindent{3}{}\<[3]%
\>[3]{}\Varid{hAbs}_{DS}\;{}\<[16]%
\>[16]{}\Varid{nv}\;\Varid{r'}\;(\Varid{k}\;(\Varid{v}\mathrel{{<\kern-1pt}{\$}}\Varid{l})){}\<[E]%
\\
\>[B]{}\Varid{hAbs}_{DS}\;\Varid{nv}\;{}\<[17]%
\>[17]{}\Varid{r}\;(\Conid{Node}\;(\Conid{Inl'}\;(\Conid{App}\;\Varid{v}_{1}\;\Varid{v}_{2}))\;\anonymous \;\anonymous \;\Varid{k})\mathrel{=}{}\<[E]%
\\
\>[B]{}\hsindent{3}{}\<[3]%
\>[3]{}\mathbf{case}\;\Varid{proj}_{\mathrm{v}}\;\Varid{v}_{1}\;\mathbf{of}{}\<[E]%
\\
\>[3]{}\hsindent{2}{}\<[5]%
\>[5]{}\Conid{Just}\;(\Conid{Clos}\;\Varid{fp}\;\Varid{nv'})\to \mathbf{do}{}\<[E]%
\\
\>[5]{}\hsindent{2}{}\<[7]%
\>[7]{}(\Varid{r'},\Varid{v})\leftarrow \Varid{hAbs}_{DS}\;(\Varid{v}_{2}\mathbin{:}\Varid{nv'})\;\Varid{r}\;(\Varid{r}\mathbin{!!}\Varid{fp}){}\<[E]%
\\
\>[5]{}\hsindent{2}{}\<[7]%
\>[7]{}\Varid{hAbs}_{DS}\;\Varid{nv}\;\Varid{r'}\;(\Varid{k}\;\Varid{v}){}\<[E]%
\\
\>[3]{}\hsindent{2}{}\<[5]%
\>[5]{}\Conid{Nothing}\to \Varid{error}\;\text{\tt \char34 application~error\char34}{}\<[E]%
\\
\>[B]{}\Varid{hAbs}_{DS}\;\Varid{nv}\;{}\<[17]%
\>[17]{}\Varid{r}\;(\Conid{Node}\;(\Conid{Inl'}\;(\Conid{Var}\;\Varid{n}))\;\Varid{l}\;\anonymous \;\Varid{k})\mathrel{=}{}\<[E]%
\\
\>[B]{}\hsindent{3}{}\<[3]%
\>[3]{}\Varid{hAbs}_{DS}\;\Varid{nv}\;\Varid{r}\;(\Varid{k}\;(\Varid{nv}\mathbin{!!}\Varid{n}\mathrel{{<\kern-1pt}{\$}}\Varid{l})){}\<[E]%
\\
\>[B]{}\Varid{hAbs}_{DS}\;\Varid{nv}\;{}\<[17]%
\>[17]{}\Varid{r}\;(\Conid{Node}\;(\Conid{Inr'}\;\Varid{op})\;\Varid{l}\;\Varid{st}\;\Varid{k})\mathrel{=}{}\<[E]%
\\
\>[B]{}\hsindent{3}{}\<[3]%
\>[3]{}\Conid{Node}\;\Varid{op}\;(\Conid{StateL}\;(\Varid{r},\Varid{l}))\;{}\<[E]%
\\
\>[3]{}\hsindent{5}{}\<[8]%
\>[8]{}(\lambda \Varid{c}\;(\Conid{StateL}\;(\Varid{r'},\Varid{l}))\to {}\<[E]%
\\
\>[8]{}\hsindent{3}{}\<[11]%
\>[11]{}\Conid{StateL}\mathrel{{<\kern-1pt}{\$}{\kern-1pt>}}\Varid{hAbs}_{DS}\;\Varid{nv}\;\Varid{r'}\;(\Varid{st}\;\Varid{c}\;\Varid{l}))\;{}\<[E]%
\\
\>[3]{}\hsindent{5}{}\<[8]%
\>[8]{}(\lambda (\Conid{StateL}\;(\Varid{r'},\Varid{lv})){}\<[29]%
\>[29]{}\to \Varid{hAbs}_{DS}\;\Varid{nv}\;\Varid{r'}\;(\Varid{k}\;\Varid{lv})){}\<[E]%
\ColumnHook
\end{hscode}\resethooks
\begin{hscode}\SaveRestoreHook
\column{B}{@{}>{\hspre}l<{\hspost}@{}}%
\column{12}{@{}>{\hspre}l<{\hspost}@{}}%
\column{E}{@{}>{\hspre}l<{\hspost}@{}}%
\>[B]{}\Varid{var}{}\<[12]%
\>[12]{}\mathbin{::}(\Conid{Abstracting}\;\Varid{v}\mathbin{<}\Varid{\sigma})\Rightarrow \Conid{Int}\to \Conid{Tree}\;\Varid{\sigma}\;\Conid{Id}\;\Varid{v}{}\<[E]%
\\
\>[B]{}\Varid{var}\;\Varid{n}{}\<[12]%
\>[12]{}\mathrel{=}\Varid{noSubNode}\;(\Conid{Var}\;\Varid{n}){}\<[E]%
\\[\blanklineskip]%
\>[B]{}\Varid{app}{}\<[12]%
\>[12]{}\mathbin{::}(\Conid{Abstracting}\;\Varid{v}\mathbin{<}\Varid{\sigma})\Rightarrow \Varid{v}\to \Varid{v}\to \Conid{Tree}\;\Varid{\sigma}\;\Conid{Id}\;\Varid{v}{}\<[E]%
\\
\>[B]{}\Varid{app}\;\Varid{v}_{1}\;\Varid{v}_{2}{}\<[12]%
\>[12]{}\mathrel{=}\Varid{noSubNode}\;(\Conid{App}\;\Varid{v}_{1}\;\Varid{v}_{2}){}\<[E]%
\\[\blanklineskip]%
\>[B]{}\Varid{abs}{}\<[12]%
\>[12]{}\mathbin{::}(\Conid{Abstracting}\;\Varid{v}\mathbin{<}\Varid{\sigma})\Rightarrow \Conid{Tree}\;\Varid{\sigma}\;\Conid{Id}\;\Varid{v}\to \Conid{Tree}\;\Varid{\sigma}\;\Conid{Id}\;\Varid{v}{}\<[E]%
\\
\>[B]{}\Varid{abs}\;\Varid{t}{}\<[12]%
\>[12]{}\mathrel{=}\Varid{oneSubNode}\;\Conid{Abs}\;\Varid{t}{}\<[E]%
\ColumnHook
\end{hscode}\resethooks

\subsection{Language Features}
\label{app:language-features}

This section introduces language features that can be used as individual components to modularly construct languages.
We provide a syntax for these features, as well as a mapping to latent effects,
by means of an algebra.

\subsubsection{Arithmetics}
We make a simple arithmetic language with numbers and addition.

\begin{hscode}\SaveRestoreHook
\column{B}{@{}>{\hspre}l<{\hspost}@{}}%
\column{21}{@{}>{\hspre}l<{\hspost}@{}}%
\column{E}{@{}>{\hspre}l<{\hspost}@{}}%
\>[B]{}\mathbf{data}\;\Conid{Arith}\;\Varid{n}\mathrel{=}\Conid{Num}\;\Conid{Int}\mid \Conid{Add}\;\Varid{n}\;\Varid{n}{}\<[E]%
\\[\blanklineskip]%
\>[B]{}\Varid{arithAlg}\mathbin{::}(\Conid{Adding}\;\Varid{v}\mathbin{<}\Varid{\sigma})\Rightarrow \Conid{Arith}\;\Varid{v}\to \Conid{Tree}\;\Varid{\sigma}\;\Conid{Id}\;\Varid{v}{}\<[E]%
\\
\>[B]{}\Varid{arithAlg}\;(\Conid{Num}\;\Varid{n}){}\<[21]%
\>[21]{}\mathrel{=}\Varid{nat}\;(\Varid{toInteger}\;\Varid{n}){}\<[E]%
\\
\>[B]{}\Varid{arithAlg}\;(\Conid{Add}\;\Varid{x}\;\Varid{y}){}\<[21]%
\>[21]{}\mathrel{=}\Varid{plus}\;\Varid{x}\;\Varid{y}{}\<[E]%
\ColumnHook
\end{hscode}\resethooks

\subsubsection{Lambda calculus}
The lambda calculus has constructs for variables, abstraction and application.

\begin{hscode}\SaveRestoreHook
\column{B}{@{}>{\hspre}l<{\hspost}@{}}%
\column{E}{@{}>{\hspre}l<{\hspost}@{}}%
\>[B]{}\mathbf{data}\;\Conid{LamExpr}\;\Varid{e}\mathrel{=}\Conid{VarExpr}\;\Conid{Int}\mid \Conid{AbsExpr}\;\Varid{e}\mid \Conid{AppExpr}\;\Varid{e}\;\Varid{e}{}\<[E]%
\ColumnHook
\end{hscode}\resethooks

We can define different algebras, depending on the evaluation strategy:
\begin{hscode}\SaveRestoreHook
\column{B}{@{}>{\hspre}l<{\hspost}@{}}%
\column{3}{@{}>{\hspre}l<{\hspost}@{}}%
\column{25}{@{}>{\hspre}l<{\hspost}@{}}%
\column{E}{@{}>{\hspre}l<{\hspost}@{}}%
\>[B]{}\Varid{lamAlg}\mathbin{::}(\Conid{Abstracting}\;\Varid{v}\mathbin{<}\Varid{\sigma})\Rightarrow \Conid{LamExpr}\;(\Conid{Tree}\;\Varid{\sigma}\;\Conid{Id}\;\Varid{v})\to \Conid{Tree}\;\Varid{\sigma}\;\Conid{Id}\;\Varid{v}{}\<[E]%
\\
\>[B]{}\Varid{lamAlg}\;(\Conid{VarExpr}\;\Varid{n}){}\<[25]%
\>[25]{}\mathrel{=}\Varid{var}\;\Varid{n}{}\<[E]%
\\
\>[B]{}\Varid{lamAlg}\;(\Conid{AbsExpr}\;\Varid{e}){}\<[25]%
\>[25]{}\mathrel{=}\Varid{abs}\;\Varid{e}{}\<[E]%
\\
\>[B]{}\Varid{lamAlg}\;(\Conid{AppExpr}\;\Varid{e}_{1}\;\Varid{e}_{2}){}\<[25]%
\>[25]{}\mathrel{=}\mathbf{do}{}\<[E]%
\\
\>[B]{}\hsindent{3}{}\<[3]%
\>[3]{}\Varid{v}_{1}\leftarrow \Varid{e}_{1}{}\<[E]%
\\
\>[B]{}\hsindent{3}{}\<[3]%
\>[3]{}\Varid{v}_{2}\leftarrow \Varid{e}_{2}{}\<[E]%
\\
\>[B]{}\hsindent{3}{}\<[3]%
\>[3]{}\Varid{app}\;\Varid{v}_{1}\;\Varid{v}_{2}{}\<[E]%
\ColumnHook
\end{hscode}\resethooks

Call-by-need evaluation can be expressed through the following algebra, where
\ensuremath{\Varid{abs}}, \ensuremath{\Varid{app}} and \ensuremath{\Varid{var}} are defined in Section \ref{app:cbn-eval}.

\begin{hscode}\SaveRestoreHook
\column{B}{@{}>{\hspre}l<{\hspost}@{}}%
\column{10}{@{}>{\hspre}l<{\hspost}@{}}%
\column{16}{@{}>{\hspre}l<{\hspost}@{}}%
\column{26}{@{}>{\hspre}l<{\hspost}@{}}%
\column{E}{@{}>{\hspre}l<{\hspost}@{}}%
\>[B]{}\Varid{lazyAlg}{}\<[10]%
\>[10]{}\mathbin{::}({}\<[16]%
\>[16]{}\Conid{Reading}\;[\mskip1.5mu \Varid{v}\mskip1.5mu]\mathbin{<}\sigma,\Conid{Thunking}\;\Varid{v}\mathbin{<}\sigma,\Conid{Suspending}\;\Varid{v}\mathbin{<}\sigma,{}\<[E]%
\\
\>[16]{}\Conid{Thunked}\;[\mskip1.5mu \Varid{v}\mskip1.5mu]\mathbin{<:}\Varid{v},\Conid{Closure}\;\Varid{v}\mathbin{<:}\Varid{v}){}\<[E]%
\\
\>[10]{}\Rightarrow \Conid{LamExpr}\;(\Conid{Tree}\;\sigma\;\Conid{Id}\;\Varid{v})\to \Conid{Tree}\;\sigma\;\Conid{Id}\;\Varid{v}{}\<[E]%
\\
\>[B]{}\Varid{lazyAlg}\;(\Conid{VarExpr}\;\Varid{n}){}\<[26]%
\>[26]{}\mathrel{=}\Varid{var}_{lazy}\;\Varid{n}{}\<[E]%
\\
\>[B]{}\Varid{lazyAlg}\;(\Conid{AbsExpr}\;\Varid{e}){}\<[26]%
\>[26]{}\mathrel{=}\Varid{abs}_{lazy}\;\Varid{e}{}\<[E]%
\\
\>[B]{}\Varid{lazyAlg}\;(\Conid{AppExpr}\;\Varid{e}_{1}\;\Varid{e}_{2}){}\<[26]%
\>[26]{}\mathrel{=}\Varid{app}_{lazy}\;\Varid{e}_{1}\;\Varid{e}_{2}{}\<[E]%
\ColumnHook
\end{hscode}\resethooks

Call-by-name evaluation is similar, but without memoizing (intermediate) results.

\begin{hscode}\SaveRestoreHook
\column{B}{@{}>{\hspre}l<{\hspost}@{}}%
\column{9}{@{}>{\hspre}l<{\hspost}@{}}%
\column{10}{@{}>{\hspre}l<{\hspost}@{}}%
\column{15}{@{}>{\hspre}l<{\hspost}@{}}%
\column{16}{@{}>{\hspre}l<{\hspost}@{}}%
\column{25}{@{}>{\hspre}l<{\hspost}@{}}%
\column{E}{@{}>{\hspre}l<{\hspost}@{}}%
\>[B]{}\Varid{cbnAlg}{}\<[9]%
\>[9]{}\mathbin{::}({}\<[15]%
\>[15]{}\Conid{Reading}\;[\mskip1.5mu \Varid{v}\mskip1.5mu]\mathbin{<}\sigma,\Conid{Suspending}\;\Varid{v}\mathbin{<}\sigma,{}\<[E]%
\\
\>[15]{}\hsindent{1}{}\<[16]%
\>[16]{}\Conid{Suspended}\;[\mskip1.5mu \Varid{v}\mskip1.5mu]\mathbin{<:}\Varid{v},\Conid{Closure}\;\Varid{v}\mathbin{<:}\Varid{v}){}\<[E]%
\\
\>[9]{}\hsindent{1}{}\<[10]%
\>[10]{}\Rightarrow \Conid{LamExpr}\;(\Conid{Tree}\;\sigma\;\Conid{Id}\;\Varid{v})\to \Conid{Tree}\;\sigma\;\Conid{Id}\;\Varid{v}{}\<[E]%
\\
\>[B]{}\Varid{cbnAlg}\;(\Conid{VarExpr}\;\Varid{n}){}\<[25]%
\>[25]{}\mathrel{=}\Varid{var}_\textsc{cbn}\;\Varid{n}{}\<[E]%
\\
\>[B]{}\Varid{cbnAlg}\;(\Conid{AbsExpr}\;\Varid{e}){}\<[25]%
\>[25]{}\mathrel{=}\Varid{abs}_\textsc{cbn}\;\Varid{e}{}\<[E]%
\\
\>[B]{}\Varid{cbnAlg}\;(\Conid{AppExpr}\;\Varid{e}_{1}\;\Varid{e}_{2}){}\<[25]%
\>[25]{}\mathrel{=}\Varid{app}_\textsc{cbn}\;\Varid{e}_{1}\;\Varid{e}_{2}{}\<[E]%
\ColumnHook
\end{hscode}\resethooks

\subsubsection{Staging}
For the staging expressions, we refer to Section \ref{app:latent-staging} for the 
implementations of \ensuremath{\Varid{splice}}, \ensuremath{\Varid{quote}}, \ensuremath{\Varid{shift}} and \ensuremath{\Varid{unquote}}.

\begin{hscode}\SaveRestoreHook
\column{B}{@{}>{\hspre}l<{\hspost}@{}}%
\column{3}{@{}>{\hspre}l<{\hspost}@{}}%
\column{13}{@{}>{\hspre}l<{\hspost}@{}}%
\column{16}{@{}>{\hspre}c<{\hspost}@{}}%
\column{16E}{@{}l@{}}%
\column{17}{@{}>{\hspre}l<{\hspost}@{}}%
\column{19}{@{}>{\hspre}l<{\hspost}@{}}%
\column{25}{@{}>{\hspre}l<{\hspost}@{}}%
\column{E}{@{}>{\hspre}l<{\hspost}@{}}%
\>[3]{}\mathbf{data}\;\Conid{StageExpr}\;\Varid{s}\mathrel{=}\Conid{Splice}\;\Varid{s}\mid \Conid{Quote}\;\Varid{s}\mid \Conid{Unquote}\;\Varid{s}\mid \Conid{Shift}\;\Conid{Int}\;\Varid{s}{}\<[E]%
\\[\blanklineskip]%
\>[3]{}\Varid{stageAlg}{}\<[13]%
\>[13]{}\mathbin{::}{}\<[17]%
\>[17]{}\forall \Varid{v}\hsforall \;\sigma\;\Varid{r}_{1}\;\Varid{r}_{2}\hsdot{\circ }{.}{}\<[E]%
\\
\>[13]{}\hsindent{3}{}\<[16]%
\>[16]{}({}\<[16E]%
\>[19]{}\Conid{Suspending}\;\Varid{v}\mathbin{<}\sigma,\Conid{Reading}\;\Varid{r}_{1}\mathbin{<}\sigma,\Conid{Reading}\;\Varid{r}_{2}\mathbin{<}\sigma,{}\<[E]%
\\
\>[19]{}\Conid{Reading}\;(\Conid{Env}_s\;\Varid{v})\mathbin{<}\sigma,\Conid{Suspended}\;\Varid{r}_{1}\mathbin{<:}\Varid{v},{}\<[E]%
\\
\>[19]{}\Conid{Suspended}\;\Varid{r}_{2}\mathbin{<:}\Varid{v},\Conid{Suspended}\;(\Conid{Env}_s\;\Varid{v})\mathbin{<:}\Varid{v}){}\<[E]%
\\
\>[13]{}\Rightarrow \Conid{StageExpr}\;(\Conid{Tree}\;\sigma\;\Conid{Id}\;\Varid{v})\to \Conid{Tree}\;\sigma\;\Conid{Id}\;\Varid{v}{}\<[E]%
\\
\>[3]{}\Varid{stageAlg}\;(\Conid{Splice}\;\Varid{s}){}\<[25]%
\>[25]{}\mathrel{=}\Varid{splice}\;\Varid{s}{}\<[E]%
\\
\>[3]{}\Varid{stageAlg}\;(\Conid{Quote}\;\Varid{s}){}\<[25]%
\>[25]{}\mathrel{=}\Varid{quote}\mathord{@}\Varid{r}_{1}\;\Varid{s}{}\<[E]%
\\
\>[3]{}\Varid{stageAlg}\;(\Conid{Unquote}\;\Varid{s}){}\<[25]%
\>[25]{}\mathrel{=}\Varid{unquote}\mathord{@}\Varid{r}_{2}\;\Varid{s}{}\<[E]%
\\
\>[3]{}\Varid{stageAlg}\;(\Conid{Shift}\;\Varid{n}\;\Varid{s}){}\<[25]%
\>[25]{}\mathrel{=}\Varid{shift}\;\Varid{n}\;\Varid{s}{}\<[E]%
\ColumnHook
\end{hscode}\resethooks
\subsubsection{Other}
We can define other simple language features such as printing and let-expressions.

\begin{hscode}\SaveRestoreHook
\column{B}{@{}>{\hspre}l<{\hspost}@{}}%
\column{E}{@{}>{\hspre}l<{\hspost}@{}}%
\>[B]{}\mathbf{data}\;\Conid{PrintExpr}\;\Varid{e}\mathrel{=}\Conid{Pr}\;\Conid{String}{}\<[E]%
\\[\blanklineskip]%
\>[B]{}\Varid{printAlg}\mathbin{::}(\Conid{Printing}\;\Varid{v}\mathbin{<}\Varid{\sigma})\Rightarrow \Conid{PrintExpr}\;\Varid{v}\to \Conid{Tree}\;\Varid{\sigma}\;\Conid{Id}\;\Varid{v}{}\<[E]%
\\
\>[B]{}\Varid{printAlg}\;(\Conid{Pr}\;\Varid{s})\mathrel{=}\Varid{print}\;\Varid{s}{}\<[E]%
\ColumnHook
\end{hscode}\resethooks

\begin{hscode}\SaveRestoreHook
\column{B}{@{}>{\hspre}l<{\hspost}@{}}%
\column{3}{@{}>{\hspre}l<{\hspost}@{}}%
\column{21}{@{}>{\hspre}l<{\hspost}@{}}%
\column{E}{@{}>{\hspre}l<{\hspost}@{}}%
\>[B]{}\mathbf{data}\;\Conid{LetExpr}\;\Varid{e}\mathrel{=}\Conid{Let}\;\Varid{e}\;\Varid{e}\mid \Conid{Seq}\;\Varid{e}\;\Varid{e}\mid \Conid{LetVar}\;\Conid{Int}{}\<[E]%
\\[\blanklineskip]%
\>[B]{}\Varid{letAlg}\mathbin{::}(\Conid{Abstracting}\;\Varid{v}\mathbin{<}\Varid{\sigma})\Rightarrow \Conid{LetExpr}\;(\Conid{Tree}\;\Varid{\sigma}\;\Conid{Id}\;\Varid{v})\to \Conid{Tree}\;\Varid{\sigma}\;\Conid{Id}\;\Varid{v}{}\<[E]%
\\
\>[B]{}\Varid{letAlg}\;(\Conid{Let}\;\Varid{e}_{1}\;\Varid{e}_{2}){}\<[21]%
\>[21]{}\mathrel{=}\mathbf{do}{}\<[E]%
\\
\>[B]{}\hsindent{3}{}\<[3]%
\>[3]{}\Varid{v}_{1}\leftarrow \Varid{abs}\;\Varid{e}_{1}{}\<[E]%
\\
\>[B]{}\hsindent{3}{}\<[3]%
\>[3]{}\Varid{v}_{2}\leftarrow \Varid{e}_{2}{}\<[E]%
\\
\>[B]{}\hsindent{3}{}\<[3]%
\>[3]{}\Varid{app}\;\Varid{v}_{1}\;\Varid{v}_{2}{}\<[E]%
\\
\>[B]{}\Varid{letAlg}\;(\Conid{Seq}\;\Varid{e}_{1}\;\Varid{e}_{2}){}\<[21]%
\>[21]{}\mathrel{=}\mathbf{do}\;\Varid{e}_{1};\Varid{e}_{2}{}\<[E]%
\\
\>[B]{}\Varid{letAlg}\;(\Conid{LetVar}\;\Varid{n}){}\<[21]%
\>[21]{}\mathrel{=}\Varid{var}\;\Varid{n}{}\<[E]%
\ColumnHook
\end{hscode}\resethooks

\section{Modular Carriers for Call-by-Need Lambdas}
\label{app:cbn}

This Appendix shows how we can use an adapted form 
of the modular carriers of Schrijvers et al. \cite{SchrijversPWJ19}
to define call-by-need and call-by-value evaluation for 
function abstractions.
We give semantics to call-by-need lambdas using three different effects: 
\ensuremath{\Conid{Reading}}, \ensuremath{\Conid{Suspending}} and \ensuremath{\Conid{Thunking}}.
For the definition and the handlers of these effects, we refer to Appendix~\ref{app:effect-lib}.

\subsection{Modular Carriers}
\label{modular-carriers}

The \ensuremath{\Varid{hThunk}} handler uses a technique that lets us access values wrapped in a latent effect functor:
we use a \emph{forwarding algebra}, inspired by the work of Schrijvers et al. \cite{SchrijversPWJ19}. 
They propose a modular approach to composing algebraic effects so that each handler only has to know about the part of the syntax its
effect is handling.
This is achieved through the definition of a modular carrier \ensuremath{\Conid{Modular}} that features a forwarding algebra \ensuremath{\Varid{fwd}}.
\begin{hscode}\SaveRestoreHook
\column{B}{@{}>{\hspre}l<{\hspost}@{}}%
\column{3}{@{}>{\hspre}l<{\hspost}@{}}%
\column{5}{@{}>{\hspre}l<{\hspost}@{}}%
\column{E}{@{}>{\hspre}l<{\hspost}@{}}%
\>[3]{}\mathbf{class}\;\Conid{Modular}\;(\Varid{m}_{1}\mathbin{::}\mathbin{*}\to \mathbin{*})\;(\Varid{m}_{2}\mathbin{::}(\mathbin{*}\to \mathbin{*})\to \mathbin{*}\to \mathbin{*})\;\mathbf{where}{}\<[E]%
\\
\>[3]{}\hsindent{2}{}\<[5]%
\>[5]{}\Varid{fwd}\mathbin{::}\Varid{m}_{1}\;(\Varid{m}_{2}\;\Varid{m}_{1}\;\Varid{a})\to \Varid{m}_{2}\;\Varid{m}_{1}\;\Varid{a}{}\<[E]%
\ColumnHook
\end{hscode}\resethooks

However, we use this modular carrier in a different context in the sense that 
it is now modular with respect to its subcomputations, 
for which the syntax and structure is known. 
On the contrary, the original definition is modular with respect to 
the outer computation.

\subsection{Call-by-Need Evaluation}
\label{app:cbn-eval}

The three operations for call-by-need evaluation are defined below.
\begin{hscode}\SaveRestoreHook
\column{B}{@{}>{\hspre}l<{\hspost}@{}}%
\column{3}{@{}>{\hspre}l<{\hspost}@{}}%
\column{6}{@{}>{\hspre}l<{\hspost}@{}}%
\column{11}{@{}>{\hspre}l<{\hspost}@{}}%
\column{E}{@{}>{\hspre}l<{\hspost}@{}}%
\>[B]{}\Varid{abs}_{lazy}{}\<[11]%
\>[11]{}\mathbin{::}\forall \Varid{v}\hsforall \;\sigma\hsdot{\circ }{.}(\Conid{Reading}\;[\mskip1.5mu \Varid{v}\mskip1.5mu]\mathbin{<}\sigma,\Conid{Suspending}\;\Varid{v}\mathbin{<}\sigma,\Conid{Closure}\;\Varid{v}\mathbin{<:}\Varid{v}){}\<[E]%
\\
\>[B]{}\hsindent{6}{}\<[6]%
\>[6]{}\Rightarrow \Conid{Tree}\;\sigma\;\Conid{Id}\;\Varid{v}\to \Conid{Tree}\;\sigma\;\Conid{Id}\;\Varid{v}{}\<[E]%
\\
\>[B]{}\Varid{abs}_{lazy}\;\Varid{body}\mathrel{=}\mathbf{do}{}\<[E]%
\\
\>[B]{}\hsindent{3}{}\<[3]%
\>[3]{}\Varid{nv}\leftarrow \Varid{ask}{}\<[E]%
\\
\>[B]{}\hsindent{3}{}\<[3]%
\>[3]{}\Varid{p}\leftarrow \Varid{suspend}\;\Varid{body}{}\<[E]%
\\
\>[B]{}\hsindent{3}{}\<[3]%
\>[3]{}\Varid{return}\;(\Varid{inj}_{\mathrm{v}}\;(\Conid{Clos}\;\Varid{p}\;(\Varid{nv}\mathbin{::}[\mskip1.5mu \Varid{v}\mskip1.5mu]))){}\<[E]%
\ColumnHook
\end{hscode}\resethooks
\begin{hscode}\SaveRestoreHook
\column{B}{@{}>{\hspre}l<{\hspost}@{}}%
\column{3}{@{}>{\hspre}l<{\hspost}@{}}%
\column{5}{@{}>{\hspre}l<{\hspost}@{}}%
\column{6}{@{}>{\hspre}l<{\hspost}@{}}%
\column{11}{@{}>{\hspre}l<{\hspost}@{}}%
\column{27}{@{}>{\hspre}l<{\hspost}@{}}%
\column{E}{@{}>{\hspre}l<{\hspost}@{}}%
\>[B]{}\Varid{var}_{lazy}{}\<[11]%
\>[11]{}\mathbin{::}\forall \Varid{v}\hsforall \;\sigma\hsdot{\circ }{.}(\Conid{Reading}\;[\mskip1.5mu \Varid{v}\mskip1.5mu]\mathbin{<}\sigma,\Conid{Thunking}\;\Varid{v}\mathbin{<}\sigma,\Conid{Thunked}\;[\mskip1.5mu \Varid{v}\mskip1.5mu]\mathbin{<:}\Varid{v}){}\<[E]%
\\
\>[B]{}\hsindent{6}{}\<[6]%
\>[6]{}\Rightarrow \Conid{Int}\to \Conid{Tree}\;\sigma\;\Conid{Id}\;\Varid{v}{}\<[E]%
\\
\>[B]{}\Varid{var}_{lazy}\;{}\<[11]%
\>[11]{}\Varid{x}\mathrel{=}\mathbf{do}{}\<[E]%
\\
\>[B]{}\hsindent{3}{}\<[3]%
\>[3]{}\Varid{nv}\leftarrow \Varid{ask}{}\<[E]%
\\
\>[B]{}\hsindent{3}{}\<[3]%
\>[3]{}\mathbf{let}\;\Varid{v}\mathrel{=}(\Varid{nv}\mathbin{::}[\mskip1.5mu \Varid{v}\mskip1.5mu])\mathbin{!!}\Varid{x}{}\<[E]%
\\
\>[B]{}\hsindent{3}{}\<[3]%
\>[3]{}\mathbf{case}\;\Varid{proj}_{\mathrm{v}}\;\Varid{v}\;\mathbf{of}{}\<[E]%
\\
\>[3]{}\hsindent{2}{}\<[5]%
\>[5]{}\Conid{Just}\;(\Conid{Thunked}\;\Varid{p}\;\Varid{nv'}){}\<[27]%
\>[27]{}\to \Varid{local}\;(\Varid{const}\;(\Varid{nv'}\mathbin{::}[\mskip1.5mu \Varid{v}\mskip1.5mu]))\;(\Varid{force}\;\Varid{p}){}\<[E]%
\\
\>[3]{}\hsindent{2}{}\<[5]%
\>[5]{}\Conid{Nothing}{}\<[27]%
\>[27]{}\to \Varid{return}\;\Varid{v}{}\<[E]%
\ColumnHook
\end{hscode}\resethooks
\begin{hscode}\SaveRestoreHook
\column{B}{@{}>{\hspre}l<{\hspost}@{}}%
\column{3}{@{}>{\hspre}l<{\hspost}@{}}%
\column{5}{@{}>{\hspre}l<{\hspost}@{}}%
\column{6}{@{}>{\hspre}l<{\hspost}@{}}%
\column{7}{@{}>{\hspre}l<{\hspost}@{}}%
\column{11}{@{}>{\hspre}l<{\hspost}@{}}%
\column{23}{@{}>{\hspre}l<{\hspost}@{}}%
\column{27}{@{}>{\hspre}l<{\hspost}@{}}%
\column{E}{@{}>{\hspre}l<{\hspost}@{}}%
\>[B]{}\Varid{app}_{lazy}{}\<[11]%
\>[11]{}\mathbin{::}\forall \Varid{v}\hsforall \;\sigma\hsdot{\circ }{.}{}\<[27]%
\>[27]{}(\Conid{Suspending}\;\Varid{v}\mathbin{<}\sigma,\Conid{Thunking}\;\Varid{v}\mathbin{<}\sigma,\Conid{Reading}\;[\mskip1.5mu \Varid{v}\mskip1.5mu]\mathbin{<}\sigma,{}\<[E]%
\\
\>[11]{}\hsindent{12}{}\<[23]%
\>[23]{}\Conid{Closure}\;\Varid{v}\mathbin{<:}\Varid{v},\Conid{Thunked}\;[\mskip1.5mu \Varid{v}\mskip1.5mu]\mathbin{<:}\Varid{v}){}\<[E]%
\\
\>[B]{}\hsindent{6}{}\<[6]%
\>[6]{}\Rightarrow \Conid{Tree}\;\sigma\;\Conid{Id}\;\Varid{v}\to \Conid{Tree}\;\sigma\;\Conid{Id}\;\Varid{v}\to \Conid{Tree}\;\sigma\;\Conid{Id}\;\Varid{v}{}\<[E]%
\\
\>[B]{}\Varid{app}_{lazy}\;\Varid{e}_{1}\;\Varid{e}_{2}\mathrel{=}\mathbf{do}{}\<[E]%
\\
\>[B]{}\hsindent{3}{}\<[3]%
\>[3]{}\Varid{vf}\leftarrow \Varid{e}_{1};\Varid{nv}\leftarrow \Varid{ask};\Varid{p}\leftarrow \Varid{thunk}\;\Varid{e}_{2}{}\<[E]%
\\
\>[B]{}\hsindent{3}{}\<[3]%
\>[3]{}\mathbf{let}\;\Varid{th}\mathbin{::}\Varid{v}\mathrel{=}\Varid{inj}_{\mathrm{v}}\mathbin{\$}\Conid{Thunked}\;\Varid{p}\;(\Varid{nv}\mathbin{::}[\mskip1.5mu \Varid{v}\mskip1.5mu]){}\<[E]%
\\
\>[B]{}\hsindent{3}{}\<[3]%
\>[3]{}\mathbf{case}\;\Varid{proj}_{\mathrm{v}}\;\Varid{vf}\;\mathbf{of}{}\<[E]%
\\
\>[3]{}\hsindent{2}{}\<[5]%
\>[5]{}\Conid{Just}\;(\Conid{Clos}\;\Varid{p'}\;\Varid{nv'})\to \mathbf{do}{}\<[E]%
\\
\>[5]{}\hsindent{2}{}\<[7]%
\>[7]{}\Varid{local}\;(\Varid{const}\;(\Varid{th}\mathbin{:}(\Varid{nv'}\mathbin{::}[\mskip1.5mu \Varid{v}\mskip1.5mu])))\;(\Varid{enact}\;\Varid{p'}){}\<[E]%
\ColumnHook
\end{hscode}\resethooks

\subsection{Call-by-Name}

The three operations for call-by-name evaluation are similar to those for 
call-by-need, but without memoizing the result.
\begin{hscode}\SaveRestoreHook
\column{B}{@{}>{\hspre}l<{\hspost}@{}}%
\column{3}{@{}>{\hspre}l<{\hspost}@{}}%
\column{7}{@{}>{\hspre}l<{\hspost}@{}}%
\column{10}{@{}>{\hspre}l<{\hspost}@{}}%
\column{E}{@{}>{\hspre}l<{\hspost}@{}}%
\>[B]{}\Varid{abs}_{cbn}{}\<[10]%
\>[10]{}\mathbin{::}\forall \Varid{v}\hsforall \;\sigma\hsdot{\circ }{.}(\Conid{Reading}\;[\mskip1.5mu \Varid{v}\mskip1.5mu]\mathbin{<}\sigma,\Conid{Suspending}\;\Varid{v}\mathbin{<}\sigma,\Conid{Closure}\;\Varid{v}\mathbin{<:}\Varid{v}){}\<[E]%
\\
\>[10]{}\Rightarrow \Conid{Tree}\;\sigma\;\Conid{Id}\;\Varid{v}\to \Conid{Tree}\;\sigma\;\Conid{Id}\;\Varid{v}{}\<[E]%
\\
\>[B]{}\Varid{abs}_{cbn}\;\Varid{body}\mathrel{=}\mathbf{do}{}\<[E]%
\\
\>[B]{}\hsindent{3}{}\<[3]%
\>[3]{}\Varid{nv}{}\<[7]%
\>[7]{}\leftarrow \Varid{ask}{}\<[E]%
\\
\>[B]{}\hsindent{3}{}\<[3]%
\>[3]{}\Varid{p}{}\<[7]%
\>[7]{}\leftarrow \Varid{suspend}\;\Varid{body}{}\<[E]%
\\
\>[B]{}\hsindent{3}{}\<[3]%
\>[3]{}\Varid{return}\;(\Varid{inj}_{\mathrm{v}}\;(\Conid{Clos}\;\Varid{p}\;(\Varid{nv}\mathbin{::}[\mskip1.5mu \Varid{v}\mskip1.5mu]))){}\<[E]%
\ColumnHook
\end{hscode}\resethooks
\begin{hscode}\SaveRestoreHook
\column{B}{@{}>{\hspre}l<{\hspost}@{}}%
\column{3}{@{}>{\hspre}l<{\hspost}@{}}%
\column{5}{@{}>{\hspre}l<{\hspost}@{}}%
\column{6}{@{}>{\hspre}l<{\hspost}@{}}%
\column{10}{@{}>{\hspre}l<{\hspost}@{}}%
\column{29}{@{}>{\hspre}l<{\hspost}@{}}%
\column{E}{@{}>{\hspre}l<{\hspost}@{}}%
\>[B]{}\Varid{var}_{cbn}{}\<[10]%
\>[10]{}\mathbin{::}\forall \Varid{v}\hsforall \;\sigma\hsdot{\circ }{.}(\Conid{Reading}\;[\mskip1.5mu \Varid{v}\mskip1.5mu]\mathbin{<}\sigma,\Conid{Suspending}\;\Varid{v}\mathbin{<}\sigma,\Conid{Suspended}\;[\mskip1.5mu \Varid{v}\mskip1.5mu]\mathbin{<:}\Varid{v}){}\<[E]%
\\
\>[B]{}\hsindent{6}{}\<[6]%
\>[6]{}\Rightarrow \Conid{Int}\to \Conid{Tree}\;\sigma\;\Conid{Id}\;\Varid{v}{}\<[E]%
\\
\>[B]{}\Varid{var}_{cbn}\;{}\<[10]%
\>[10]{}\Varid{x}\mathrel{=}\mathbf{do}{}\<[E]%
\\
\>[B]{}\hsindent{3}{}\<[3]%
\>[3]{}\Varid{nv}\leftarrow \Varid{ask}{}\<[E]%
\\
\>[B]{}\hsindent{3}{}\<[3]%
\>[3]{}\mathbf{let}\;\Varid{v}\mathrel{=}(\Varid{nv}\mathbin{::}[\mskip1.5mu \Varid{v}\mskip1.5mu])\mathbin{!!}\Varid{x}{}\<[E]%
\\
\>[B]{}\hsindent{3}{}\<[3]%
\>[3]{}\mathbf{case}\;\Varid{proj}_{\mathrm{v}}\;\Varid{v}\;\mathbf{of}{}\<[E]%
\\
\>[3]{}\hsindent{2}{}\<[5]%
\>[5]{}\Conid{Just}\;(\Conid{Suspended}\;\Varid{p}\;\Varid{nv'}){}\<[29]%
\>[29]{}\to \Varid{local}\;(\Varid{const}\;(\Varid{nv'}\mathbin{::}[\mskip1.5mu \Varid{v}\mskip1.5mu]))\;(\Varid{enact}\;\Varid{p}){}\<[E]%
\\
\>[3]{}\hsindent{2}{}\<[5]%
\>[5]{}\Conid{Nothing}{}\<[29]%
\>[29]{}\to \Varid{return}\;\Varid{v}{}\<[E]%
\ColumnHook
\end{hscode}\resethooks
\begin{hscode}\SaveRestoreHook
\column{B}{@{}>{\hspre}l<{\hspost}@{}}%
\column{3}{@{}>{\hspre}l<{\hspost}@{}}%
\column{5}{@{}>{\hspre}l<{\hspost}@{}}%
\column{6}{@{}>{\hspre}l<{\hspost}@{}}%
\column{7}{@{}>{\hspre}l<{\hspost}@{}}%
\column{10}{@{}>{\hspre}l<{\hspost}@{}}%
\column{26}{@{}>{\hspre}l<{\hspost}@{}}%
\column{E}{@{}>{\hspre}l<{\hspost}@{}}%
\>[B]{}\Varid{app}_{cbn}{}\<[10]%
\>[10]{}\mathbin{::}\forall \Varid{v}\hsforall \;\sigma\hsdot{\circ }{.}{}\<[26]%
\>[26]{}(\Conid{Suspending}\;\Varid{v}\mathbin{<}\sigma,\Conid{Reading}\;[\mskip1.5mu \Varid{v}\mskip1.5mu]\mathbin{<}\sigma,{}\<[E]%
\\
\>[26]{}\Conid{Closure}\;\Varid{v}\mathbin{<:}\Varid{v},\Conid{Suspended}\;[\mskip1.5mu \Varid{v}\mskip1.5mu]\mathbin{<:}\Varid{v}){}\<[E]%
\\
\>[B]{}\hsindent{6}{}\<[6]%
\>[6]{}\Rightarrow \Conid{Tree}\;\sigma\;\Conid{Id}\;\Varid{v}\to \Conid{Tree}\;\sigma\;\Conid{Id}\;\Varid{v}\to \Conid{Tree}\;\sigma\;\Conid{Id}\;\Varid{v}{}\<[E]%
\\
\>[B]{}\Varid{app}_{cbn}\;\Varid{e}_{1}\;\Varid{e}_{2}\mathrel{=}\mathbf{do}{}\<[E]%
\\
\>[B]{}\hsindent{3}{}\<[3]%
\>[3]{}\Varid{vf}\leftarrow \Varid{e}_{1};\Varid{nv}\leftarrow \Varid{ask};\Varid{p}\leftarrow \Varid{suspend}\;\Varid{e}_{2}{}\<[E]%
\\
\>[B]{}\hsindent{3}{}\<[3]%
\>[3]{}\mathbf{let}\;\Varid{th}\mathbin{::}\Varid{v}\mathrel{=}\Varid{inj}_{\mathrm{v}}\mathbin{\$}\Conid{Suspended}\;\Varid{p}\;(\Varid{nv}\mathbin{::}[\mskip1.5mu \Varid{v}\mskip1.5mu]){}\<[E]%
\\
\>[B]{}\hsindent{3}{}\<[3]%
\>[3]{}\mathbf{case}\;\Varid{proj}_{\mathrm{v}}\;\Varid{vf}\;\mathbf{of}{}\<[E]%
\\
\>[3]{}\hsindent{2}{}\<[5]%
\>[5]{}\Conid{Just}\;(\Conid{Clos}\;\Varid{p'}\;\Varid{nv'})\to \mathbf{do}{}\<[E]%
\\
\>[5]{}\hsindent{2}{}\<[7]%
\>[7]{}\Varid{local}\;(\Varid{const}\;(\Varid{th}\mathbin{:}(\Varid{nv'}\mathbin{::}[\mskip1.5mu \Varid{v}\mskip1.5mu])))\;(\Varid{enact}\;\Varid{p'}){}\<[E]%
\ColumnHook
\end{hscode}\resethooks

\section{A Latent Effect Semantics of Staging}
\label{app:latent-staging}

In this appendix we show how to model three staging operations, inspired by MetaML~\cite{TahaS2000}: \emph{quote} for staging a piece of code; \emph{unquote} for unstaging; and \emph{push} for defining code with one or more variables whose bindings only become known when the code is unquoted.
First, we define a small library of latent effects to work with.

\subsection{Reading, Suspending, and Revisiting Lambdas}
\label{sec:effect-library}

Appendix~\ref{app:effect-lib} defines the two necessary effect interfaces: 
\ensuremath{\Conid{Reading}} and \ensuremath{\Conid{Suspending}}.
Furthermore, the following \ensuremath{\Conid{Bindable}} type class has operations for binding a value in an environment, and looking up a value in an environment.
\begin{hscode}\SaveRestoreHook
\column{B}{@{}>{\hspre}l<{\hspost}@{}}%
\column{3}{@{}>{\hspre}l<{\hspost}@{}}%
\column{14}{@{}>{\hspre}l<{\hspost}@{}}%
\column{E}{@{}>{\hspre}l<{\hspost}@{}}%
\>[B]{}\mathbf{class}\;\Conid{Bindable}\;\Varid{r}\;\Varid{v}\;\mathbf{where}{}\<[E]%
\\
\>[B]{}\hsindent{3}{}\<[3]%
\>[3]{}\Varid{bindVal}{}\<[14]%
\>[14]{}\mathbin{::}\Varid{v}\to \Varid{r}\to \Varid{r}{}\<[E]%
\\
\>[B]{}\hsindent{3}{}\<[3]%
\>[3]{}\Varid{lookupVal}{}\<[14]%
\>[14]{}\mathbin{::}\Varid{r}\to \Conid{Int}\to \Varid{v}{}\<[E]%
\ColumnHook
\end{hscode}\resethooks
For plain, de Bruijn indexed environments, these functions are straightforwardly defined:
\begin{hscode}\SaveRestoreHook
\column{B}{@{}>{\hspre}l<{\hspost}@{}}%
\column{3}{@{}>{\hspre}l<{\hspost}@{}}%
\column{14}{@{}>{\hspre}l<{\hspost}@{}}%
\column{E}{@{}>{\hspre}l<{\hspost}@{}}%
\>[B]{}\mathbf{instance}\;\Conid{Bindable}\;[\mskip1.5mu \Varid{v}\mskip1.5mu]\;\Varid{v}\;\mathbf{where}{}\<[E]%
\\
\>[B]{}\hsindent{3}{}\<[3]%
\>[3]{}\Varid{bindVal}{}\<[14]%
\>[14]{}\mathrel{=}(\mathbin{:}){}\<[E]%
\\
\>[B]{}\hsindent{3}{}\<[3]%
\>[3]{}\Varid{lookupVal}{}\<[14]%
\>[14]{}\mathrel{=}(\mathbin{!!}){}\<[E]%
\ColumnHook
\end{hscode}\resethooks
The challenge with code splicing is to allow spliced code to reference variables 
from the context that code is being spliced with.
We address this challenge by using a refined type of environment and \ensuremath{\Conid{Bindable}}.

\subsection{Quote and Unquote}
The \ensuremath{\Varid{quote}} operation is defined in  \ensuremath{\Conid{Suspending}} and \ensuremath{\Conid{Reading}}.
It thunks a subcomputation, remembering the environment that the code is supposed to be evaluated under, reusing the \ensuremath{\Conid{Suspended}} value from Section \ref{lazy} as the representation of a quoted piece of code:
\begin{hscode}\SaveRestoreHook
\column{B}{@{}>{\hspre}l<{\hspost}@{}}%
\column{3}{@{}>{\hspre}l<{\hspost}@{}}%
\column{7}{@{}>{\hspre}l<{\hspost}@{}}%
\column{8}{@{}>{\hspre}l<{\hspost}@{}}%
\column{E}{@{}>{\hspre}l<{\hspost}@{}}%
\>[B]{}\Varid{quote}{}\<[8]%
\>[8]{}\mathbin{::}\forall \Varid{r}\hsforall \;\Varid{v}\;\Varid{l}\;\sigma\hsdot{\circ }{.}(\Conid{Reading}\;\Varid{r}\mathbin{<}\sigma,\Conid{Suspending}\;\Varid{v}\mathbin{<}\sigma,\Conid{Suspended}\;\Varid{r}\mathbin{<:}\Varid{v}){}\<[E]%
\\
\>[8]{}\Rightarrow \Conid{Tree}\;\sigma\;\Conid{Id}\;\Varid{v}\to \Conid{Tree}\;\sigma\;\Conid{Id}\;\Varid{v}{}\<[E]%
\\
\>[B]{}\Varid{quote}\;\Varid{m}\mathrel{=}\mathbf{do}{}\<[E]%
\\
\>[B]{}\hsindent{3}{}\<[3]%
\>[3]{}\Varid{p}{}\<[7]%
\>[7]{}\leftarrow \Varid{suspend}\;\Varid{m}{}\<[E]%
\\
\>[B]{}\hsindent{3}{}\<[3]%
\>[3]{}\Varid{nv}{}\<[7]%
\>[7]{}\leftarrow \Varid{ask}{}\<[E]%
\\
\>[B]{}\hsindent{3}{}\<[3]%
\>[3]{}\Varid{return}\;(\Varid{inj}_{\mathrm{v}}\mathbin{\$}\Conid{Suspended}\;\Varid{p}\;(\Varid{nv}\mathbin{::}\Varid{r})){}\<[E]%
\ColumnHook
\end{hscode}\resethooks
The \ensuremath{\Varid{unquote}} operation forces a quoted (thunked) computation.
\begin{hscode}\SaveRestoreHook
\column{B}{@{}>{\hspre}l<{\hspost}@{}}%
\column{3}{@{}>{\hspre}l<{\hspost}@{}}%
\column{5}{@{}>{\hspre}l<{\hspost}@{}}%
\column{10}{@{}>{\hspre}l<{\hspost}@{}}%
\column{28}{@{}>{\hspre}l<{\hspost}@{}}%
\column{E}{@{}>{\hspre}l<{\hspost}@{}}%
\>[B]{}\Varid{unquote}{}\<[10]%
\>[10]{}\mathbin{::}\forall \Varid{r}\hsforall \;\Varid{v}\;\Varid{l}\;\sigma\hsdot{\circ }{.}(\Conid{Reading}\;\Varid{r}\mathbin{<}\sigma,\Conid{Suspending}\;\Varid{v}\mathbin{<}\sigma,\Conid{Suspended}\;\Varid{r}\mathbin{<:}\Varid{v}){}\<[E]%
\\
\>[10]{}\Rightarrow \Conid{Tree}\;\sigma\;\Conid{Id}\;\Varid{v}\to \Conid{Tree}\;\sigma\;\Conid{Id}\;\Varid{v}{}\<[E]%
\\
\>[B]{}\Varid{unquote}\;\Varid{m}\mathrel{=}\mathbf{do}{}\<[E]%
\\
\>[B]{}\hsindent{3}{}\<[3]%
\>[3]{}\Varid{v}\leftarrow \Varid{m}{}\<[E]%
\\
\>[B]{}\hsindent{3}{}\<[3]%
\>[3]{}\mathbf{case}\;\Varid{proj}_{\mathrm{v}}\;\Varid{v}\;\mathbf{of}{}\<[E]%
\\
\>[3]{}\hsindent{2}{}\<[5]%
\>[5]{}\Conid{Just}\;(\Conid{Suspended}\;\Varid{p}\;\Varid{nv}){}\<[28]%
\>[28]{}\to \Varid{local}\;(\Varid{const}\;(\Varid{nv}\mathbin{::}\Varid{r}))\;(\Varid{enact}\;\Varid{p}){}\<[E]%
\\
\>[3]{}\hsindent{2}{}\<[5]%
\>[5]{}\Conid{Nothing}{}\<[28]%
\>[28]{}\to \Varid{error}\;\text{\tt \char34 bad~unquote\char34}{}\<[E]%
\ColumnHook
\end{hscode}\resethooks

\subsection{Pushing and Splicing}
We will introduce an operation for defining code with variables whose bindings are not yet known.
This operation relies on the following type of \emph{staged environment}:
\begin{hscode}\SaveRestoreHook
\column{B}{@{}>{\hspre}l<{\hspost}@{}}%
\column{E}{@{}>{\hspre}l<{\hspost}@{}}%
\>[B]{}\mathbf{newtype}\;\Conid{Env}_s\;\Varid{v}\mathrel{=}\Conid{Env}_s\;\{\mskip1.5mu \Varid{unEnv}_s\mathbin{::}[\mskip1.5mu \Conid{Maybe}\;(\Conid{Either}\;\Varid{v}\;\Conid{Int})\mskip1.5mu]\mskip1.5mu\}{}\<[E]%
\ColumnHook
\end{hscode}\resethooks
The \ensuremath{\Conid{Maybe}} type is used to distinguish positions in the environment whose bindings are not yet known.
An unknown binding is represented as \ensuremath{\Conid{Nothing}}.
A known binding is \ensuremath{\Conid{Either}} an actual value \ensuremath{\Varid{v}}, or an integer representing a \emph{forward reference} to a value that occurs \ensuremath{\Varid{n}} positions further down the environment.
We use forward references because our staged environments behave in a way that is reminiscent of a telescope~\cite{DeBruijn1991189}: for a bound value \ensuremath{\Varid{v}} occurring in a staged environment such as \ensuremath{\Conid{Env}_s\;(\Conid{Just}\;(\Conid{Left}\;\Varid{v})\mathbin{:}\Varid{nv})}, the value \ensuremath{\Varid{v}} is closed under the substitutions in the environment \ensuremath{\Varid{nv}}, since \ensuremath{\Varid{nv}} contains ``contextual'' bindings that potential unknown bindings in \ensuremath{\Varid{v}} must be instantiated with.

The implementation of the \ensuremath{\Conid{Bindable}} interface below illustrates how this ``closing'' of values of type \ensuremath{\Conid{Val}} works:
\begin{hscode}\SaveRestoreHook
\column{B}{@{}>{\hspre}l<{\hspost}@{}}%
\column{11}{@{}>{\hspre}l<{\hspost}@{}}%
\column{E}{@{}>{\hspre}l<{\hspost}@{}}%
\>[B]{}\mathbf{data}\;\Conid{Val}{}\<[11]%
\>[11]{}\mathrel{=}\Conid{CloV}\;(\Conid{Closure}\;(\Conid{Env}_s\;\Conid{Val})){}\<[E]%
\\
\>[11]{}\mid \Conid{CodV}\;(\Conid{Suspended}\;(\Conid{Env}_s\;\Conid{Val})){}\<[E]%
\\
\>[11]{}\mid \Conid{IntV}\;\Conid{Int}{}\<[E]%
\ColumnHook
\end{hscode}\resethooks
\begin{hscode}\SaveRestoreHook
\column{B}{@{}>{\hspre}l<{\hspost}@{}}%
\column{3}{@{}>{\hspre}l<{\hspost}@{}}%
\column{5}{@{}>{\hspre}l<{\hspost}@{}}%
\column{12}{@{}>{\hspre}l<{\hspost}@{}}%
\column{14}{@{}>{\hspre}l<{\hspost}@{}}%
\column{26}{@{}>{\hspre}l<{\hspost}@{}}%
\column{32}{@{}>{\hspre}l<{\hspost}@{}}%
\column{39}{@{}>{\hspre}l<{\hspost}@{}}%
\column{42}{@{}>{\hspre}l<{\hspost}@{}}%
\column{E}{@{}>{\hspre}l<{\hspost}@{}}%
\>[B]{}\mathbf{instance}\;\Conid{Bindable}\;(\Conid{Env}_s\;\Conid{Val})\;\Conid{Val}\;\mathbf{where}{}\<[E]%
\\
\>[B]{}\hsindent{3}{}\<[3]%
\>[3]{}\Varid{bindVal}\;{}\<[14]%
\>[14]{}\Varid{v}\;{}\<[26]%
\>[26]{}(\Conid{Env}_s\;\Varid{nv})\mathrel{=}\Conid{Env}_s\mathbin{\$}\Conid{Just}{}\<[E]%
\\
\>[3]{}\hsindent{9}{}\<[12]%
\>[12]{}(\Conid{Right}\;(\Varid{length}\;\Varid{nv}))\mathbin{:}\Varid{nv}\plus [\mskip1.5mu \Conid{Just}\;(\Conid{Left}\;\Varid{v})\mskip1.5mu]{}\<[E]%
\\[\blanklineskip]%
\>[B]{}\hsindent{3}{}\<[3]%
\>[3]{}\Varid{lookupVal}\;{}\<[14]%
\>[14]{}(\Conid{Env}_s\;\Varid{nv})\;{}\<[26]%
\>[26]{}\Varid{x}\mathrel{=}\Varid{go}\;\Varid{nv}\;\Varid{x}{}\<[E]%
\\
\>[3]{}\hsindent{2}{}\<[5]%
\>[5]{}\mathbf{where}\;{}\<[12]%
\>[12]{}\Varid{go}\;(\Conid{Nothing}{}\<[32]%
\>[32]{}\mathbin{:}\Varid{nv})\;{}\<[39]%
\>[39]{}\mathrm{0}{}\<[42]%
\>[42]{}\mathrel{=}\Varid{error}\;\text{\tt \char34 quote~error\char34}{}\<[E]%
\\
\>[12]{}\Varid{go}\;(\Conid{Just}\;(\Conid{Right}\;\Varid{y}){}\<[32]%
\>[32]{}\mathbin{:}\Varid{nv})\;{}\<[39]%
\>[39]{}\mathrm{0}{}\<[42]%
\>[42]{}\mathrel{=}\Varid{lookupVal}\;(\Conid{Env}_s\;\Varid{nv})\;\Varid{y}{}\<[E]%
\\
\>[12]{}\Varid{go}\;(\Conid{Just}\;(\Conid{Left}\;\Varid{v}){}\<[32]%
\>[32]{}\mathbin{:}\Varid{nv})\;{}\<[39]%
\>[39]{}\mathrm{0}{}\<[42]%
\>[42]{}\mathrel{=}\Varid{cover}\;\Varid{v}\;(\Conid{Env}_s\;\Varid{nv}){}\<[E]%
\\
\>[12]{}\Varid{go}\;(\anonymous {}\<[32]%
\>[32]{}\mathbin{:}\Varid{nv})\;{}\<[39]%
\>[39]{}\Varid{x}{}\<[42]%
\>[42]{}\mid \Varid{x}\geq \mathrm{0}\mathrel{=}\Varid{lookupVal}\;(\Conid{Env}_s\;\Varid{nv})\;(\Varid{x}\mathbin{-}\mathrm{1}){}\<[E]%
\\
\>[12]{}\Varid{go}\;\Varid{nv}\;{}\<[39]%
\>[39]{}\Varid{x}{}\<[42]%
\>[42]{}\mathrel{=}\Varid{error}\;\text{\tt \char34 bad~index\char34}{}\<[E]%
\ColumnHook
\end{hscode}\resethooks
The \ensuremath{\Varid{bindVal}} function prepends an environment by a forward reference to a value that is appended to the back of the environment.
The \ensuremath{\Varid{lookupVal}} function follows forward references, and when it finds a value, it uses the \ensuremath{\Varid{cover}} function to ``close'' a value over the bindings in the tail of the environment (i.e., the bindings from the context of the bound value):
\begin{hscode}\SaveRestoreHook
\column{B}{@{}>{\hspre}l<{\hspost}@{}}%
\column{33}{@{}>{\hspre}l<{\hspost}@{}}%
\column{37}{@{}>{\hspre}l<{\hspost}@{}}%
\column{56}{@{}>{\hspre}l<{\hspost}@{}}%
\column{59}{@{}>{\hspre}l<{\hspost}@{}}%
\column{E}{@{}>{\hspre}l<{\hspost}@{}}%
\>[B]{}\Varid{cover}\mathbin{::}\Conid{Val}\to \Conid{Env}_s\;\Conid{Val}\to \Conid{Val}{}\<[E]%
\\
\>[B]{}\Varid{cover}\;(\Conid{CloV}\;(\Conid{Clos}\;\Varid{p}\;\Varid{nv'}))\;{}\<[33]%
\>[33]{}\Varid{nv}{}\<[37]%
\>[37]{}\mathrel{=}\Conid{CloV}\;(\Conid{Clos}\;{}\<[56]%
\>[56]{}\Varid{p}\;{}\<[59]%
\>[59]{}(\Varid{combine}\;\Varid{nv'}\;\Varid{nv})){}\<[E]%
\\
\>[B]{}\Varid{cover}\;(\Conid{CodV}\;(\Conid{Suspended}\;\Varid{p}\;\Varid{nv'}))\;{}\<[33]%
\>[33]{}\Varid{nv}{}\<[37]%
\>[37]{}\mathrel{=}\Conid{CodV}\;(\Conid{Suspended}\;{}\<[56]%
\>[56]{}\Varid{p}\;{}\<[59]%
\>[59]{}(\Varid{combine}\;\Varid{nv'}\;\Varid{nv})){}\<[E]%
\\
\>[B]{}\Varid{cover}\;(\Conid{IntV}\;\Varid{i})\;{}\<[33]%
\>[33]{}\Varid{nv}{}\<[37]%
\>[37]{}\mathrel{=}\Conid{IntV}\;\Varid{i}{}\<[E]%
\ColumnHook
\end{hscode}\resethooks
The \ensuremath{\Varid{combine}} function used by this \ensuremath{\Varid{cover}} function instantiates unknown bindings by bindings coming from the context:
\begin{hscode}\SaveRestoreHook
\column{B}{@{}>{\hspre}l<{\hspost}@{}}%
\column{3}{@{}>{\hspre}l<{\hspost}@{}}%
\column{5}{@{}>{\hspre}l<{\hspost}@{}}%
\column{18}{@{}>{\hspre}l<{\hspost}@{}}%
\column{26}{@{}>{\hspre}l<{\hspost}@{}}%
\column{43}{@{}>{\hspre}l<{\hspost}@{}}%
\column{51}{@{}>{\hspre}l<{\hspost}@{}}%
\column{E}{@{}>{\hspre}l<{\hspost}@{}}%
\>[B]{}\Varid{combine}\mathbin{::}\Conid{Env}_s\;\Varid{v}\to \Conid{Env}_s\;\Varid{v}\to \Conid{Env}_s\;\Varid{v}{}\<[E]%
\\
\>[B]{}\Varid{combine}\;(\Conid{Env}_s\;\Varid{nv}_{1})\;(\Conid{Env}_s\;\Varid{nv}_{2})\mathrel{=}\Conid{Env}_s\;(\Varid{go}\;\Varid{nv}_{1}\;\Varid{nv}_{2}){}\<[E]%
\\
\>[B]{}\hsindent{3}{}\<[3]%
\>[3]{}\mathbf{where}{}\<[E]%
\\
\>[3]{}\hsindent{2}{}\<[5]%
\>[5]{}\Varid{go}\;[\mskip1.5mu \mskip1.5mu]\;{}\<[26]%
\>[26]{}\Varid{nv}{}\<[51]%
\>[51]{}\mathrel{=}\Varid{nv}{}\<[E]%
\\
\>[3]{}\hsindent{2}{}\<[5]%
\>[5]{}\Varid{go}\;(\Conid{Nothing}{}\<[18]%
\>[18]{}\mathbin{:}\Varid{nv}_{1})\;{}\<[26]%
\>[26]{}[\mskip1.5mu \mskip1.5mu]{}\<[51]%
\>[51]{}\mathrel{=}\Conid{Nothing}\mathbin{:}\Varid{nv}_{1}{}\<[E]%
\\
\>[3]{}\hsindent{2}{}\<[5]%
\>[5]{}\Varid{go}\;(\Conid{Nothing}{}\<[18]%
\>[18]{}\mathbin{:}\Varid{nv}_{1})\;{}\<[26]%
\>[26]{}(\Conid{Nothing}{}\<[43]%
\>[43]{}\mathbin{:}\Varid{nv}_{2}){}\<[51]%
\>[51]{}\mathrel{=}\Conid{Nothing}\mathbin{:}\Varid{go}\;\Varid{nv}_{1}\;\Varid{nv}_{2}{}\<[E]%
\\
\>[3]{}\hsindent{2}{}\<[5]%
\>[5]{}\Varid{go}\;(\Conid{Nothing}{}\<[18]%
\>[18]{}\mathbin{:}\Varid{nv}_{1})\;{}\<[26]%
\>[26]{}(\Conid{Just}\;(\Conid{Left}\;\Varid{v}){}\<[43]%
\>[43]{}\mathbin{:}\Varid{nv}_{2}){}\<[51]%
\>[51]{}\mathrel{=}\Conid{Just}\;(\Conid{Left}\;\Varid{v})\mathbin{:}\Varid{go}\;\Varid{nv}_{1}\;\Varid{nv}_{2}{}\<[E]%
\\
\>[3]{}\hsindent{2}{}\<[5]%
\>[5]{}\Varid{go}\;(\Conid{Nothing}{}\<[18]%
\>[18]{}\mathbin{:}\Varid{nv}_{1})\;{}\<[26]%
\>[26]{}(\Conid{Just}\;(\Conid{Right}\;\Varid{n}){}\<[43]%
\>[43]{}\mathbin{:}\Varid{nv}_{2}){}\<[51]%
\>[51]{}\mathrel{=}\Conid{Just}\;(\Conid{Right}\;(\Varid{n}\mathbin{+}\Varid{length}\;\Varid{nv}_{1})){}\<[E]%
\\
\>[43]{}\mathbin{:}\Varid{go}\;\Varid{nv}_{1}\;\Varid{nv}_{2}{}\<[E]%
\\
\>[3]{}\hsindent{2}{}\<[5]%
\>[5]{}\Varid{go}\;(\Conid{Just}\;\Varid{v}{}\<[18]%
\>[18]{}\mathbin{:}\Varid{nv}_{1})\;{}\<[26]%
\>[26]{}\Varid{nv}_{2}{}\<[51]%
\>[51]{}\mathrel{=}\Conid{Just}\;\Varid{v}\mathbin{:}\Varid{go}\;\Varid{nv}_{1}\;\Varid{nv}_{2}{}\<[E]%
\ColumnHook
\end{hscode}\resethooks
The \ensuremath{\Varid{push}} operation introduces unknown bindings in the context of a subcomputation.
\begin{hscode}\SaveRestoreHook
\column{B}{@{}>{\hspre}l<{\hspost}@{}}%
\column{7}{@{}>{\hspre}l<{\hspost}@{}}%
\column{8}{@{}>{\hspre}l<{\hspost}@{}}%
\column{19}{@{}>{\hspre}l<{\hspost}@{}}%
\column{20}{@{}>{\hspre}l<{\hspost}@{}}%
\column{26}{@{}>{\hspre}l<{\hspost}@{}}%
\column{E}{@{}>{\hspre}l<{\hspost}@{}}%
\>[B]{}\Varid{push}{}\<[7]%
\>[7]{}\mathbin{::}\forall \Varid{v}\hsforall \;\sigma\hsdot{\circ }{.}(\Conid{Reading}\;(\Conid{Env}_s\;\Varid{v})\mathbin{<}\sigma){}\<[E]%
\\
\>[7]{}\hsindent{1}{}\<[8]%
\>[8]{}\Rightarrow \Conid{Int}\to \Conid{Tree}\;\sigma\;\Conid{Id}\;\Varid{v}\to \Conid{Tree}\;\sigma\;\Conid{Id}\;\Varid{v}{}\<[E]%
\\
\>[B]{}\Varid{push}\;\Varid{n}\;\Varid{m}\mathrel{=}\Varid{local}\;{}\<[19]%
\>[19]{}(\lambda (\Conid{Env}_s\;\Varid{nv}\mathbin{::}\Conid{Env}_s\;\Varid{v})\to {}\<[E]%
\\
\>[19]{}\hsindent{7}{}\<[26]%
\>[26]{}\Conid{Env}_s\mathbin{\$}\Varid{replicate}\;\Varid{n}\;\Conid{Nothing}\plus \Varid{nv})\;{}\<[E]%
\\
\>[19]{}\hsindent{1}{}\<[20]%
\>[20]{}\Varid{m}{}\<[E]%
\ColumnHook
\end{hscode}\resethooks
Finally, the \ensuremath{\Varid{splice}} operation is like \ensuremath{\Varid{unquote}}, but it \emph{combines} the environment in a \ensuremath{\Conid{Suspended}} value with the contextual environment of the \ensuremath{\Varid{splice}} operation, thereby instantiating variables whose bindings were unknown when the thunk was created.
\begin{hscode}\SaveRestoreHook
\column{B}{@{}>{\hspre}l<{\hspost}@{}}%
\column{3}{@{}>{\hspre}l<{\hspost}@{}}%
\column{5}{@{}>{\hspre}l<{\hspost}@{}}%
\column{9}{@{}>{\hspre}l<{\hspost}@{}}%
\column{28}{@{}>{\hspre}l<{\hspost}@{}}%
\column{31}{@{}>{\hspre}l<{\hspost}@{}}%
\column{E}{@{}>{\hspre}l<{\hspost}@{}}%
\>[B]{}\Varid{splice}{}\<[9]%
\>[9]{}\mathbin{::}\forall \Varid{r}\hsforall \;\Varid{v}\;\Varid{l}\;\sigma\hsdot{\circ }{.}({}\<[31]%
\>[31]{}\Conid{Reading}\;(\Conid{Env}_s\;\Varid{v})\mathbin{<}\sigma,\Conid{Suspending}\;\Varid{v}\mathbin{<}\sigma,{}\<[E]%
\\
\>[31]{}\Conid{Suspended}\;(\Conid{Env}_s\;\Varid{v})\mathbin{<:}\Varid{v}){}\<[E]%
\\
\>[9]{}\Rightarrow \Conid{Tree}\;\sigma\;\Conid{Id}\;\Varid{v}\to \Conid{Tree}\;\sigma\;\Conid{Id}\;\Varid{v}{}\<[E]%
\\
\>[B]{}\Varid{splice}\;\Varid{m}\mathrel{=}\mathbf{do}{}\<[E]%
\\
\>[B]{}\hsindent{3}{}\<[3]%
\>[3]{}\Varid{v}\leftarrow \Varid{m}{}\<[E]%
\\
\>[B]{}\hsindent{3}{}\<[3]%
\>[3]{}\mathbf{case}\;\Varid{proj}_{\mathrm{v}}\;\Varid{v}\;\mathbf{of}{}\<[E]%
\\
\>[3]{}\hsindent{2}{}\<[5]%
\>[5]{}\Conid{Just}\;(\Conid{Suspended}\;\Varid{p}\;\Varid{nv}){}\<[28]%
\>[28]{}\to \Varid{local}\;(\Varid{combine}\;(\Varid{nv}\mathbin{::}\Conid{Env}_s\;\Varid{v}))\;(\Varid{enact}\;\Varid{p}){}\<[E]%
\\
\>[3]{}\hsindent{2}{}\<[5]%
\>[5]{}\Conid{Nothing}{}\<[28]%
\>[28]{}\to \Varid{error}\;\text{\tt \char34 bad~unquote\char34}{}\<[E]%
\ColumnHook
\end{hscode}\resethooks

\subsection{The Difference between \ensuremath{\Varid{splice}} and \ensuremath{\Varid{push}}}

While it is straightforward to define latent effect handlers that \emph{delay} the evaluation of subtrees (which is exactly what \ensuremath{\Conid{Suspending}} and \ensuremath{\Conid{Thunking}} do), it is problematic to define handlers that ``hoist'' code that occurs in the computational subtree or the continuation of a tree.
The crux of the problem is that subtrees and continuations require a value and a latent effect context before we can inspect their \ensuremath{\Conid{Tree}}s.
Escape expressions in MetaML represent code that must be run \emph{before} returning a code value, and if escape expressions were to occur in computational subtrees of a \ensuremath{\Varid{quote}} operation, we would have to hoist the code out of the subtree.
Our \ensuremath{\Varid{push}} and \ensuremath{\Varid{splice}} operations work around this problem.
They allow us to write splicing code so that the effectful operations that compute code to be spliced occur in the ``right'' place in the latent effect tree.

The program in Figre \ref{fig:metaml-puzzle} justifies the need for the two different constructs to mimic escape operations.

\begin{figure}
\begin{minipage}{0.28\linewidth}
\begin{lstlisting}[language=ml,numbers=left,basicstyle=\footnotesize\ttfamily]
let T = fn e =>
   <fn x => x + ~e>
in <fn x => ~(T <x>)>
\end{lstlisting}
\end{minipage}
\hfill
\hspace{-30pt}%
\begin{minipage}{0.71\linewidth}
\centering\begin{hscode}\SaveRestoreHook
\column{B}{@{}>{\hspre}l<{\hspost}@{}}%
\column{3}{@{}>{\hspre}l<{\hspost}@{}}%
\column{5}{@{}>{\hspre}l<{\hspost}@{}}%
\column{7}{@{}>{\hspre}l<{\hspost}@{}}%
\column{9}{@{}>{\hspre}l<{\hspost}@{}}%
\column{13}{@{}>{\hspre}l<{\hspost}@{}}%
\column{15}{@{}>{\hspre}l<{\hspost}@{}}%
\column{E}{@{}>{\hspre}l<{\hspost}@{}}%
\>[3]{}\Varid{letbind}\;\mbox{\commentbegin T \commentend}(\Varid{abs'}\mbox{\commentbegin e \commentend}{}\<[E]%
\\
\>[3]{}\hsindent{4}{}\<[7]%
\>[7]{}(\Varid{letbind}\;\mbox{\commentbegin d \commentend}{}\<[E]%
\\
\>[7]{}\hsindent{2}{}\<[9]%
\>[9]{}(\Varid{push}\;\mbox{\commentbegin x \commentend}\mathrm{1}\;(\Varid{var}\;\mbox{\commentbegin e \commentend}\mathrm{1}))\;{}\<[E]%
\\
\>[7]{}\hsindent{2}{}\<[9]%
\>[9]{}(\Varid{quote}\;(\Varid{abs'}\mbox{\commentbegin x \commentend}{}\<[E]%
\\
\>[9]{}\hsindent{4}{}\<[13]%
\>[13]{}(\Varid{add}\;(\Varid{var}\;\mbox{\commentbegin x \commentend}\mathrm{0})\;(\Varid{splice}\;(\Varid{var}\;\mbox{\commentbegin d \commentend}\mathrm{1})))))))\;{}\<[E]%
\\
\>[3]{}\hsindent{2}{}\<[5]%
\>[5]{}(\Varid{letbind}\;\mbox{\commentbegin d' \commentend}(\Varid{push}\;\mbox{\commentbegin x \commentend}\mathrm{1}{}\<[E]%
\\
\>[5]{}\hsindent{4}{}\<[9]%
\>[9]{}(\Varid{app}\;{}\<[15]%
\>[15]{}(\Varid{var}\;\mbox{\commentbegin T \commentend}\mathrm{1})\;(\Varid{quote}\;(\Varid{var}\;\mbox{\commentbegin x \commentend}\mathrm{0}))))\;{}\<[E]%
\\
\>[5]{}\hsindent{2}{}\<[7]%
\>[7]{}(\Varid{quote}\;(\Varid{abs'}\;\mbox{\commentbegin x \commentend}(\Varid{splice}\;(\Varid{var}\;\mbox{\commentbegin d' \commentend}\mathrm{1}))))){}\<[E]%
\ColumnHook
\end{hscode}\resethooks
\end{minipage}
\caption{A MetaML program and its counterpart with \ensuremath{\Varid{quote}}, \ensuremath{\Varid{unquote}}, \ensuremath{\Varid{push}}, and \ensuremath{\Varid{splice}}.}
\label{fig:metaml-puzzle}
\end{figure}

The program in Figure \ref{fig:metaml-basic} of Section \ref{call-by-need-evaluation} only uses the \ensuremath{\Varid{splice}} construct.
In order to write programs that behave similarly to MetaML programs with escape expressions that occur under a binder inside staged code, we also need \ensuremath{\Varid{push}}.
In Figure \ref{fig:metaml-puzzle}, the MetaML program on the left (due to Taha \cite{Taha99thesis}) contains two escape expressions that both occur under function binders inside staged code.
The result of running the program is a value that is alpha-equivalent to the code value \lstinline[language=ml]{<fn x1 => (fn x2 => x2 + x1)>}.
The program on the right in Figure \ref{fig:metaml-puzzle} uses \ensuremath{\Varid{push}} to indicate how many quoted binders a given computation is scoped by.\footnote{We have also added variable names in \ensuremath{\mbox{\commentbegin comments \commentend}} in Figure \ref{fig:metaml-puzzle} to increase the readability of the de Bruijn indexed program.}
The \ensuremath{\Varid{push}} on line 3 on the right in Figure \ref{fig:metaml-puzzle} corresponds to the escape expression in line 2 on the left.
Likewise, the \ensuremath{\Varid{push}} on line 6 on the right corresponds to the escape expression in line 3 on the left.
The dynamic semantics of \ensuremath{\Varid{push}} creates an environment with ``holes'' that represent unknown bindings.
The dynamic semantics of \ensuremath{\Varid{splice}} automatically fills in the holes of environments in \ensuremath{\Varid{push}}ed code values, with bindings from the dynamic context of the \ensuremath{\Varid{splice}} expression.
This pushing and filling of holes makes the splicing in the program on the right behave similar to the (lexically-scoped) escape expressions in the program on the left.
This is to yield a code value similar to the code value returned by the MetaML program.

\end{document}